\documentclass[reprint,amsmath,amssymb,aps,prb,floats,english,showpacs]{revtex4-1}
\usepackage[T1]{fontenc}
\usepackage[latin9]{inputenc}
\usepackage{babel}
\usepackage{times}
\usepackage{mathptmx}
\usepackage{graphicx}% Include figure files
\usepackage{dcolumn}% Align TABLE columns on decimal point
\usepackage{bm}% bold math
\usepackage{multirow}
\usepackage{booktabs}

\begin{document}
%\begin{CJK*}[HL]{KS}{}

\preprint{PRB}

\title{Modulation frequency dependence of continuous-wave optically/electrically detected magnetic resonance}

%\author{Sang-Yun Lee (ÀÌ»óÀ±)}
\author{Sang-Yun Lee}
 \email{s.lee@physik.uni-stuttgart.edu}
 \altaffiliation[currently at: ]{3. Institute of Physics, University of Stuttgart, Pfaffenwaldring 57, 70569 Stuttgart, Germany}
%\author{Seoyoung Paik (¹éŒ­¿µ)}
\author{Seoyoung Paik}
\author{Dane R. McCamey}
\author{Christoph Boehme}
 \email{boehme@physics.utah.edu}
\affiliation{Department of Physics and Astronomy, University of Utah, 115 South
1400 East Rm 201, Salt Lake City, Utah 84112}
\date{\today}

\begin{abstract}
Continuous wave optically and electrically detected magnetic resonance spectroscopy (cwODMR/cwEDMR) allow the investigation of paramagnetic states involved in spin-dependent transitions, like recombination and transport. Although experimentally similar to conventional electron spin resonance (ESR), there exist limitations when applying models originally developed for ESR to observables (luminescence and electric current) of cwODMR and cwEDMR. Here we present closed-form solutions for the modulation frequency dependence of cwODMR and cwEDMR based on an intermediate pair recombination model and discuss ambiguities which arise when attempting to distinguish the dominant spin-dependent processes underlying experimental data. These include: 1) a large number of quantitatively different models cannot be differentiated, 2) signs of signals are determined not only by recombination, but also by other processes like dissociation, intersystem-crossing, pair generation, and even experimental parameter such as, modulation frequency, microwave power, and temperature, 3) radiative and non-radiative recombination cannot be distinguished due to the observed signs of cwODMR and cwEDMR experiments.
\end{abstract}
\maketitle
%\end{CJK*}
\section{Introduction}\label{cw_intro}
\begin{figure}
\centering\includegraphics[width=0.8\columnwidth]{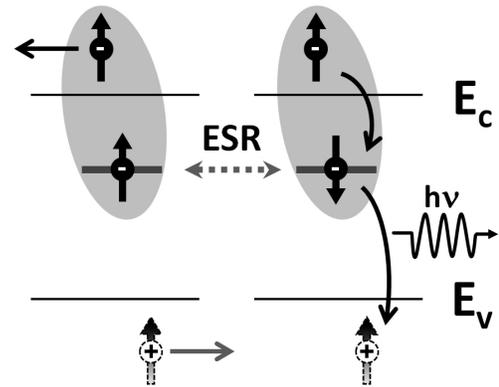}
\caption{\label{fig:fig0}Spin-dependent recombination via localized paramagnetic bandgap states. Excess charge carriers, electrons and holes can recombine via a localized paramagnetic state which acts as a recombination center. If a conduction electron and a unpaired electron at a paramagnetic recombination center form a spin singlet pair, the conduction electron can be captured by the recombination center. An electron at the recombination center can eventually recombine with a hole and create a photon. When they form a spin triplet pair, the capture probability of the conduction electron by the recombination center is low and excess carriers contribute to photocurrent. Because recombination process is dependent on mutual spin orientation, this recombination rate can be altered by ESR when they are weakly coupled. Thus ESR can alter recombination rate which results in change in photoluminescence and photocurrent, and they can be detected optically and electrically, respectively.}
\end{figure}

Electron spin resonance (ESR) is a useful tool for the investigation of microscopic properties of paramagnetic states in a wide variety of materials. In conventional ESR experiments, the total polarization of the investigated spin ensemble is observed by the measurement of microwave absorption. In some materials, there are other observables which can be used to detect electron spin states. For instance, when electron spins control electronic transitions such as transport or recombination, macroscopic materials properties such as photoluminescence, electroluminescence or conductivity can change under spin resonance. Fig.~\ref{fig:fig0} depicts a conceptual process of spin-dependent recombination~\cite{Lepine:prb1972,Stutzmann:jncs2000} which can be detected by ODMR and EDMR. The advantage of these electrically detected magnetic resonance (EDMR) and optically detected magnetic resonance (ODMR) spectroscopies is that they are significantly more sensitive than conventional ESR (spin polarization is usually low), and provide direct insight regarding how paramagnetic states in semiconductors affect some of the technologically most widely used electrical and optical materials properties. ODMR has been used in a wide range of research areas since its first invention~\cite{Geschwind:prl1959,Brossel:prl1959}. ODMR and EDMR are about 8 to 9 orders more sensitive than ESR, they both are proven to have single spin sensitivity~\cite{Cavenett:ap1981,Street:prb1982,Depinna:pmb1982_1st,Lifshitz:arpc2004,Stutzmann:jncs2000}, and they both can directly link a paramagnetic center to a specific luminescence center~\cite{Cavenett:ap1981,Street:prb1982,Depinna:pmb1982_1st,Chen:tsf2000}. Thanks to these advantages, ODMR can be used to deconvolute unresolved, overlapping luminescence bands in semiconductors~\cite{Boulitrop:prb1983}. EDMR provides information about electronically active paramagnetic centers in a similar way, again with higher sensitivity than ESR~\cite{Stutzmann:jncs2000, McCamey:APL2006}. In the early stage (until about the 1980's), ODMR was mainly conducted on inorganic semiconductors to identify paramagnetic recombination centers and to investigate their spin-dependent processes~\cite{Cavenett:ap1981,Dunstan:jpc1979}. It played an important role in investigating spin-dependent processes especially in amorphous silicon (a-Si) and revealed a variety of defect states which influence recombination in a-Si~\cite{Boulitrop:prb1983,Street:prb1982,Depinna:pmb1982_1st,Depinna:pmb1982_2nd,Morigaki:ssc1978,Lenahan:prb1984,Dersch:prb1983}.

Continuous wave ODMR and EDMR (cwODMR and cwEDMR) have been used in a wide range of research fields: they have been used to investigate spin-dependent transitions involving phosphorous donors in crystalline silicon~\cite{McCamey:APL2006,Morishita:prb2009}, trapping centers and their recombination dynamics in nanocrystals~\cite{Lifshitz:cpl1997,Lifshitz:jpc2000,Langof:jpcb2002,Lifshitz:arpc2004}, transport and recombination in microcrystalline hydrogenated silicon~\cite{Kanschat:jncs2000}, GaN~\cite{Glaser:prb1995,Glaser:prb2003}, and SiC~\cite{Son:prb1997}, and spin-dependent recombination in nitrogen vacancy centers in diamond~\cite{Oort:jpc1988,Nazare:prb1995,Nizovtsev:pb2001}. Because  cwODMR and cwEDMR can be used to distinguish overlapping recombination bands and their dynamics in disordered materials, they have also been used to investigate (usually amorphous) organic semiconductors: cwODMR and cwEDMR have provided information about spin-pairs dominating electronic processes and their transitions in conducting polymers~\cite{Swanson:prl1990,Swanson:prb1991,Swanson:prb1992,Graupner:prl1996,Greenham:prb1996,Lane:prb1997,Silva:pb2001,List:prb2002,Osterbacka:jcp2003,lee:prl2005,Segal:prb2005,Yang:prb2008}, small molecules~\cite{Castro:oe2007,Li:prb2004,Castro:jncs2004}, and polymer or small molecule/fullerene blends~\cite{Hiromitsu:prb1999,Scharber:prb2003}. The effect of isotopic modification on magnetic field effects in organic semiconductors also has been observed by ODMR~\cite{Nguyen:nm2010}, and the intersystem-crossing time has been extracted from the modulation frequency dependence~\cite{Yang:prl2007}.

Experimentally, cwODMR and cwEDMR are similar to conventional ESR except that luminescence intensity and electric current are picked up instead of the microwave absorption. Two magnetic fields, a static field $B_{0}$ and oscillating field $B_{1}$, are applied to a sample with $B_{0} \perp B_{1}$. The frequency of the sinusoidal $B_{1}$ field is matched with the Larmor frequency of the paramagnetic center to satisfy the resonance condition. As for most ESR spectrometers, X-band ($\approx 9.7~GHz$) is used, a frequency in the microwave (MW) range. In the case of cwODMR, to allow for optical detection, optical or electrical excitation of electronic states is necessary. Depending on the excitation method, photoluminescence detected magnetic resonance (PLDMR) or electroluminescence detected magnetic resonance (ELDMR) can be performed. In the case of PLDMR, constant optical excitation is applied using, for example, a Laser, and the resulting photoluminescence (PL) is detected. To increase the signal to noise ratio, lock-in detection is oftentimes employed. Two different modulation methods can be used. One method involves modulation of the static magnetic field, $B_{0}$, as used for conventional cwESR, the other approach is based on the modulation of the  MW amplitude. Experimentally,  $B_{0}$ modulation has been found to give weaker signals than MW amplitude modulation~\cite{Cavenett:ap1981}. Square modulation of the microwaves at a fixed reference frequency is generally used. The PL intensity reflecting the varying MW amplitude is then fed into a lock-in amplifier, and both in-phase and out-of-phase signals are obtained. In some studies found in the literature~\cite{Swanson:prb1992,Kanschat:jncs2000,Segal:prb2005,Li:prb2004,Dyakonov:cp1998}, the out-of-phase signal is ignored, however, doing so can result in the loss of important information, as will be explained later.

\begin{figure}
\centering\includegraphics[width=1\columnwidth]{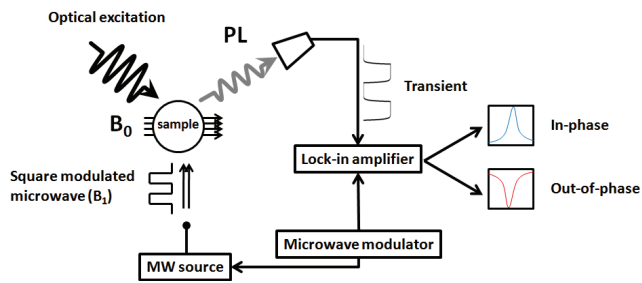}
\caption{\label{fig:setup}(Color online) Sketch of a cwODMR setup. The basic principle of cwODMR is the same as that of conventional ESR. Square microwave modulation can be used instead of $B_{0}$ field modulation and a lock-in amplifier is employed to increase the signal-to-noise ratio.}
\end{figure}

When the optical excitation is also modulated, a double modulated PLDMR (DMPLDMR) becomes possible~\cite{Segal:prb2005}. An experimental setup for a MW modulated ODMR experiment is shown in Fig.~\ref{fig:setup}. For EDMR, the optical detection is replaced by a current measurement. The metallic contacts needed for this, require a design that prevents the distortion of the MW field.

For both cwEDMR and cwODMR, the responses of the observables to the induced magnetic resonances are determined by the underlying electronic processes. The time scales on which these processes occur depend on various experimental parameters, such as excitation density~\cite{Depinna:pmb1982_2nd,Street:prb1982, Depinna:pmb1982_1st,Lifshitz:jpc1995,Lifshitz:cpl1997,List:prb2001,Li:prb2004,Dyakonov:prb1997} (or an injection current for EDMR~\cite{Swanson:prb1992,Lifshitz:jpc1993,Li:prb2004}), temperature~\cite{Street:prb1982,Boulitrop:prb1983,Swanson:prb1992,Li:prb2004}, and MW power (equivalently $B_{1}$ field strength)~\cite{Street:prb1982, Stich:jap1995,Hiromitsu:prb1999,Dunstan:jpc1979,Depinna:pmb1982_1st, Biegelsen:pmb1978, Xiong:apl1993,Langof:jpcb2002,Morishita:prb2009,Lifshitz:jpc2000,Lifshitz:cpl1997,Nazare:prb1995,Castro:oe2007,Yang:prl2007,Kawachi:jjap1997}. The dependencies of cwODMR and cwEDMR signals on these parameters can allow us to distinguish overlapping transitions and to understand their dynamics. For cwODMR, spectral information also can provide additional information for distinguishing overlapping luminescence bands~\cite{Street:prb1982, Boulitrop:prb1983,Nazare:prb1995}.

Another experimental parameter that can influence the observed cwODMR and cwEDMR signals is the modulation frequency, as the lock-in detected signals depend on the transient responses to the modulated MW~\cite{Street:prb1982,Boulitrop:prb1983,Depinna:pmb1982_2nd,Langof:jpcb2002,Baldo:prb2007}. Although its importance has been sometime discussed in conventional ESR studies~\cite{Murany:jmr2004,Simon:prb2007}, modulation frequency effects on cwODMR and cwEDMR have often been ignored in the literature, and, as a result, studies often reported results obtained using only one (or a small number) of modulation frequencies (usually the one which maximized the obtained signal). One can, however, find a number of reports showing modulation frequency dependencies. Different signals at different modulation frequencies were reported for the first time by Biegelsen et al~\cite{Biegelsen:pmb1978}. Other investigators have noticed that modulation frequency effects play an important role in the observed signal, which can change drastically as a function of the modulation frequency~\cite{Depinna:pmb1982_2nd,Boulitrop:prb1983,Cavenett:ap1981}. Qualitative reports of modulation frequency dependencies can be found in the early ODMR and EDMR literature~\cite{Cavenett:ap1981,Boulitrop:prb1983,Morigaki:ssc1978} which were sometimes used to identify the overlap of separate spin-dependent signals~\cite{Depinna:pmb1982_1st}. Even so, very little systematic research into modulation frequency effects was undertaken before the late 1990's, when research into this question became more common ~\cite{Lifshitz:jpc1993,Dyakonov:prb1997,List:prb2001,Langof:jpcb2002,Lifshitz:jpc2000,lee:prl2005,Nazare:prb1995,Graupner:prl1996,Yang:prl2007,Yang:prb2008,Segal:prb2005}.

A number of researchers have attempted to understand modulation frequency effects by developing rate models. Dunstan and Davies were the first to develop solutions for ODMR transients~\cite{Dunstan:jpc1979}. Next, Street and Depinna \textit{et al.} developed rate models and found transient solutions~\cite{Street:prb1982,Depinna:pmb1982_2nd}. Lenahan \textit{et al.} explained their observed modulation frequency dependence using a simple rate model described by only one time constant~\cite{Lenahan:prb1984}. A number of studies based on the steady-state solutions of such rate models have been reported~\cite{Xiong:apl1993,Stich:jap1995,Lifshitz:cpl1997,List:prb2001,Morishita:prb2009}. However, to understand the modulation frequency effects the exact solutions for the frequency dependence are necessary. There has been a number of efforts to find the solutions for modulation frequency dependence~\cite{Graupner:prl1996,Dyakonov:prb1997,Hiromitsu:prb1999,Langof:jpcb2002,Lifshitz:arpc2004,lee:prl2005,Segal:prb2005,Yang:prl2007,Yang:prb2008}. However, no closed form analytical solutions have been reported, and important aspects of modulation frequency effects remain not well understood. This has led to a number of debates regarding the underlying physical mechanisms of cwODMR and cwEDMR signals, because modulation frequency dependencies observed by different groups on similar systems have sometime led to completely different spin-dependent transition models. For example, the source of EDMR and ODMR signals seen in organic semiconductors has been attributed to both a spin-dependent polaron pair model~\cite{Yang:prl2007,Yang:prb2007,Yang:prb2008} and a triplet exciton-polaron quenching model~\cite{Segal:prb2005,lee:prl2005,Baldo:prb2007}.

Lock-in detected cwODMR and cwEDMR signals can be either positive or negative depending on the shapes of transient responses~\cite{Street:prb1982,Boulitrop:prb1983,Depinna:pmb1982_2nd,Langof:jpcb2002,Baldo:prb2007}. A variety of spin-dependent models have been developed based on the observed signs of cwODMR and cwEDMR signals as well as experimental parameters, like pair generation rates, temperature, MW power, and modulation frequency. Examples for such studies exist for a-si~\cite{Dersch:prb1983,Boulitrop:prb1983,Depinna:pmb1982_1st,Depinna:pmb1982_2nd,Biegelsen:pmb1978,Street:prb1982,Dunstan:jpc1979,Stutzmann:jncs2000,Morigaki:ssc1978,Depinna:prb1983,Boulitrop:prb1983}, InP nanocrystals~\cite{Langof:jpcb2002}, II-IV semiconductors~\cite{Lifshitz:cpl1998,Lifshitz:jpc2000}, $\mathrm{PbI_{2}}$ films~\cite{Lifshitz:jpc1995}, nanoparticles~\cite{Lifshitz:cpl1997}, and organic semiconductors~\cite{Graupner:prl1996,List:prb2001,Dyakonov:prb1997,Swanson:prb1992,lee:prl2005,List:prb2002,Silva:pb2001,Castro:jncs2004,Scharber:prb2003,Yang:prb2008}. For instance, it has been generally accepted that radiative and non-radiative recombination results in enhancement and quenching of cwODMR signal respectively~\cite{Depinna:pmb1982_1st,Lifshitz:jpc1995,Carlos:jcg1998,Lifshitz:cpl1997,Stutzmann:jncs2000}, and all recombination processes and all de-trapping processes result in quenching and enhancement of cwEDMR signals, respectively~\cite{Carlos:jcg1998,Stutzmann:jncs2000}. The qualitative explanation for signs of cwODMR signals is as following: spin resonance induces mixing between triplet and singlet pairs, and because initial states are generally dominated by triplet pairs due to the fast recombination of singlet pairs, the number of singlet pairs is increased at resonance. Thus, the  overall transition rate increases~\cite{Stutzmann:jncs2000}. Some studies even concluded that a certain channel is radiative or non-radiative, based on the sign of the ODMR signal~\cite{Street:prb1982,Depinna:pmb1982_2nd,Depinna:prb1983,Boulitrop:prb1983}. The idea here is that when a non-radiative recombination process is enhanced under spin-resonance, the competing optically detected radiative channels must be quenched.

The above examples show how critical it is to understand how MW modulation affects the observed cwODMR and cwEDMR signals. In this report, we employ the widely accepted spin-dependent transition model based on weakly coupled electron-hole pairs~\cite{Kaplan:jpl1978}, and find its closed-form analytical solutions. We then use this solution to explain how a broad range of electronic transitions, including recombination, dissociation, intersystem-crossing, pair generation, and spin-flips can affect the cwODMR and cwEDMR signals. We show how serious ambiguities related to the modulation frequency dependencies can arise, which make it difficult to determine the fundamental physical processes responsible for the observed cwEDMR or cwODMR frequency dependence. For example, extensive ODMR studies have been conducted on organic semiconductors to determine their dominant recombination processes. A variety of models have been suggested based on the observations of the signs of cwODMR and cwEDMR such as the singlet exciton-quenching model~\cite{List:prb2001,lee:prl2005,List:prb2002}, the triplet-triplet annihilation model~\cite{Dyakonov:prb1997}, the polaron-to-bipolaron decay~\cite{Swanson:prb1992,Silva:pb2001}, and the polaron pair recombination~\cite{Yang:prb2008}. We show that in many cases, the modulation frequency dependence cannot be used for such assignments, since the sign of these signals can be negative or positive for both, radiative or non-radiative processes.

\section{Models for the description of spin-dependent transition rates}\label{sec:rate}
The first quantitative model explaining spin-dependent recombination was suggested by Lepin~\cite{Lepine:prb1972} who described a thermal polarization model which predicted a relative change in photoconductivity of less than $10^{-6}$ at 300 K for X-band ESR. Microwave frequency and temperature dependencies were also predicted. However, it turned out that this model could neither explain the signal intensity of more than $10^{-3}$ that was observed in undoped a-Si:H at R.T.~\cite{Lepine:prb1972}, and the very weak dependencies on microwave frequency~\cite{Brandt:jncs1998} and temperature~\cite{Dersch:prb1983,Stutzmann:jncs1992}. These problems were soon resolved by another model developed by Kaplan, Solomon, and Mott (KSM model)~\cite{Kaplan:jpl1978}. In the KSM model, intermediate pair states exist prior to a spin-dependent transition and the spin pair states may recombine or dissociate. In addition, it is assumed that spin pairs in the triplet state can be annihilated only when one of pair partners is flipped by the spin-lattice relaxation process or the induced ESR, pairs dissociate otherwise. Thus, the recombination of triplet pairs happens only when they experience a transition to the singlet state.

In the past decades, a number of refinements were introduced to the KSM model, in which spin-spin interactions such as exchange and dipolar interactions exist within the pair, and spin-orbit coupling that is weak but not negligible is permitted such that weak triplet transitions become possible~\cite{Boehme:prb2003}. Because the intermediate pairs, consisting of two spins with s=1/2, can experience spin-spin interactions, the pair eigenbasis consists in general of two parallel states ($|T_{+}\rangle$ and $|T_{-}\rangle$) and two mixed states ($|2\rangle$ and $|3\rangle$) which change continuously from $|\uparrow\downarrow\rangle$ and $|\downarrow\uparrow\rangle$ to $|S\rangle$ and $|T_{0}\rangle$ respectively as the spin-spin interactions increase. ESR can induce transitions between the eigenstates of weakly coupled pairs such as $|T_{+}\rangle\leftrightarrow|\uparrow\downarrow\rangle$, $|T_{-}\rangle\leftrightarrow|\uparrow\downarrow\rangle$, $|T_{+}\rangle\leftrightarrow|\downarrow\uparrow\rangle$, and $|T_{-}\rangle\leftrightarrow|\downarrow\uparrow\rangle$. Thus, when the spin-spin interaction is weak, there can appear transitions among all four eigenstates and the transition probabilities are functions of the spin-spin interaction strength. Note that transitions of $|\downarrow\uparrow\rangle\leftrightarrow|\uparrow\downarrow\rangle$ are ESR forbidden but possible due to $T_{1}$ relaxation, and $|2\rangle\leftrightarrow|3\rangle$ transitions are possible via mixed relaxation processes. To understand the change of spin pair densities by ESR induced transitions, a mathematical approach will be given. Boehme and Lips have found the effective changes of spin densities by solving Louville equations describing the propagation of a spin ensemble during an ESR excitation~\cite{Boehme:prb2003}. The corresponding Hamiltonian is
\begin{equation}
\hat{H}=\mu_{B} g_{a} \hat{S}_{a}+\mu_{B} g_{b} \hat{S}_{b}-J \hat{S}_{b} \cdot \hat{S}_{b}-D^{d} [3 S^{z}_{a} S^{z}_{b} -\hat{S}_{b} \cdot \hat{S}_{b}]+\hat{H}_{1}
\end{equation}
where the first two terms correspond to the Zeeman terms of two pair partners, the third and fourth represent the exchange and dipolar couplings, respectively, and the last term is the alternating magnetic field. To describe the weakly coupled spin pair, the exchange and dipolar coupling constant, $J$ and $D^{d}$ respectively, are assumed to be smaller than the Larmor separation. The solutions (density matrix elements) for the corresponding Liouville equation can be found elsewhere~\cite{Boehme:prb2003}. The density changes of each spin state are then given by~\cite{Boehme:prb2003},
\begin{eqnarray}
\rho_{1,4}(\tau)&=&\rho^{0}_{1,4}\Delta^{\mathrm{u}}(\tau),\nonumber \\
\rho_{2,3}(\tau)&=&\rho^{0}_{2,3}\Delta^{\mathrm{v}}(\tau)\pm\rho^{0}_{2,3} \frac{J+D}{\hbar \omega_{\mathrm{\Delta}}} \Delta^{\mathrm{w}}(\tau)
\end{eqnarray}
where indices 1 and 4 represent the states $|T_{+}\rangle$ and $|T_{-}\rangle$ respectively, $\rho^{0}_{i}$ is the initial density, $J$ and $D$ are the exchange and dipolar coupling constants respectively, $\omega_{\Delta}$ represents the half of the frequency separation between the states $|2\rangle$ and $|3\rangle$. $\Delta^{u}(\tau)$, $\Delta^{v}(\tau)$, and $\Delta^{w}(\tau)$ represent the ESR duration time ($\tau$) dependencies. When the Larmor separation (which is the difference of the two Larmor frequencies within a pair) is larger than the applied $B_{1}$ field strength, only one pair partner can be flipped. In this case the $\tau$-dependencies become,
\begin{eqnarray}\label{pulse_time_dependence_parameter}
\Delta^{v}(\tau)&=&\frac{\gamma^2 B_{1}^{2}}{\Omega^{2}} \mathrm{sin^{2}}(\frac{\Omega \tau}{2}) \equiv \Delta(\tau),\nonumber \\
\Delta^{u}(\tau)&=&1-\Delta(\tau),\nonumber \\
\Delta^{w}(\tau)&=&0
\end{eqnarray}
where $\Omega=2\pi f_{Rabi}$ represents the Rabi frequency of the flipped pair partner. Therefore, the density changes of each eigenstates become
\begin{eqnarray}\label{pulse_density_change}
\rho_{1,4}(\tau)&=&\rho^{0}_{1,4} (1-\Delta(\tau)),\nonumber \\
\rho_{2,3}(\tau)&=&\rho^{0}_{2,3} \Delta(\tau).
\end{eqnarray}
Because either one of the states 2 or 3 is always involved in a possible transition among four eigenstates, any transition will cause a decrease or increase of $\rho_{2}$ or $\rho_{3}$. Density changes in state 2 and 3 are equivalent to density changes of singlet and triplet pair states. Therefore we don't need to deal with four state problems, instead two pair densities of singlet and triplet pairs are enough to describe recombination processes as long as any coherent spin motion is not of interest. Note that this is a valid statement because modulation frequency is typically not faster than the time scale of coherent spin motion so that all coherent phenomena will be averaged out. This is also the reason why all off-diagonal elements $\rho_{ij}$ for $i\neq j$ of the Louville density matrix can be neglected. Therefore, only the singlet and triplet pair densities, $n_{s}$ and $n_{t}$, will be considered in the following section.

\begin{figure}
\centering\includegraphics[width=1\columnwidth]{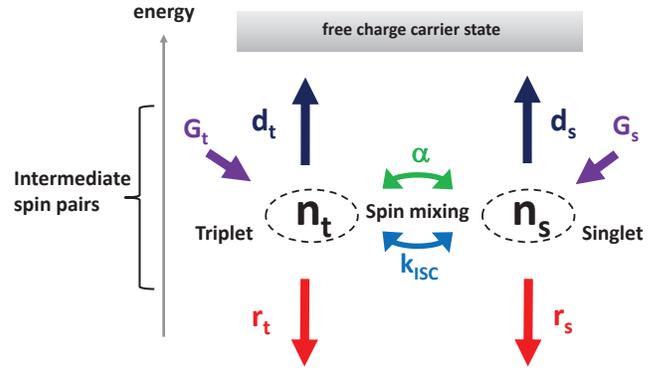}
\caption{\label{fig:ratemodel}(Color online) The intermediate pair recombination model (KSM) as relevant for cwODMR and cwEDMR. Triplet and singlet pairs are formed with two constant generation rates $G_{t}$ and $G_{s}$ respectively. Those pairs can dissociate into free charge carrier states with certain probabilities $d_{t}$ and $d_{s}$ (dissociation rates) or can recombine to singlet ground state with recombination rates $r_{t}$ and $r_{s}$. A spin mixing process can be introduced by ESR externally and this rate is described by $\alpha$. Another spin mixing process, intersystem-crossing is described by $k_{ISC}$. Note that $n_{t}$ and $n_{s}$ represent triplet and singlet pair densities, respectively. They do no necessarily correspond to eigenstate densities.}
\end{figure}

An illustration of the resulting spin pair rate model is given in Fig.~\ref{fig:ratemodel}. Prior to a spin pair transition to a singlet state, it is in the intermediate pair state. This pair is created with a certain rate, $G_{s}$ for a singlet pair and $G_{t}$ for a triplet pair. If this process is due to optical generation of electron-hole pairs and spin-orbit coupling is infinitely small, $G_{t}$ can be considered to be infinitely small. In the other case, if pair generation is achieved due to electrical injection of an electron and hole, $G_{t}/G_{s}$ becomes three, because a pair will be created with a random spin configuration. The pair can recombine to a singlet ground state with a recombination rate, $r_{s}$ for a singlet pair and $r_{t}$ for a triplet pair. This pair may dissociate into two free charge carriers without recombination. This happens at a dissociation rate, $d_{s}$ for a singlet pair and $d_{t}$ for a triplet pair. Before a pair recombines or dissociates, it can change its spin configuration from singlet to triplet or vice versa. This transition is possible via two spin mixing processes. One is intersystem-crossing, which is equivalent to a longitudinal spin relaxation process which can be defined as a ``radiationless transition between two electronic states having different spin multiplicities''~\cite{McNaught:book1997}. Among many processes, the spin-lattice relaxation is one of them which can cause the intersystem-crossing. The intersystem-crossing rate is described by $k_{ISC}$. The other process is ESR induced spin-mixing as can be seen from eqs.~(\ref{pulse_time_dependence_parameter}) and (\ref{pulse_density_change}). This ESR-induced transition rate is given by $\alpha$ which is proportional to the microwave power ($\propto B_{1}^{2}$) and dependent on the spin-spin interaction controlled oscillator strength of  the pair~\cite{Eickelkamp:mp1998}.

In the following section, a large number of quantitative models will be tested with analytical solutions for the observables of cwEDMR and cwODMR. Using realistic values for each transition probability, we consider experimentally relevant values for the cwODMR experiment. A wide range of transition rates have been reported. Examples include PL lifetimes in a-Si which span 11 orders of magnitude from $10^{-9}\,s$ to $10^{2}\,s$ ~\cite{Aoki:jncs2006}; bound pair decay (e-h pair dissociation) life times of $5\times 10^{-5}\,s$ in polymer-fullerene blends ~\cite{Mihailetch:afm2006}; fluorescence life times of $2\times 10^{-7}\,s$ and phosphorescence life times of $10^{-4}\,s$ in conjugated polymers~\cite{Romanovski:cp1999}; microsecond-millisecond time scales of recombination in nano-crystalline $\mathrm{TiO_{2}}$ thin films ~\cite{Tachibana:jpc1996}; radiative decay rates of $10^{6}\sim10^{7}\,s^{-1}$, non-radiative decay rates of $10^{9}\sim10^{10}\,s^{-1}$, dissociation rates of $10^{7}\,s^{-1}$ in organic semiconductors ~\cite{Nollau:tsf2000}, and a lower limit for the intersystem-crossing time of $10^{-5}\,s$ in organic semiconductors ~\cite{McCamey:prb2010}. In the following work, we vary the electronic transition rates, including recombination, dissociation, intersystem-crossing, and flip-flop, in the range between $10^{-4}$ and $10^{9} s^{-1}$ to cover as wide a range of experimentally observed parameters as possible.

\subsection{Rate equations\label{sec:rate_eq}}
CwODMR is fundamentally similar to conventional ESR spectroscopy - the one major modification is that the observable of ODMR is not the magnetization but the change in the number of photons induced by ESR. Generally, lock-in detected modulation of the $B_{0}$ or the $B_{1}$ field is used to enhance the resulting ODMR signal. For $B_{1}$ field modulation, square modulated microwaves are continuously applied, and the response to this excitation contains various harmonic frequency components. In the following we will focus on this kind of experiment.

Based on the rate model described in Section~\ref{sec:rate}, two coupled rate equations for the singlet and triplet pair densities can be written,
\begin{widetext}
\begin{equation}\label{eq:singlet_rate}
{\frac{dn_{s}}{dt}} = G_{\mathrm{s}}-C_{\mathrm{s}} n_{\mathrm{s}}+\alpha  ( n_{\mathrm{t}}-n_{\mathrm{s}} )-{k_{ISC} (n_{\mathrm{s}}- F n_{\mathrm{s}} )} +{k_{\mathrm{ISC}} (n_{\mathrm{t}}- (1-F) n_{\mathrm{t}}   )},
\end{equation}
\begin{equation}\label{eq:triplet_rate}
\frac {dn_{t}}{dt}=G_{\mathrm{t}}-C_{\mathrm{t}} n_{\mathrm{t}}+\alpha (n_{\mathrm{s}}-n_{\mathrm{t}} )-{k_{\mathrm{ISC}} (n_{\mathrm{t}}- (1-F) n_{\mathrm{t}} )} +{k_{\mathrm{ISC}} (n_{\mathrm{s}}- F n_{\mathrm{s}}   )},
\end{equation}
\end{widetext}
where $F$ is the Fermi-Dirac distribution function, $F=(1+e^{\frac{\Delta E}{k T }})^{-1}$, which approaches one at low temperature and 1/2 at high temperature and used to consider thermalization~\cite{Lifshitz:cpl1997,Langof:jpcb2002,Yang:prl2007}. $\Delta E$ has the order of Zeeman splitting. We chose F to be 0.25 in all numerical calculations to describe the two-level spin system which represents neither a complete thermalization nor a complete non-thermalization. It should be noted that $\alpha$ is turned on and off for each half cycle because of the square modulated microwave with frequency of 1/T. $C_{\textrm{s}}$ and $C_{\textrm{t}}$ are singlet and triplet pair annihilation rate coefficients respectively. They consist of recombination and dissociation rate coefficients, $C_{\textrm{s,t}}=r_{\textrm{s,t}}+d_{\textrm{s,t}}$. Some aspects with regard to radiative and non-radiative recombination rate coefficients should be mentioned: For radiative recombination, the spatial correlation between the electron and the hole influence the transition probability, so $r_{\textrm{t}}$ and $r_{\textrm{s}}$ depend on the separation between an electron and hole~\cite{Street1991,Kanschat:jncs2000}. Therefore, because the higher generation rate results in less separation, the radiative recombination probability is also a function of the generation rate. However, this effect will not be considered in this study, as we assume that the average separation is larger than the localization radii of electrons and holes. Note that this transition corresponds to the radiative tunneling in hydrogenated amorphous silicon~\cite{Street1991}. Non-radiative recombination includes all recombination processes which are not mediated by emission of photons, but phonons and hot carriers: phonon emission, Auger processes, surface and interface recombination, and recombination through defect states~\cite{Pankove:1971}. Non-radiative processes quench radiation efficiency in both organic semiconductors~\cite{Bradley:jpcm1981} and inorganic semiconductors~\cite{Pankove:1971}. As treated by List \textit{et al.}~\cite{List:prb2001} and Dyakonov et al.~\cite{Dyakonov:cp1998}, we consider both radiative and non-radiative recombination processes, and thus $C_{\textrm{s}}=(r_{\textrm{s}}+r_{\textrm{s,nr}} +d_{\textrm{s}})$ and $C_{\textrm{t}}=(r_{\textrm{t}}+r_{\textrm{t,nr}}+d_{\textrm{t}})$ where the subscript nr indicates non-radiative recombination.

Given the above definitions, the luminescent intensity and electric conductivity become
\begin{equation}\label{eq:PL}
I\propto r_{\mathrm{s}} n_{\mathrm{s}} +r_{\mathrm{t}} n_{\mathrm{t}},
\end{equation}
\begin{center}
and
\end{center}
\begin{equation}\label{eq:PC}
\sigma\propto d_{\mathrm{s}} n_{\mathrm{s}} +d_{\mathrm{t}} n_{\mathrm{t}},
\end{equation}
respectively. We note that electric conductivity is also determined by the carrier life time and mobility but ignored them because they are merely multiplied to the total dissociation rate (right term in eq.~(\ref{eq:PC}))~\cite{McCamey:prb2010} so that will not affect the time dependence nor modulation frequency dependence. Non-radiative recombination behaves as a pair annihilation process as other radiative recombination and dissociation, but it does not appear as proportionality constants in eq.~(\ref{eq:PL}) and (\ref{eq:PC}). In the following sections, only radiative recombination will be considered ($r_{\textrm{s,nr}}$, $r_{\textrm{t,nr}}=0$) for simplicity and the contributions of non-radiative recombination will be discussed in Section~\ref{sec:signs}. It shall be noted that there are many more complicated scenario for ODMR detected spin-dependent transitions conceivable, including ODMR signals due to non-radiative spin-dependent transitions which compete with non-spin dependent radiative processes. The stochastical description of these processes with rate equations is more complex but leaves the conclusions made in the following for directly detected radiative spin-dependent processes unchanged.

Rate equations similar to eq.~(\ref{eq:singlet_rate}) and (\ref{eq:triplet_rate}) can be found throughout the literature. However, usually only steady state solutions were found for the consideration of cwODMR and cwEDMR experiments~\cite{Wohlgenannt:prb2002,List:prb2001,Morishita:prb2009}. In some cases, only the time dependence was considered~\cite{Dunstan:jpc1979,Hiromitsu:prb1999,Depinna:pmb1982_2nd}. Modulation frequency dependence solutions have also been reported, but there have been no reports of closed-form analytical solutions. Some solutions reported in the literature were obtained from a simplified rate model~\cite{Lenahan:prb1984,Dyakonov:prb1997,Graupner:prl1996}, some solutions were based on the steady state~\cite{Segal:prb2005,lee:prl2005}, some solutions based on the rate model reported here were solely reported as numerical solutions~\cite{Yang:prb2008,Yang:prl2007,Langof:jpcb2002,Yang:prb2007,Lifshitz:arpc2004}, or the described observable was not the number of photons or electrons but total spin densities~\cite{Graupner:prl1996,Yang:prl2007,Yang:prb2008}. One solution given by Hiromitsu \textit{et al.} was based on an assumed steady state for the half cycle where the MW is off~\cite{Hiromitsu:prb1999}.

The rate equations corresponding to eq.~(\ref{eq:singlet_rate}) and (\ref{eq:triplet_rate}) are solved for the two separated time regions where the pulse is on and off, and the closed-form solutions can be explicitly expressed as:
\begin{equation}\label{eq:ns1}
n_{\mathrm{s}1}   ( t   ) =A_{11}{e}^{-m_{11}t}+A_{21}{e}^{-m_{21}t}+n_{\mathrm{s1}}^{0},
\end{equation}
\begin{equation}\label{eq:nt1}
n_{\mathrm{t}1}   ( t   ) =B_{11}{e}^{-m_{11}t}+B_{21}{e}^{-m_{21}t}+n_{\mathrm{t1}}^{0},
\end{equation}
\begin{equation}\label{eq:ns2}
n_{ \mathrm{s}2}   ( t   ) =A_{12}{e}^{-m_{12}   (t-{\frac{T}{2}}  )}+A_{22}{e}^{-m_{22}   (t-{\frac{T}{2}}  )}+n_{\mathrm{s2}}^{0},
\end{equation}
\begin{equation}\label{eq:nt2}
n_{ \mathrm{t}2}   ( t   ) =B_{12}{e}^{-m_{12}   (t-{\frac{T}{2}}  )}+B_{22}{e}^{-m_{22}   (t-{\frac{T}{2}}  )}+n_{\mathrm{t2}}^{0},
\end{equation}
where $n_{\textrm{s1}}$ and $n_{\textrm{t1}}$ are the singlet and triplet populations when the MW pulse is on, and $n_{\textrm{s2}}$ and $n_{\textrm{t2}}$ are the singlet and triplet populations when the MW pulse is off. Those solutions consist of double exponential functions as is often found in the literatures regarding pulsed experiments~\cite{McCamey:nm2008,Boehme:prb2003,Boehme:apl2001,McCamey:prb2010}.

The introduced constants in the above solutions are summarized below,
\begin{widetext}
\begin{eqnarray}
m_{1j}=\frac{C_{\mathrm{s}}+w_{1j}+C_{\mathrm{t}}+w_{2j}-\sqrt {   ( C_{\mathrm{s}}+w_{1j}-C_{\mathrm{t}}-w_{2j}   ) ^{2}+4w_{1j}w_{2j}}}{2},
\end{eqnarray}
\begin{eqnarray}\label{eq:cw_m2j}
m_{2j}=\frac{C_{\mathrm{s}}+w_{1j}+C_{\mathrm{t}}+w_{2j}+\sqrt {   ( C_{\mathrm{s}}+w_{1j}-C_{\mathrm{t}}-w_{2j}   ) ^{2}+4w_{1j}w_{2j}}}{2},
\end{eqnarray}
\end{widetext}
\begin{eqnarray}\label{eq:cw_ns0}
n_{\mathrm{s}j}^{0}={\frac {w_{2j}G_{\mathrm{t}}+   ( C_{\mathrm{t}}+w_{2j}   ) G_{\mathrm{s}}}{  ( C_{\mathrm{s}}+w_{1j}   )    ( C_{\mathrm{t}}+w_{2j}   ) -w_{1j}w_{2j}}},
\end{eqnarray}
\begin{eqnarray}\label{eq:cw_nt0}
n_{\mathrm{t}j}^{0}={\frac {w_{1j}G_{\mathrm{s}}+   ( C_{\mathrm{s}}+w_{1j}   ) G_{\mathrm{t}}}{  ( C_{\mathrm{s}}+w_{1j}   )    ( C_{\mathrm{t}}+w_{2j}   ) -w_{1j}w_{2j}}},
\end{eqnarray}
\begin{eqnarray}\label{eq:cw_wij}
w_{11}=\alpha+{k_{\mathrm{ISC}} (1-F)},\,w_{21}=\alpha+{k_{\mathrm{ISC}}\cdot F}
,\,\nonumber \\w_{12}={k_{\mathrm{ISC}} (1-F)},\,w_{22}={k_{\mathrm{ISC}} \cdot F},
\end{eqnarray}
where j=1 or 2. It should be noted that the exponents, $m_{1j}$ and $m_{2j}$, are independent on either the generation rates or the modulation frequency. It can be easily seen that $m_{2j}$ is decided by the fastest rate coefficient, but it is difficult to predict $m_{1j}$. However, it is clear that $m_{2j}$ is always larger than $m_{1j}$. Two constant terms, $n_{\textrm{s}j}^{0}$ and $n_{\textrm{t}j}^{0}$, are the steady-state solutions which the system assumes for very low modulation frequency~\cite{Wohlgenannt:prb2002,List:prb2001,Morishita:prb2009,Yang:prb2008,Yang:prl2007}. It should also be noted that the singlet and triplet pair populations will approach values at the end of each half cycle which are at the same time the initial values of the following half cycle. Therefore, the frequency dependence can be explained in terms of the differences between the populations at the end of each half cycle~\cite{Yang:prb2008,Yang:prl2007}, $n_{\textrm{s}1}(T/2)-n_{\textrm{s}2}(T)$ and $n_{\textrm{t}1}(T/2)-n_{\textrm{t}2}(T)$. However, lock-in detected signals are not simply decided by these quantities. The observables are not the population changes, but the changes in the number of photons, which incoprorates both the population change and the recombination probability.

\subsection{Boundary conditions}\label{BC_cw}
Because the spin populations assume the steady state only as the modulation frequency $f\rightarrow0$, the time dependent solutions must be solved to explain the transient behavior at arbitrary modulation frequencies. To find the exact solution, the expressions for the eight unknown coefficients $A_{ij}$ and $B_{ij}$ ($i,\, j=1\, or\, 2$) in eq.~(\ref{eq:ns1}), (\ref{eq:nt1}), (\ref{eq:ns2}), and (\ref{eq:nt2}) must be derived by application of eight boundary conditions. The boundary conditions used as well as the subsequent derivation of the analytic form of the coefficients are given in Appendix~\ref{app_BC}.

Equations (\ref{eq:cw_a22}), (\ref{eq:cw_a12}), (\ref{eq:a21}), (\ref{eq:a11}) represent exact and general analytical solutions for the singlet and triplet density functions during a cwODMR modulation cycle. We are thus in a position to determine the temporal evolution of the cwODMR observable.

\section{Transient behavior of cwODMR}
The observable in cwODMR is the emission rate of photons, and, as described in eq.~(\ref{eq:PL}), the time dependence can be obtained by adding the contribution from the singlet and triplet pair populations multiplied by the singlet and triplet recombination rate coefficients respectively. Thus,
\begin{eqnarray}\label{eq:pl1}
{ I}_{1}&=&  ( r_{\mathrm{s}}A_{11}+r_{\mathrm{t}}B_{11}  ) {e}^{-m_{11}t} \nonumber \\
&&+  ( r_{\mathrm{s}}A_{21}+r_{\mathrm{t}}B_{21}  ){e}^{-m_{21}t} \nonumber \\
&&+r_{\mathrm{s}}n_{\mathrm{s}1}^{0}+r_{\mathrm{t}}n_{\mathrm{t}1}^{0},
\end{eqnarray}
\begin{eqnarray}\label{eq:pl2}
{ I}_{2}&=&  ( r_{\mathrm{s}}A_{12}+r_{\mathrm{t}}B_{12}  ) {e}^{-m_{12} (t-\frac{T}{2} )}\nonumber \\
&&+  ( r_{\mathrm{s}}A_{22}+r_{\mathrm{t}}B_{22}  ){e}^{-m_{22} (t-\frac{T}{2} )}\nonumber \\
&&+r_{\mathrm{s}}n_{\mathrm{s}2}^{0}+r_{\mathrm{t}}n_{\mathrm{t}2}^{0}
\end{eqnarray}
\begin{figure}
\centering\includegraphics[width=1\columnwidth]{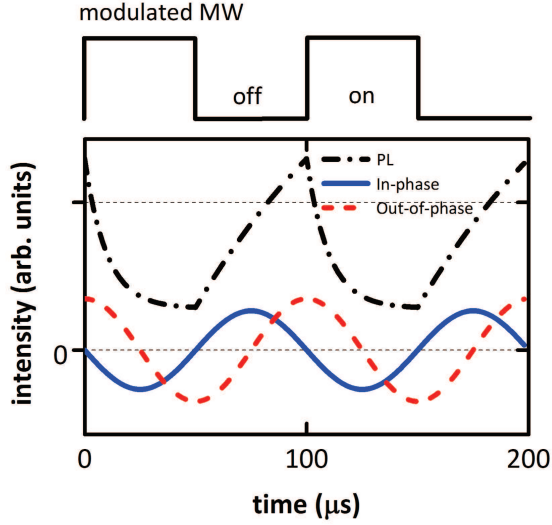}
\caption{(Color online) A time transient calculated from a numerical model described by a combination of parameters as $r_{\textrm{s}}=10^{4}~s^{-1}$, $r_{\textrm{t}}=10^{0}~s^{-1}$, $d_{\textrm{s}}=10^{2}~s^{-1}$, $d_{\textrm{t}}=10^{6}~s^{-1}$, $k_{\textrm{ISC}}=10^{-2}~s^{-1}$, $\alpha=10^{5}~s^{-1}$, $F=0.25$, $G_{\textrm{s}}=10^{23}~s^{-1}$, and $G_{\textrm{t}}=10^{20}~s^{-1}$. The dash-dotted curve shows the overall response obtained from eq.~(\ref{eq:pl1}) and (\ref{eq:pl2}). The blue solid and red dashed curves are the in-phase and the out-of-phase components described by $I_{\textrm{s}1}\sin( \frac {2\pi }{T} t)$ and $I_{\textrm{c}1}\cos(\frac {2\pi}{T} t )$, respectively. See detail in text.}
\label{fig:decomposition}
\end{figure}
where, $I_{1}$ and $I_{2}$ are the photon emission rates due to recombination of both singlet and triplets pairs when the pulse is on and off, respectively. The dash-dotted curve in Fig.~\ref{fig:decomposition} is a numerical example of the time dependence. Because $m_{1j}$ and $m_{2j}$ are always positive and $m_{2j}>m_{1j}$, the first and second terms in both eq.~(\ref{eq:pl1}) and (\ref{eq:pl2}) determine the slower and faster decay, respectively. It is difficult to predict which response will show an enhancement or quenching behavior because the overall response depends not only on $m_{1j}$ and $m_{2j}$ but also on $r_{s}A_{ij}+r_{t}B_{ij}$. Since the coefficients of all exponential terms have very complicated dependencies on a variety of parameters (see eq.~(\ref{eq:cw_a22}), (\ref{eq:cw_a12}), (\ref{eq:a21}), and (\ref{eq:a11})), it is clear that sign predictions depend on the magnitudes of many parameters at the same time. Using the above solution, we have been able to reproduce a wide variety of cwODMR transients reported in the literature~\cite{Dunstan:jpc1979,Langof:jpcb2002,Depinna:pmb1982_2nd,Street:prb1982,Boulitrop:prb1983}.

\section{Modulation frequency dependence}\label{cw_modulation_frequency_dependence}
The time dependence solutions, eq.~(\ref{eq:pl1}) and (\ref{eq:pl2}), are the collective responses to the modulated $B_{1}$ field over all frequency ranges. However, in experimental implementations which utilize a lock-in technique, only the component of the transient signal  which has the same frequency as the reference will be obtained. With lock-in quadrature detection, both the in- an out-of-phase components are available. While the out-of-phase components have often been ignored in the literature~\cite{Swanson:prb1992,Kanschat:jncs2000,Segal:prb2005,Li:prb2004,Dyakonov:cp1998}, we note that the out-of-phase components contain important information, which has been sometime addressed in longitudinally detected electron spin resonance~\cite{Murany:jmr2004,Simon:prb2007}.

The details of the modulation frequency dependence solutions are given in Appendix~\ref{app_frequency}. We are thus able to find an analytic expression for the in-phase and out-of-phase components of the transient during a MW modulated cwODMR experiment. These are given by
\begin{equation}\label{eq:in}
V_{\mathrm{in}}=\frac{V_{01}}{2}  \cos  ( \varphi_{1}  )=\frac{1}{2} I_{\mathrm{s}1},
\end{equation}
\begin{equation}\label{eq:out}
V_{\mathrm{out}}=\frac{V_{01}}{2}  \sin  ( \varphi_{1}  )=\frac{1}{2} I_{\mathrm{c}1}
\end{equation}
where $V_{01}$ is the magnitude of the first harmonic component, $I_{\textrm{s}1}$ and $I_{\textrm{c}1}$ are the amplitudes of the first sine and cosine components, and $\varphi_{1}=\tan ^{-1}( {\frac {{ I}_{{\textrm{c}1}}}{{ I}_{{\textrm{s}1}}}} )$ (see Appendix~\ref{app_frequency}).

Thus the in-phase and out-of-phase cwODMR signals are the Fourier coefficients of the lowest frequency sine and cosine terms of the Fourier series solution (eq.~(\ref{eq:fs})), respectively. Examples are shown in Fig.~\ref{fig:decomposition} to explain the decomposed in-phase and out-of-phase components of the time response. It should be noted that the cwEDMR solutions can also be obtained by replacing $r_{\textrm{s}}$ and $r_{\textrm{t}}$ in front of the exponential functions in eq.~(\ref{eq:pl1}) and (\ref{eq:pl2}) with $d_{\textrm{s}}$ and $d_{\textrm{t}}$ respectively as shown in eq.~(\ref{eq:PC})~\cite{McCamey:nm2008}.

Similarly the solutions for $B_{0}$-field modulated cwODMR and cwEDMR can be found in the same way as for microwave modulated cwODMR and cwEDMR. While the difference between these two modulation techniques is that the spin resonance is modulated by a square function and a harmonic function respectively, the lock-in detected observables are identical since the lock-in technique is sensitive to the lowest harmonic component only in either case.

\subsection{At low modulation frequency}
We use the low modulation frequency limit to check the solution of our model, by varifying that these solutions can explain the cwODMR response. From the solutions above, the low frequency behavior is seen to be
\begin{widetext}
\begin{eqnarray}\label{eq:low_freq_in}
V_{\mathrm{in, lf}}&=&\frac{(r_{\mathrm{s}}+r_{\mathrm{t}})(G_{\mathrm{t}}+G_{\mathrm{s}})\alpha+(r_{\mathrm{t}} r_{\mathrm{s}} + r_{\mathrm{s}} w_{22} + r_{\mathrm{t}} w_{12} ) (G_{\mathrm{t}} +G_{\mathrm{s}}) +r_{\mathrm{t}} d_{\mathrm{s}} G_{\mathrm{t}} + r_{\mathrm{s}} d_{\mathrm{t}} G_{\mathrm{s}}}{(C_{\mathrm{s}}+C_{\mathrm{t}})\alpha+(C_{\mathrm{s}} +w_{12} )(C_{\mathrm{t}} +w_{22})-w_{12} w_{22}}\cdot \frac{2}{\pi}
\nonumber \\
&&-\frac{(r_{\mathrm{t}} r_{\mathrm{s}} + r_{\mathrm{s}} w_{22} + r_{\mathrm{t}} w_{12} ) (G_{\mathrm{t}} +G_{\mathrm{s}}) +r_{\mathrm{t}} d_{\mathrm{s}} G_{\mathrm{t}} + r_{\mathrm{s}} d_{\mathrm{t}} G_{\mathrm{s}}}{(C_{\mathrm{s}} +w_{12} )(C_{\mathrm{t}} +w_{22})-w_{12} w_{22}}\cdot \frac{2}{\pi},
\end{eqnarray}
\end{widetext}
%\begin{widetext}
\begin{equation}\label{eq:low_freq_out}
V_{\mathrm{out, lf}}=0.
\end{equation}
%\end{widetext}
The out-of-phase component vanishes since the transient response can easily follow the slow modulation. The in-phase component shows a typical microwave power dependence: it vanishes at small power (when $\alpha\rightarrow0$) and it becomes saturated at high power (i.e. it has a non-zero constant value). The MW power dependencies of eqs.~(\ref{eq:low_freq_in}) and (\ref{eq:low_freq_out}) will be explained in the Section~\ref{cw_power}.

\subsection{Ambiguity of cwODMR measurements}
To understand the modulation frequency dependence of cwODMR, we inspected a large number of quantitative models.  There is an extremely large number of possible qualitative and quantitative relationships betwen the model parameters. To limit the number of cases that we inspected, we choose a number of relationships between these parameters. We considered that i) the triplet recombination coefficient is the smallest one among all the recombination and dissociation rate coefficients ($r_{\textrm{t}}<r_{\textrm{s}}$, $d_{\textrm{s}}$, $d_{\textrm{t}}$) (unless otherwise noted), and ii) the singlet dissociation rate coefficient is smaller than the triplet dissociation rate coefficient ($d_{\textrm{s}}<d_{\textrm{t}}$) which means that the singlet intermediate state is assumed to be energetically lower than the triplet intermediate state (unless otherwise noted). Under these assumptions, a large number of quantitative models were investigated by varying $r_{\textrm{t}}$, $r_{\textrm{s}}$, $d_{\textrm{s}}$, $d_{\textrm{t}}$, $k_{\textrm{ISC}}$, and $\alpha$ in the range from $10^{-4}$ to $10^{9}$ $s^{-1}$. We investigated almost a thousand different variations of the relationship between those parameters.
\begin{figure}
\centering\includegraphics[width=1\columnwidth]{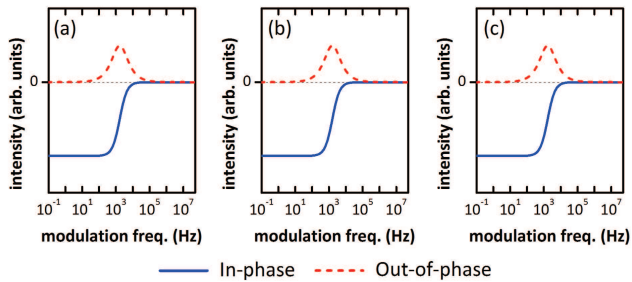}
\caption{\label{fig:ambiguity}(Color online) Three different quantitative models result in indistinguishable frequency dependencies. Each quantitative model is determined by a different set of parameters.
%(a) $r_{s}=10^{2}$, $r_{t}=10^{0}$, $d_{s}=10^{4}$, $d_{t}=10^{6}$, $k_{ISC}=10^{-2}$, $\alpha=10^{3}$, $F=0.25$, $G_{s}=10^{23}$, $G_{t}=10^{20}$, (b) $r_{s}=10^{4}$, $r_{t}=10^{-1}$, $d_{s}=10^{1}$, $d_{t}=10^{2}$, $k_{ISC}=10^{-2}$, $\alpha=10^{-1}$, $F=0.25$, $G_{s}=10^{25}$, $G_{t}=10^{20}$, (c) $r_{s}=10^{4}$, $r_{t}=10^{0}$, $d_{s}=10^{2}$, $d_{t}=10^{6}$, $k_{ISC}=10^{-2}$, $\alpha=10^{3}$, $F=0.25$, $G_{s}=1/3 \times 10^{20}$, $G_{t}=10^{20}$.
Refer to Table~\ref{tab:all_parameters} for all used values.}
\end{figure}

After looking through these cases, we find that it is almost impossible to distinguish some of the quantitative models based on their modulation frequency behaviors. Fig.~\ref{fig:ambiguity} illustrates this ambiguity. Figure~\ref{fig:ambiguity} (a), (b), and (c) show nearly identical frequency dependencies of three very different quantitative models. The frequencies at which the in-phase signals have their maximum slope and the out-of-phase signals show their local maximum values are almost identical, and their shapes are also indistinguishable. The patterns shown in Fig.~\ref{fig:ambiguity} represent in fact the most common frequency dependency that we have found out by the tested quantitative models. This illustrates the difficulty in extracting correct values for the corresponding coefficients from a simple frequency dependence - one can find a wide range of values which can reproduce it. This ambiguity is one of the most significant disadvantages of cwODMR or cwEDMR. It puts many interpretations of cwODMR data reported in the literature in question.

\begin{figure}
\centering\includegraphics[width=1\columnwidth]{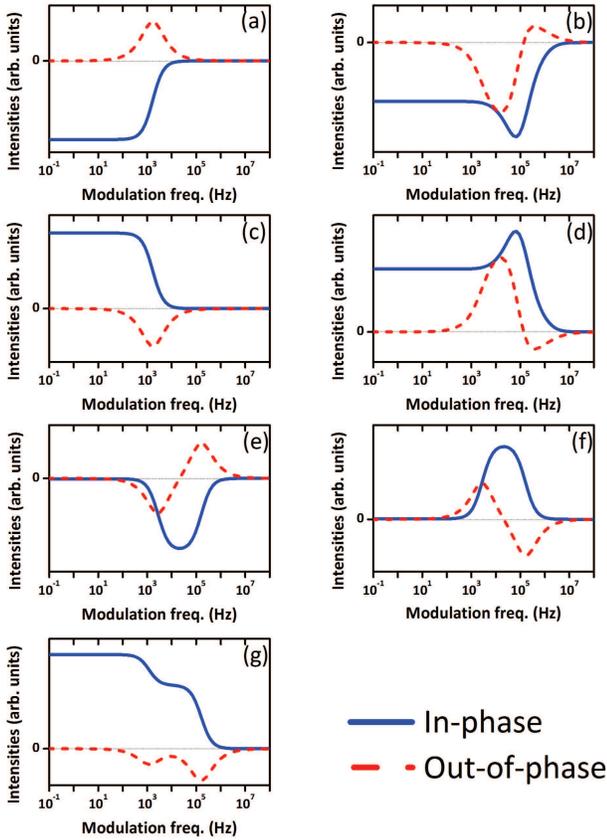}
\caption{\label{fig:patterns}(Color online) Seven distinguishable patterns of the modulation frequency dependence of cwODMR have been found out of almost a thousand quantitative models. (c), (d), and (f) are equivalent with (a), (b), and (e), respectively, but with opposite signs. The parameters used for this data are listed in Table~\ref{tab:all_parameters}.}
\end{figure}

Of the nearly thousand models we tested, we were able to describe them all with only seven frequency dependency patterns. These are shown in Fig.~\ref{fig:patterns}. We find that those patterns are determined mostly by the recombination rate coefficients, the microwave power, the spin mixing rates, as well as the generation rates. How each parameter influences the frequency dependence will be discussed in the following sections. The most trivial cases, seen in Fig.~\ref{fig:patterns} (a) and (c), will be discussed first.

\subsection{Trivial case (small spin mixing rates)\label{sec:simple}}
To understand the behavior of the response to the modulation frequency, the trivial patterns will be discussed. ``Trivial'' means that the spin mixing rates, both $k_{\textrm{ISC}}$ and $\alpha$ are negligible when compared to all the other rates. In this case, only the spin pair annihilation processes determined by the recombination and dissociation rate coefficients become dominant.  All the patterns in Fig.~\ref{fig:ambiguity} as well as the patterns in Fig.~\ref{fig:patterns} (a) and (c) are obtained under the assumption of insignificant spin mixing rates, $k_{\textrm{ISC}}$ and $\alpha$. The pattern in Fig.~\ref{fig:patterns} (c) is identical to the one in (a), but inverted due to different ratios between $G_{s}$ and $G_{t}$. We found that the sign of the lock-in detected signal depends on almost all transition processes as one can deduce from Table~\ref{tab:all_parameters}.

\begin{figure}
\centering\includegraphics[width=1\columnwidth]{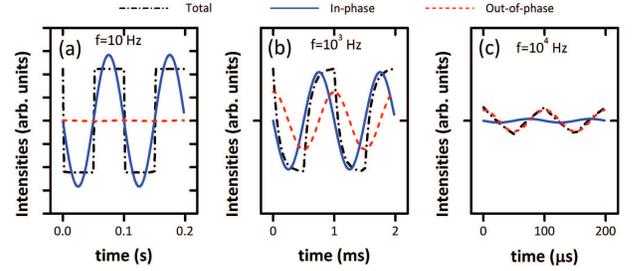}
\caption{\label{fig:transient_change}(Color online) Calculated transient behaviors at different modulation frequencies. Black dash-dot line is overall response and blue solid line and red dashed line are in-phase and out-of phase components of it. Parameters are the same with those in Fig.~\ref{fig:ambiguity} (a). The three graphs are normalized by the same scaling factor. Thus the relative intensities among three graphs can be compared.}
\end{figure}

The most often seen patterns displayed in Fig.~\ref{fig:patterns} (a) and (c) can be described qualitatively as following: at low frequencies, the in-phase signal has a constant non-zero value  with no out-of-phase component. This is because the approach to the steady-state takes place on a time scale much faster than the modulation period, and the recorded transient response looks like the applied microwave pulse train shown in Fig.~\ref{fig:transient_change} (a). The in-phase and out-of-phase responses are not significantly changed until the modulation frequency approaches the slowest time constant, $m_{1j}^{-1}$, as one can see from the low-frequency responses in Fig.~\ref{fig:ambiguity}. For all cases in Fig.~\ref{fig:ambiguity} and in Fig.~\ref{fig:transient_change}, $m_{1j}$ and $m_{2j}$ are in the ranges of $10^{2}~s^{-1}\sim10^{6}~s^{-1}$ and $10^{4}~s^{-1}\sim10^{6}~s^{-1}$, respectively. As the modulation frequency approaches $m_{1j}$, the system begins to lag behind the applied MW modulation, and the overall response ceases to resemble the simple harmonic function. This results in a decrease of the in-phase signal and an increase of the out-of-phase signal as seen in Fig.~\ref{fig:transient_change}  (b). At very high frequencies, much faster than than the fastest time constant, $m_{2j}^{-1}\sim10^{-6}~s^{-1}$, both the in- and out-of-phase components tend to approach zero. This behavior is explained by the exponential decay functions which become linear with small arguments and thus, they become constants (no change) when the period, $T\rightarrow0$~\cite{Lenahan:prb1984,Dyakonov:prb1997}.

\subsection{Recombination, dissociation, and flip-flop}
\begin{figure}
\centering\includegraphics[width=1\columnwidth]{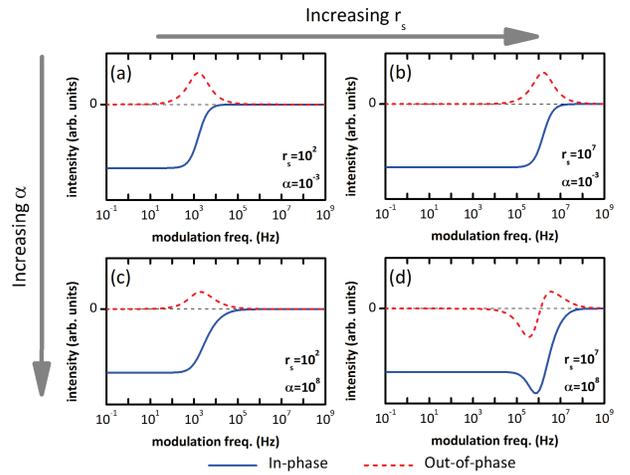}
\caption{\label{fig:rs}(Color online) Role of the singlet recombination rate, $r_{\textrm{s}}$. When $r_{\textrm{s}}$ is small, no significant change in the frequency dependence pattern is found when $\alpha$ is increased (from (a) to (c)). But for large $r_{\textrm{s}}$, a pattern change is observed when  $\alpha$ is increased (from (b) to (d)). All four quantitative models have the same parameters except (a) $r_{\textrm{s}}=10^{2}~s^{-1}$, $\alpha=10^{-3}~s^{-1}$, (b) $r_{\textrm{s}}=10^{7}~s^{-1}$, $\alpha=10^{-3}~s^{-1}$, (c) $r_{\textrm{s}}=10^{2}~s^{-1}$, $\alpha=10^{8}~s^{-1}$, and (d) $r_{\textrm{s}}=10^{7}~s^{-1}$, $\alpha=10^{8}~s^{-1}$. The values for the other parameters used for this data are listed in Table \ref{tab:all_parameters}.}
\end{figure}

Because cwODMR measures emission rates of photons, which are usually determined by the dominant singlet recombination rate $r_{\textrm{s}} n_{\textrm{s}}$, one might expect that $r_{\textrm{s}}$ has a dominant role in determining the frequency dependence pattern. In general, this is not the case though: other rate coefficients, especially spin mixing rates, can dominate the behavior of a cwODMR signal. Fig.~\ref{fig:rs} shows one of the most frequently observed examples of the frequency dependence patterns influenced by both $r_{\textrm{s}}$ and $\alpha$.

When $\alpha$ is small, increasing $r_{\textrm{s}}$ has little impact on the observed frequency dependence (Fig.~\ref{fig:rs} (a) and (b)). The most significant effect is a shift of the frequencies where both the in-phase and the out-of-phase components show their maximum rate changes. This is due to the increase of the time constants, $m_{ij}^{-1}$, from $m_{1j}\sim10^{4}~s^{-1}$ and $m_{2j}\sim10^{6}~s^{-1}$ to $m_{1j}\sim10^{6}~s^{-1}$ and $m_{2j}\sim10^{7}~s^{-1}$, due to very fast $r_{s}$. It should be noted that $d_{t}$ is $10^{6}~s^{-1}$ in all examples in Fig.~\ref{fig:rs} and $r_{s}$ is $10^{7}~s^{-1}$ in Fig.~\ref{fig:rs} (b) and (d).  The frequency dependence also shows little change when $r_{\textrm{s}}$ remains small and $\alpha$ is increased (Fig.~\ref{fig:rs} (c)). This corresponds to Fig.~\ref{fig:patterns} (a) and (c). However, when $\alpha$ becomes fast enough to compete with the slower time constant, $m_{1j}^{-1}$, (or even faster than $m_{2j}^{-1}$), and $r_{s}$ is faster than any dissociation rate coefficients, a more complicated frequency dependence emerges. The in-phase signal now has a local extremum. The out-of-phase signal not only shows the local extremum (as in the  simple pattern) but also a zero-crossing point, due to a sign change (Fig.~\ref{fig:rs} (d)). This pattern corresponds to Fig.~\ref{fig:patterns} (b) and (d). In this section the intersystem-crossing rate, $k_{\textrm{ISC}}$, was chosen to be small to investigate the influence of $\alpha$. We note that this pattern also appears when $k_{\textrm{ISC}}$ becomes large with a small $\alpha$, as explained further in the following section. Note that for cwODMR experiments this pattern appears only when $r_{\textrm{s}}$ becomes faster than any dissociation rate coefficient and $\alpha$ or $k_{\textrm{ISC}}$ is fast, too. It can also be seen for cwEDMR experiments when the dissociation rate coefficients and $\alpha$ or $k_{\textrm{ISC}}$ are fast (not shown here). We can thus infer that dissociation has a similar effect on cwEDMR experiments as recombination on cwODMR experiments.

\subsection{The influence of intersystem-crossing on cwODMR experiments}
Because the intersystem-crossing rate, $k_{\textrm{ISC}}$, represents a spin mixing process, it acts in a similar way as $\alpha$, even though $\alpha$ is modulated in time. To investigate the influence of $k_{\textrm{ISC}}$, $\alpha$ is chosen to be small in this section.
\begin{figure}
\centering\includegraphics[width=1\columnwidth]{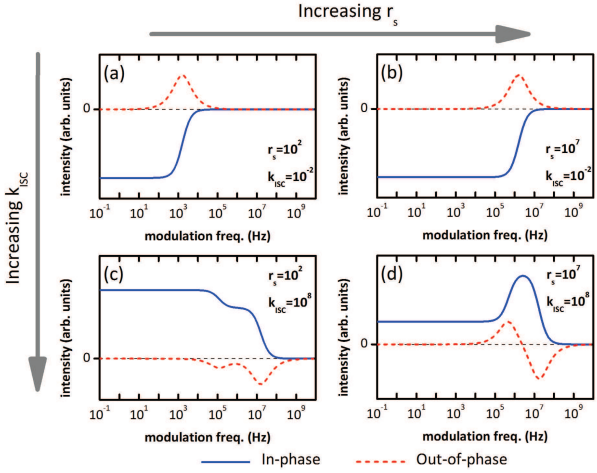}
\caption{\label{fig:kSL}(Color online) Role of the intersystem-crossing rate, $k_{\textrm{ISC}}$. At small $r_{s}$, local extrema appear on both, the in-phase and the out-of-phase signal at the high frequency region, when $k_{\textrm{ISC}}$ becomes large (from (a) to (c)). At large $r_{\textrm{s}}$, the in-phase signal shows local extrema and the out-of-phase signal shows sign change as $k_{\textrm{ISC}}$ is increased (from (b) to (d)). All four quantitative models have the same combinations of parameters but (a) $r_{\textrm{s}}=10^{2}~s^{-1}$, $k_{\textrm{ISC}}=10^{-2}~s^{-1}$, (b) $r_{\textrm{s}}=10^{7}~s^{-1}$, $k_{\textrm{ISC}}=10^{-2}~s^{-1}$, (c) $r_{\textrm{s}}=10^{2}~s^{-1}$, $k_{\textrm{ISC}}=10^{8}~s^{-1}$, (d) $r_{\textrm{s}}=10^{7}~s^{-1}$, $k_{\textrm{ISC}}=10^{8}~s^{-1}$. The other parameters used for this data are listed in Table \ref{tab:all_parameters}.}
\end{figure}
When $k_{\textrm{ISC}}$ is slow, very little change of the frequency dependence as a function of $r_{\textrm{s}}$ is seen, similar to the behavior described in the previous section (Fig.~\ref{fig:kSL} (a), (b)). In contrast to the case of large $\alpha$ and small $r_{\textrm{s}}$, a major change in the frequency dependence can be seen at fast $k_{\textrm{ISC}}$ and slow $r_{\textrm{s}}$ (Fig.~\ref{fig:kSL} (c)). A second local extremum appears in the out-of-phase component and a small bump at high frequency in the in-phase component. When both $k_{\textrm{ISC}}$ and $r_{\textrm{s}}$ compete with each other, a new pattern appears  (Fig.~\ref{fig:kSL} (d)). This pattern is similar to Fig.~\ref{fig:rs} (d) and similar to the pattern in Fig.~\ref{fig:patterns} (e) and (f) when $V_{\textrm{in, l f}}\rightarrow 0$ at small $\alpha$ (eq.~(\ref{eq:low_freq_in})). The other important observation is that the sign changes from positive for both in-phase and out-of-phase components ((a) and (b)) to negative ((c) and (d)). These sign changes due to $k_{\textrm{ISC}}$ is explained in sec.~\ref{sec:signs}.

\subsection{Pair generation}
Due to spin-selection rules, optically generated electron-hole pairs (the geminate state) are formed in singlet states and remain in this configuration unless strong spin-orbit coupling is present~\cite{MQM:Atkins}. Thus, we can assume $G_{\textrm{s}} \gg G_{\textrm{t}}$. Figure~\ref{fig:patterns} (a) corresponds to this case in which the in-phase and the out-of-phase components are always negative and positive, respectively. This case represents the frequency dependence of photoluminescence detected ODMR (PLDMR). In contrast to optical generation, the spin configuration of electron-hole pairs formed electrically, i.e. via electrical injection, is determined by spin statistics and we can assume $3 G_{\textrm{s}} \approx G_{\textrm{t}}$. All parameters in Fig.~\ref{fig:patterns} (a) and (c) are the same except that $G_{\textrm{s}}=10^{4}\times G_{\textrm{t}}$ in Fig.~\ref{fig:patterns} (a) and $3 G_{\textrm{s}} = G_{\textrm{t}}$ in Fig.~\ref{fig:patterns} (c). We can see from these calculations that electroluminescence detected ODMR (also called ELDMR) can show the opposite sign compared to PLDMR, for very similar underlying physical processes. It should be noted that this inversion could be found only for certain parameter sets, and this inversion can also happen when $3G_{\textrm{s}}\neq G_{\textrm{t}}$. For example, the sign of the in-phase component also becomes positive (not shown here) if every parameter remains the same except for $G_{\textrm{s}}=10\times G_{\textrm{t}}$. Thus, cwODMR can result in a positive in-phase and a negative out-of phase signal even though $G_{\textrm{t}}$ is smaller than $G_{\textrm{s}}$. This is because the sign inversion is also determined by rate coefficients and not just the generation rates. These cases will be discussed in Section~\ref{sec:signs}.

\section{Power dependence}\label{cw_power}
\begin{figure}
\centering\includegraphics[width=1\columnwidth]{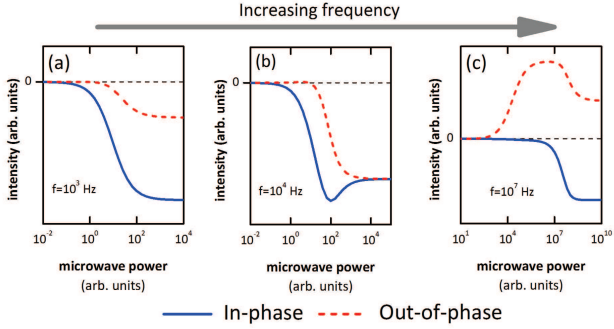}
\caption{\label{fig:power}(Color online) MW power dependence. All four quantitative models have the same combination of parameters except for (a) $f=10^{3}~Hz$, (b) $f=10^{4}~Hz$, (c) $f=10^{7}~Hz$. At low modulation frequencies, typical saturation curves can be found. At high modulation frequency, a non-trivial saturation behavior occurs. Refer to Table (\ref{tab:all_parameters}) for the values used for the other parameters.}
\end{figure}

The spin flip rate coefficient, $\alpha$, is proportional to the applied microwave power~\cite{Eickelkamp:mp1998}. Thus we can calculate the power dependence of cwODMR signals. Examples are shown in Fig.~\ref{fig:power}. For low modulation frequencies, (see Fig.~\ref{fig:power} (a)), a simple saturation behavior is predicted by eq.~(\ref{eq:low_freq_in}) and (\ref{eq:low_freq_out}). Note that the out-of-phase component is not always zero, but approaches zero at low frequencies, as expected from eq.~(\ref{eq:low_freq_out}). The saturation characteristics becomes more complicated as the modulation frequency increases. At $10^{4}$ Hz, the in-phase component shows a local extremum before it returns to a saturation value (Fig.~\ref{fig:power} (b)). Experimentally this behavior has been reported recently for low magnetic field cwEDMR on crystalline silicon interface defects~\cite{Morishita:prb2009}. At high modulation frequency, the in-phase component shows the usual saturation behavior (even though its saturation occurs at very higher power) but the out-of-phase component shows a local extremum before it approaches a saturation value. It also has a different sign than at lower frequencies (Fig.~\ref{fig:power} (c)). This shows that one can find opposite signs of in-phase and out-of-phase signals at high MW power and high MW modulation frequencies.

\section{Signal sign dependencies on the modulation frequency}\label{sec:zero}
Sign changes of cwEDMR and cwODMR signal have been found in InP nanoparticles~\cite{Langof:jpcb2002} and organic semiconductors~\cite{Yang:prl2007,Yang:prb2008}. The sign change of the cwODMR response in organic semiconductor has been attributed to the imbalance between changes in the numbers of singlet and triplet pairs when the pulse is on and off, which are equivalent to  $n_{\textrm{s}1}(T/2)-n_{\textrm{s}2}(T)$ and $n_{\textrm{t}1}(T/2)-n_{\textrm{t}2}(T)$ in our model. The zero-crossing point of the modulation frequency dependence function has also been used to estimate the intersystem-crossing time~\cite{Yang:prl2007,Yang:prb2008}. According to those reports, the zero-crossing can appear at a certain frequency where the increase of the number of singlet pairs is matched with the decrease of the number of triplet pairs so that the change in the total number of pairs is zero. However, we show here that the zero-crossing can be due to not only the imbalance of changes between singlet and triplet pairs but also to other more complicated relationships betwen physical parameters.

\begin{figure}
\centering\includegraphics[width=1\columnwidth]{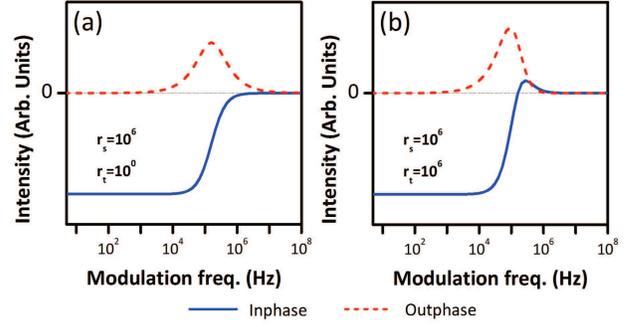}
\caption{\label{fig:zero}(Color online) Example of a modulation frequency dependence function showing a change from non-zero-crossing pattern to a zero-crossing pattern. The only difference between the two quantitative models can be found in the triplet recombination rate coefficients. (a) $r_{\textrm{t}}=10^{0}~s^{-1}$, (b) $r_{\textrm{t}}=10^{6}~s^{-1}$. Values for the other parameters are listed in Table \ref{tab:all_parameters}.}
\end{figure}

As can be seen in the solutions of the rate equations given above, the frequency dependence is not simply obtained from $n_{\textrm{s}1}(T/2)-n_{\textrm{s}2}(T)$ and $n_{\textrm{t}1}(T/2)-n_{\textrm{t}2}(T)$, but has a complicated dependence on various parameters. Among the quantitative models tested here, zero-crossing behavior is rarely seen. Fig.~\ref{fig:zero} shows one example: no zero-crossing is observed for small $r_{\textrm{t}}$, but when $r_{\textrm{t}}$ becomes larger and very close to $r_{\textrm{s}}$, zero-crossing is observed (Fig.~\ref{fig:zero} (a), (b)). It should be noted that the origin of this zero-crossing is not obvious because of the complexity of the solutions, although we note that $n_{\textrm{s}1}(T/2)-n_{\textrm{s}2}(T)$ and $n_{\textrm{t}1}(T/2)-n_{\textrm{t}2}(T)$ do not meet each other at the zero-crossing point in this case, in contrast to the model described elsewhere~\cite{Yang:prl2007,Yang:prb2008}. Thus the imbalance between changes in $n_{\textrm{s}}$ and $n_{\textrm{t}}$ cannot be the reason for the observed zero-crossing. We note that zero-crossing also can appear due to an overlap of two different spin-dependent mechanisms whose signs are opposite (e.g. in cwODMR of a radiative and a non-radiative channel). Note however that all zero-crossing effect demonstrated here resulted from a single spin-dependent process. The existence of zero-crossing indicates that one can observe different signs of cwODMR and cwEDMR signals from identical samples at different modulation frequencies.
\section{The interpretation of cwEDMR and cwODMR signal signs}\label{sec:signs}
\begin{figure}
\centering\includegraphics[width=1\columnwidth]{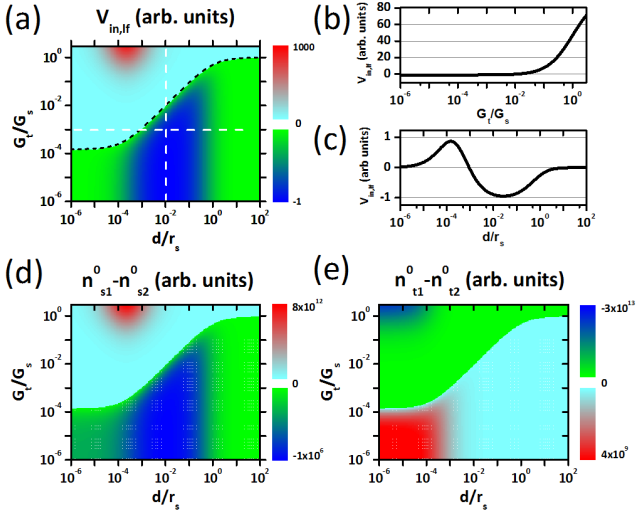}
\caption{\label{fig:signs}(Color online) Sign changes due to various rate coefficients. (a) In-phase intensities of the zero modulation frequency component as a function of $G_{\textrm{t}}/G_{\textrm{s}}$ and $d/r_{\textrm{s}}$. To distinguish positive values and negative values, different color scales are used (positive in upper left corner, and negative in lower right corner). The black dotted line describes the boundary separating positive values and negative values. (b) and (c) are two randomly chosen two dimensional subsets of the data in (a) representing a generation rate ratio dependence and dissociation rate ratio dependence. These slices are shown as white dashed lines in (a). Intensities in (a), (b), and (c) are normalized but in the same scale. (d) Changes in the numbers of singlet pairs, $n_{\textrm{s}1}^{0}-n_{\textrm{s}2}^{0}$ as a function of the same parameters as in (a). (e) Changes in the number of triplets pairs, $n_{\textrm{t}1}^{0}-n_{\textrm{t}2}^{0}$ as a function of the same parameters as in (a). Intensities in (d) and (e) are normalized but in the same scale.  All calculations in this figure are obtained from the same condition of $r_{\textrm{s}}=10^{4}~s^{-1}$, $r_{\textrm{t}}=1~s^{-1}$, $k_{\textrm{ISC}}=1~s^{-1}$, $\alpha=1~s^{-1}$, $F=0.25$, $G_{\textrm{s}}+G_{\textrm{t}}=10^{16}~s^{-1}$.}
\end{figure}
The signs of the cwEDMR and cwODMR signals have long been considered important indicators for the nature of electronic transitions. For example, it has been generally accepted that radiative recombination results in positive in-phase ODMR signals~\cite{Stutzmann:jncs2000,Depinna:pmb1982_1st,Lifshitz:cpl1997}. However, the recent observations of sign changes~\cite{Langof:jpcb2002,Yang:prl2007,Yang:prb2008} at certain frequencies suggest that signs may depend on complicated processes and the interpretation based exclusively on the sign of a modulated cwODMR or cwEDMR signal is not possible.

CwEDMR and cwODMR signal signs are determined by the transient responses of optical or electrical observables to a repeated change between on- and off-resonance, as described in Section~\ref{cw_modulation_frequency_dependence}. Because the time constants and pre-factors of the double exponential functions in eq.~(\ref{eq:ns1}), (\ref{eq:nt1}), (\ref{eq:ns2}), and (\ref{eq:nt2}) are functions of all the transition rate coefficients, there are many scenarios which can produce quenching and enhancement signals for both radiative and non-radiative ODMR signals as well as for EDMR signals. Many transitions compete with each other. For instance, recombination as well as dissociation are pair annihilation processes but only recombination causes PL while dissociation does not. Thus, when a radiative recombination process is slow and dissociation is fast, the resonant response may lead to quenching. This example shows that the following qualitative description of the sign of cwODMR signals is important.

The study of the sign change of cwODMR signals as functions of all individual parameters is beyond the scope of this work. Instead, only the low modulation frequency behavior will be discussed.

\subsection{For the case of radiative recombination}
As mentioned in Section~\ref{sec:rate_eq}, only radiative recombination has been considered so far and the solution for the in-phase cwODMR signal when radiative recombination is dominant at low modulation frequency is given in eq.~(\ref{eq:low_freq_in}). A quantitative analysis has been done by calculating $V_{\textrm{in, lf}}$ while changing some parameters, for the example shown in Fig.~\ref{fig:signs} we assumed that both singlet and triplet dissociation probabilities are not distinguishable, two mixing rate coefficients, $k_{\textrm{ISC}}$ and $\alpha$, are slower than any other recombination and dissociation, and total generation rate, $G_{\textrm{s}}+G_{\textrm{t}}$ is fixed to $10^{16}~s^{-1}$, $r_{\textrm{t}}$ to $1~s^{-1}$, and $F$ to 0.25. Fig.~\ref{fig:signs} (a) shows the zero frequency in-phase cwODMR signal, $V_{\textrm{in, l f}}$, as a function of the relative ratio of the triplet generation rate to the singlet generation rate, $G_{\textrm{t}}/G_{\textrm{s}}$, and the ratio of the dissociation rate coefficient to the singlet recombination rate coefficient which is fixed to $r_{\textrm{s}}=10^{4}~s^{-1}$. Color reflects the normalized intensity of $V_{\textrm{in, lf}}$. It should be noted that positive and negative values are intentionally placed in different scales to make them clearly distinguishable. One can find two noticeable features. (i) The intensity tends to increase as $G_{\textrm{t}}/G_{\textrm{s}}$ becomes larger and becomes negative at low $G_{\textrm{t}}/G_{\textrm{s}}$, as in Fig.~\ref{fig:signs} (b). (ii) The intensity also depends on the dissociation rate coefficients: when d is larger or smaller than the singlet recombination rate coefficient $r_{\textrm{s}}$, $V_{\textrm{in,lf}}$ becomes very small, and shows an extremum and sign change. Fig.~\ref{fig:signs} (a), (b), and (c) show that the signs are positive at high triplet generation rates and low dissociation rates or, equivalently, high recombination rates. When dissociation is not fast, signs are positive as long as triplet generation is not slower than singlet generation rate. This means that changing pair generation between optical and electrical can induce a sign change in cwODMR. This behavior can be more easily understood by consideration of competing singlet and triplet pairs. In Fig.~\ref{fig:signs} (d) and (e), the differences $n_{\textrm{s}1}^{0}-n_{\textrm{s}2}^{0}$ and $n_{\textrm{t}1}^{0}-n_{\textrm{t}2}^{0}$, are plotted for the same parameters as (a). Note that the low-frequency solution for the in-phase cwODMR signal, $V_{\textrm{in,lf}}$, is proportional to $r_{\textrm{t}}(n_{\textrm{s}1}^{0}-n_{\textrm{s}2}^{0})+r_{\textrm{s}} (n_{\textrm{t}1}^{0}-n_{\textrm{t}2}^{0})$. Both plots show different behavior compared to $V_{\textrm{in,lf}}$ but the boundaries dividing positive and negative values are very similar. When the pair annihilation is dominated only by singlet recombination, one can infer that the number of singlet pairs quickly decreases in the steady-state off-resonance condition. Thus, the steady-state is dominated by triplet pairs. Consequently a resonant MW converts triplet pairs to singlet pairs, it increases the number of singlet pairs which results in an enhancement of cwODMR signal.

This qualitative pictures applies to the region where $n_{\textrm{s}1}^{0}-n_{\textrm{s}2}^{0}$ is positive and $n_{\textrm{t}1}^{0}-n_{\textrm{t}2}^{0}$ is negative, in the upper left regions in Fig.~\ref{fig:signs} (a), (d), and (e) for example. In contrast, if the triplet generation is too low ($G_{\textrm{t}}<\frac{r_{\textrm{t}}+d_{\textrm{t}}}{r_{\textrm{s}}+d_{\textrm{s}}}G_{\textrm{s}}$), (lower-left corners in Fig.~\ref{fig:signs} (a), (d), and (e)), only a small number of triplet pairs forms during the off resonance steady-state, and the steady-state at off-resonance is dominated by singlet pairs. In this case spin resonance induced changes to the number of singlet pairs can become negative.

The statements above are based on the assumption of low $k_{\textrm{ISC}}$ and $\alpha$. When $k_{\textrm{ISC}}$ becomes larger than the other rates, sign changes are observed as in Fig.~\ref{fig:kSL} and patterns of $V_{\textrm{in,lf}}$ (not shown here) similar to the pattern in Fig.~\ref{fig:signs} are found, although slight shifts of boundaries dividing positive and negative are seen. Similar shifts have been found at different $F$ and $\alpha$. These shifts can be explained by an expression,
\begin{equation}\label{boundary_for_V_in,lf}
\frac{G_{\mathrm{s}}}{G_{t}}=-\frac{(r_{\mathrm{s}}+d_{\mathrm{s}})(w_{22}-w_{21})+w_{22}w_{11}-w_{21}w_{12}}{(r_{\mathrm{t}}+d_{\mathrm{t}})(w_{11}-w_{12})+w_{22}w_{11}-w_{21}w_{12}}
\end{equation}
which is obtained from eq. (\ref{eq:low_freq_in}) by setting $V_{\textrm{in,lf}}=0$. This formula explains that the boundary separating the positive and negative values in Fig.~\ref{fig:signs} (a) is dependent on all rate coefficients. Consequently, cwODMR and cwEDMR signs also depend on intersystem-crossing rate $k_{\textrm{ISC}}$, the temperature (note that $F$ is a function of temperature), and the MW power $\alpha$. We note again that sign changes can also occur at a certain modulation frequency as already explained above.
\begin{figure}
\centering\includegraphics[width=1\columnwidth]{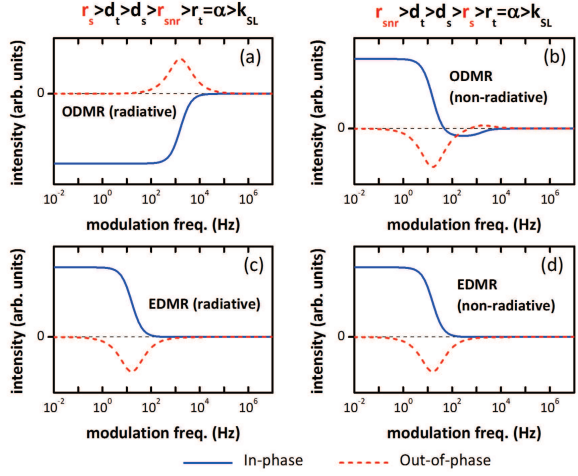}
\caption{\label{fig:radiative}(Color online) The sign of cwODMR signals can be negative when radiative recombination is dominant as in (a), and positive when non-radiative recombination is dominant as in (b). In contrast the signs of cwEDMR are not different, (c) and (d). Used common values for each rate parameters can be found in Table~\ref{tab:all_parameters}.
%are $r_{s}=10^{4}$, $r_{t}=10^{-1}$, $d_{s}=10$, $d_{t}=10^{2}$, $k_{ISC}=10^{-2}$,  $F=0.25$, $\alpha=10^{-1}$, $G_{s}= 10^{25}$, $G_{t}=10^{20}$.
(a) and (c) $r_{\textrm{s}}=10^{4}$, $r_{\textrm{s,nr}}=1$. (b) and (d) $r_{\textrm{s}}=1$, $r_{\textrm{s,nr}}=10^{4}$.}
\end{figure}

\subsection{For the case of non-radiative recombination}
Finally, we want to address the question of whether radiative and non-radiative recombination results in opposite cwODMR signal signs. We have checked a number of quantitative models and two examples are shown in Fig.~\ref{fig:radiative}. In contrast to all other cases discussed above, the non-radiative singlet recombination coefficients, $r_{\textrm{s,nr}}$ is taken into account. In Fig.~\ref{fig:radiative} (a) and (c), $r_{\textrm{s,nr}}$ is assumed to be smaller than $r_{\textrm{s}}$ to simulate the modulation frequency dependence in which radiative recombination is dominant. In Fig.~\ref{fig:radiative} (b) and (d), $r_{\textrm{s,nr}}$ is assumed to be the larger than $r_{\textrm{s}}$ to investigate the non-radiative process. It should be mentioned again that $r_{\textrm{s,nr}}$ contributes to the pair annihilation process but it does not contribute to the radiative emission rate term as explained in Section~\ref{sec:rate_eq}. Note that Fig.~\ref{fig:radiative} (a) shows one of the modulation frequency dependence patterns that are discussed above. The in-phase signal is negative even though $r_{\textrm{s}}$ is most dominant because $G_{\textrm{s}}\gg G_{\textrm{t}}$.

Fig.~\ref{fig:radiative} (b) shows a zero-crossing behavior, thus, the in-phase component can be positive and negative even though $r_{\textrm{s,nr}}$ is dominant. In contrast to the cwODMR cases, the signs of the cwEDMR in-phase signals are positive in both cases as shown in (c) and (d). To summarize, our results show that cwODMR signals can be negative and positive for both radiative and non-radiative recombination processes. Any conclusion about the nature of a spin-dependent recombination process from the sign of an observed cwODMR signal is therefore speculative, and should be confirmed with additional evidence.

We conclude that dissociation, recombination, ratio between singlet and triplet generation, intersystem-crossing, temperature, modulation frequency, MW power, and the nature of recombination (radiative or non-radiative) can all determine the sign of cwODMR signals.

\section{Summary and conclusion}
A set of rate equations based on an intermediate pair recombination model are presented and generalized analytical solutions have been obtained. These solutions have been used to calculate modulation frequency dependencies of cwEDMR and cwODMR signals. We have investigated how experimental parameters affect these modulation frequency dependencies which revealed that a large number of quantitatively different models show non-distinguishable modulation frequency dependence patterns. This implies that the interpretation of cwODMR and cwEDMR experiments can be very ambiguous. We further showed that the sign of cwODMR and cwEDMR signals depend on most of the rate coefficients, as well as experimental parameters such as temperature, MW power, and modulation frequency. Thus, there are many variables which can reverse the sign of cwEDMR and cwODMR signals and consequently, conclusions about the radiative or non-radiative nature of an observed spin-dependent transition based solely on the sign of an observed spin-dependent process or its modulation frequency dependence is not possible.

\begin{acknowledgments}
Acknowledgment is made to the DoE (Grant DESC0000909), the donors of the American Chemical Society Petroleum
Research Fund (Grant PRF 48916-DNI10), and the NSF for a CAREER Award ($\#$0953225).
\end{acknowledgments}

\appendix

\section{Boundary conditions and exact solutions for the pair densities}\label{app_BC}
Four of the boundary conditions can be easily found from the periodicity of the solution: $n_{ s1}   ( 0   ) =n_{ s2}   ( T   )$, $n_{ t1}   ( 0   ) =n_{ t2}   ( T   )$, $n_{ s1}   ( \frac{T}{2}   ) =n_{ s2}   (\frac{T}{2}   )$, and $n_{ t1} ( \frac{T}{2}   ) =n_{ t2}   (\frac{T}{2}   )$. From these boundary conditions, we obtain
\begin{eqnarray}
A_{1}+A_{2}+n_{s1}^{0}=A_{3}e^{(-m_{12}T/2)} +A_{4}e^{(-m_{22}T/2)}+n_{s2}^{0}\label{bc1}\\
B_{1}+B_{2}+n_{t1}^{0}=B_{3}e^{(-m_{12}T/2)} +B_{4}e^{(-m_{22}T/2)}+n_{t2}^{0}\label{bc2}\\
A_{1}e^{(-m_{11}T/2)} +A_{2}e^{(-m_{21}T/2)}+n_{s1}^{0}=A_{3}+A_{4}+n_{s2}^{0}\label{bc3}\\
B_{1}e^{(-m_{11}T/2)} +B_{2}e^{(-m_{21}T/2)}+n_{t1}^{0}=B_{3}+B_{4}+n_{t2}^{0}\label{bc4}
\end{eqnarray}
After each half cycle, the number of singlet and triplet pairs are decreased or increased. These changes depend on the given rate coefficients: the number of singlet or triplet pairs are increased by pair generation, decreased by the dissociation and recombination processes, or either decreased or increased by spin mixing. From this condition, the other four equations can be found as
\begin{eqnarray}\label{app_bc5}
n_{ s1}   (\frac{T}{2}   ) -n_{ s1}   ( 0   ) =G_{s}\frac{T}{2}+\int _{0}^{\frac{T}{2}}  (w_{21}n_{ t1}-   ( C_{s}+w_{11}   ) n_{ s1}  ){dt},\nonumber \\
\end{eqnarray}
\begin{eqnarray}\label{app_bc6}
n_{ s2}   (T   ) -n_{ s2}   ( \frac{T}{2}   ) =G_{s}\frac{T}{2}+\int _{\frac{T}{2}}^{T}  (w_{22} n_{ t2}-   ( C_{s}+w_{12}   ) n_{ s2}  ){dt},\nonumber \\
\end{eqnarray}
\begin{eqnarray}\label{app_bc7}
n_{ t1}   (\frac{T}{2}   ) -n_{ t1}   ( 0   ) =G_{t}\frac{T}{2}+\int _{0}^{\frac{T}{2}}  (w_{11}n_{ t1}-   ( C_{t}+w_{21}   ) n_{ t1}  ){dt},\nonumber \\
\end{eqnarray}
\begin{eqnarray}\label{app_bc8}
n_{ t2}   (T   ) -n_{ t2}   ( \frac{T}{2}   ) =G_{t}\frac{T}{2}+\int _{\frac{T}{2}}^{T}  (w_{12} n_{ t2}-   ( C_{t}+w_{22}   ) n_{ t2}  ){dt}.\nonumber \\
\end{eqnarray}
By plugging eqs.~(\ref{eq:ns1})--(\ref{eq:nt2}) into (\ref{app_bc5})--(\ref{app_bc8}), we obtain
\begin{eqnarray}\label{bc5}
A_{1} (e^{(-m_{11}T/2)}-1) +A_{2} (e^{(-m_{21}T/2)}-1)\nonumber \\
=-\frac{w_{21}B_{1}-(C_{s}+w_{11})A_{1}}{m_{11}} (e^{(-m_{11}T/2)}-1) \nonumber \\
-\frac{w_{21}B_{2}-(C_{s}+w_{11})A_{2}}{m_{21}} (e^{(-m_{21}T/2)}-1),
\end{eqnarray}

\begin{eqnarray}\label{bc6}
A_{3} (e^{(-m_{12}T/2)}-1) +A_{4} (e^{(-m_{22}T/2)}-1)\nonumber \\
=-\frac{w_{22}B_{3}-(C_{s}+w_{11})A_{3}}{m_{12}} (e^{(-m_{12}T/2)}-1) \nonumber \\
-\frac{w_{22}B_{4}-(C_{s}+w_{12})A_{4}}{m_{22}} (e^{(-m_{22}T/2)}-1),
\end{eqnarray}

\begin{eqnarray}\label{bc7}
B_{1} (e^{(-m_{11}T/2)}-1) +B_{2} (e^{(-m_{21}T/2)}-1)\nonumber \\
=-\frac{w_{11}A_{1}-(C_{t}+w_{21})B_{1}}{m_{11}} (e^{(-m_{11}T/2)}-1) \nonumber \\
-\frac{w_{11}A_{2}-(C_{t}+w_{21})B_{2}}{m_{21}} (e^{(-m_{21}T/2)}-1),
\end{eqnarray}

\begin{eqnarray}\label{bc8}
B_{3} (e^{(-m_{12}T/2)}-1) +B_{4} (e^{(-m_{22}T/2)}-1)\nonumber \\
=-\frac{w_{12}A_{3}-(C_{t}+w_{22})B_{3}}{m_{12}} (e^{(-m_{12}T/2)}-1) \nonumber \\
-\frac{w_{12}A_{4}-(C_{t}+w_{22})B_{4}}{m_{22}} (e^{(-m_{22}T/2)}-1).
\end{eqnarray}

Note that $G_{s}+w_{21} n_{t1}^{0}-(C_{s}+w_{11})n_{s1}^{0}=0,\,G_{s}+w_{22} n_{t2}^{0}-(C_{s}+w_{12})n_{s2}^{0}=0,\,G_{t}+w_{11} n_{s1}^{0}-(C_{t}+w_{21})n_{t1}^{0}=0,\,G_{t}+w_{12} n_{s2}^{0}-(C_{t}+w_{22})n_{t2}^{0}=0$ are used here, which are obtained from eq.~(\ref{eq:cw_ns0}) and (\ref{eq:cw_nt0}).

Solving eq.~(\ref{bc1})--(\ref{bc4}), (\ref{bc5})--(\ref{bc8}), and by introducing the parameters $\beta_{ij}={\frac {C_{s}+w_{1j}-m_{1j}}{w_{2j}}}$, $\Delta n_{s}^{0} = n_{s2}^{0}-n_{s1}^{0}$, $\Delta n_{t}^{0} = n_{t2}^{0}-n_{t1}^{0}$, and $\gamma_{ij} = e^{-m_{ij} \frac{T}{2}}$, we realize that $B_{ij}=A_{ij}\beta_{ij}$ and four simplified equations
\begin{eqnarray}\label{eq:matrix_eq}
\left ( \begin{array}{cccc} 1&1&-\gamma_{12}&-\gamma_{22}\\ \beta_{11}&\beta_{21}&-\beta_{12}\gamma_{12}&-\beta_{22}\gamma_{22}\\ \gamma_{11}&\gamma_{21}&-1&-1\\ \beta_{11}\gamma_{11}&\beta_{21}\gamma_{21}&-\beta_{12}&-\beta_{22}\end {array}\right )
\left ( \begin {array}{c} A_{11}\\A_{21}\\A_{12}\\A_{22}\end {array} \right )
=\left ( \begin {array}{c} \Delta n_{s}^{0}\\ \Delta n_{t}^{0}\\ \Delta n_{s}^{0}\\ \Delta n_{t}^{0}\end {array} \right )\nonumber\\
\end{eqnarray}
are obtained for $A_{ij}$.

Equation~(\ref{eq:matrix_eq}) is a fully determined system of linear equations which can be solved. This leads to the solution,
\begin{widetext}
\begin{eqnarray}\label{eq:cw_a22}
A_{22} &=&(((\beta_{21}-\beta_{11})\cdot(\Delta n_{s}^{0}-\gamma_{11} \Delta n_{s}^{0})-(\Delta n_{t}^{0}-\beta_{11}\Delta n_{s}^{0})\cdot(\gamma_{21}-\gamma_{11}))
\nonumber \\
&&\cdot ((\beta_{21}-\beta_{11})\cdot(\beta_{11}\gamma_{11}\gamma_{12}-\beta_{12})-(\beta_{11}\gamma_{12}-\beta_{12}\gamma_{12})\cdot(\beta_{21}\gamma_{21}-\beta_{11}\gamma_{11})) \nonumber \\
&&-((\beta_{21}-\beta_{11})\cdot(\gamma_{11}\gamma_{12}-1)-(\beta_{11}\gamma_{12}-\beta_{12}\gamma_{12})\cdot(\gamma_{21}-\gamma_{11}))
\nonumber \\
&&\cdot
((\beta_{21}-\beta_{11})\cdot(\Delta n_{t}^{0}-\beta_{11}\gamma_{11}\Delta n_{s}^{0})-(\Delta n_{t}^{0}-\beta_{11}\Delta n_{s}^{0})\cdot(\beta_{21}\gamma_{21}-\beta_{11}\gamma_{11})))
\nonumber \\
&&/(((\beta_{21}-\beta_{11})\cdot(\gamma_{11}\gamma_{22}-1)-(\beta_{11}\gamma_{22}-\beta_{22}\gamma_{22})\cdot(\gamma_{21}-\gamma_{11}))
\nonumber \\
&&\cdot
((\beta_{21}-\beta_{11})\cdot(\beta_{11}\gamma_{11}\gamma_{12}-\beta_{12})-(\beta_{11}\gamma_{12}-\beta_{12}\gamma_{12})\cdot(\beta_{21}\gamma_{21}-\beta_{11}\gamma_{11})) \nonumber \\
&&-((\beta_{21}-\beta_{11})\cdot(\gamma_{11}\gamma_{12}-1)-(\beta_{11}\gamma_{12}-\beta_{12}\gamma_{12})\cdot(\gamma_{21}-\gamma_{11}))
\nonumber \\
&&\cdot
((\beta_{21}-\beta_{11})\cdot(\beta_{11}\gamma_{11}\gamma_{22}-\beta_{22})-(\beta_{11}\gamma_{22}-\beta_{22}\gamma_{22})\cdot(\beta_{21}\gamma_{21}-\beta_{11}\gamma_{11}))),\nonumber \\
\end{eqnarray}
\end{widetext}
\begin{widetext}
\begin{eqnarray}\label{eq:cw_a12}
A_{12} &=& ((\beta_{21}-\beta_{11})\cdot(\Delta n_{s}^{0}-\gamma_{11} \Delta n_{s}^{0})-(\Delta n_{t}^{0}-\beta_{11} \Delta n_{s}^{0})\cdot(\gamma_{21}-\gamma_{11})
\nonumber \\
&&-((\beta_{21}-\beta_{11})\cdot(\gamma_{11}\gamma_{22}-1)
-(\beta_{11}\gamma_{22}-\beta_{22}\gamma_{22})\cdot(\gamma_{21}-\gamma_{11}))\cdot A_{22})
\nonumber \\
&&/((\beta_{2}-\beta_{11})\cdot(\gamma_{11}\gamma_{12}-1)-(\beta_{11}\gamma_{12}-\beta_{12}\gamma_{12})\cdot(\gamma_{21}-\gamma_{11})),
\end{eqnarray}
\end{widetext}
%\begin{widetext}
\begin{eqnarray}\label{eq:a21}
A_{21}&=&((\Delta n_{t}^{0}-\beta_{11}\Delta n_{s}^{0})\cdot(\beta_{21}\gamma_{21}-\beta_{11}\gamma_{11})
\nonumber \\
&&-(\beta_{11}\gamma_{12}-\beta_{12}\gamma_{12})\cdot(\beta_{21}\gamma_{21}-\beta_{11}\gamma_{11})\cdot A_{12}
\nonumber \\
&&-(\beta_{11}\gamma_{22}-\beta_{22}\gamma_{22})\cdot(\beta_{21}\gamma_{21}-\beta_{11}\gamma_{11})\cdot A_{22})
\nonumber \\
&&/((\beta_{21}-\beta_{11})\cdot(\beta_{21}\gamma_{21}-\beta_{11}\gamma_{11})),
\end{eqnarray}
%\end{widetext}
%\begin{widetext}
\begin{eqnarray}\label{eq:a11}
A_{11} = \Delta n_{s}^{0} - A_{21} + \gamma_{12} \cdot A_{12} + \gamma_{22}\cdot A_{22}.
\end{eqnarray}
%\end{widetext}

\section{Modulation frequency dependence solutions}\label{app_frequency}
To find the in-phase and out-of-phase components at a given modulation frequency, it is better to find the Fourier series of eq.~(\ref{eq:pl1}) and (\ref{eq:pl2}), and the frequency responses will be decided from the Fourier coefficients according to the definition of the Fourier series,
\begin{eqnarray}\label{eq:fs}
{I}_{Fs}   ( t   ) =\frac{ {I}_{0}}{2}+\sum _{l=1}^{\infty }  ({ I}_{c}\cos   ( \frac {2 l \pi }{T} t  ) +{ I}_{s}\sin ( \frac {2 l \pi }{T} t  )  ),\nonumber\\
\end{eqnarray}
\begin{equation}
{ I}_{c}=\frac{2}{T} \int _{0}^{T}{ I}  ( t  ) \cos  ( \frac {2 l \pi }{T} t  ) {dt},
\end{equation}
\begin{equation}
{ I}_{s}=\frac{2}{T} \int _{0}^{T}{ I}  ( t  ) \sin  ( \frac {2 l \pi }{T} t  ) {dt}.
\end{equation}
Then the obtained two coefficients as well as the zero frequency component are:
%\begin{widetext}
\begin{eqnarray}\label{eq:PLc}
{ I}_{c}&=&\frac{2 m_{11}}{T}  ( r_{s}A_{11}+r_{t}B_{11}  )   (\frac{ 1-{e}^{-m_{11}T/2}\cos  ( l\pi   )  }{ {m_{11}}^{2}+4{l}^{2}{\pi }^{2}/{{T}^{2}}  } )\nonumber \\
&& +\frac{2 m_{21}}{T}  ( r_{s}A_{21}+r_{t}B_{21}  )   (\frac{ 1-{e}^{-m_{21}T/2}\cos  ( l\pi   )  }{  {m_{21}}^{2}+4{l}^{2}{\pi }^{2}/{{T}^{2}} } )\nonumber \\
&&+\frac{2 m_{12}}{T}  ( r_{s}A_{12}+r_{t}B_{12}  )   (\frac{ \cos  ( l\pi   ) -{e}^{- m_{12}T/2}  }{   {m_{12}}^{2}+4{l}^{2}{\pi }^{2}/{{T}^{2}}  } )\nonumber \\
&&+\frac{2 m_{22}}{T}  ( r_{s}A_{22}+r_{t}B_{22}  )   (\frac{   \cos  ( l\pi   ) -{e}^{-m_{22}T/2} }{ {m_{22}}^{2}+4{l}^{2}{\pi }^{2}/{{T}^{2}} } ),\nonumber\\
\end{eqnarray}
%\end{widetext}
%\begin{widetext}
\begin{eqnarray}\label{eq:PLs}
{ I}_{s}&=&\frac{4l\pi}{{T}^{2}}  ( r_{s}A_{11}+r_{t}B_{11}  )   (\frac{1-{e}^{-m_{11}T/2}\cos  ( l\pi   )   }{ {m_{11}}^{2}+4{l}^{2}{\pi }^{2}/{{T}^{2}}  } ) \nonumber \\
&&+\frac{4l\pi}{{T}^{2}}  ( r_{s}A_{21}+r_{t}B_{21}  )  ( \frac{  1-{e}^{-m_{21}T/2}\cos  ( l\pi   )  }{ {m_{21}}^{2}+4{l}^{2}{\pi }^{2}/{{T}^{2}}  } )  \nonumber \\
&&+\frac{4l\pi}{{T}^{2}}  ( r_{s}A_{12}+r_{t}B_{12}  )  ( \frac{  \cos  ( l\pi   ) -{e}^{-m_{12}T/2} }{  {m_{12}}^{2}+4{l}^{2}{\pi }^{2}/{{T}^{2}} } )\nonumber \\
&&+\frac{4l\pi}{{T}^{2}}  ( r_{s}A_{22}+r_{t}B_{22}  )  ( \frac{   \cos  ( l\pi   ) -{e}^{-m_{22}^{2}T/2} }{  {m_{22}}^{2}+4{l}^{2}{\pi }^{2}/{{T}^{2}}  }  )\nonumber \\
&&+ ( r_{s} \Delta n_{s}^{0}+r_{t} \Delta n_{t}^{0}  )   (\frac {    \cos  ( l\pi   ) -1  }{l\pi } ),\nonumber\\
\end{eqnarray}
%\end{widetext}
%\begin{widetext}
\begin{eqnarray}
{ I}_{{0}}&=&\frac{2 }{T}  ( r_{s}A_{11}+r_{t}B_{11}  )   (\frac{ 1-{e}^{-m_{11}T/2}}{ {m_{11}}  } )\nonumber\\
 &&+\frac{2 }{T}  ( r_{s}A_{21}+r_{t}B_{21}  )   (\frac{ 1-{e}^{-m_{21}T/2}  }{  {m_{21}} } )\nonumber \\
 &&+ \frac{2 }{T}  ( r_{s}A_{12}+r_{t}B_{12}  )   (\frac{ 1 -{e}^{- m_{12}T/2}  }{   {m_{12}} } )\nonumber\\
 &&+\frac{2 }{T}  ( r_{s}A_{22}+r_{t}B_{22}  )   (\frac{   1 -{e}^{-m_{22}T/2} }{ {m_{22}} } )\nonumber \\
 &&+r_{s}(n_{s1}^{0}+n_{s2}^{0})+r_{t}(n_{t1}^{0}+n_{t2}^{0}).
\end{eqnarray}
%\end{widetext}
The Fourier series in eq.~(\ref{eq:fs}) can be simplified by introducing $V_{0}= \sqrt{{{ I}_{c}}^{2}+{{ I}_{s}}^{2}}$ and $\varphi =\tan ^{-1}  ( {\frac {{ I}_{{c}}}{{ I}_{{s}}}}  )$ as below,
\begin{equation}
{ I}_{Fs}  ( t  ) =\frac{{ I}_{0}}{2}+\sum _{l=1}^{\infty }V_{0}\sin  ( 2 l \pi f t+\varphi   ),
\end{equation}
where f=1/T is the frequency of the square modulation. A Lock-in amplifier multiplies the input signal by its own internal reference signals, $\sin (\omega _{L} t +\theta _{L})$ and $\cos (\omega _{L} t +\theta _{L})$, to detect in-phase and out-of-phase signals, respectively. Thus, the in-phase $V_{in}$ and out-of-phase $V_{out}$ signals are
\begin{eqnarray}
V_{in}&=&\frac{{ I}_{0}}{2} V_{L} \sin  ( \omega _{L} t +\theta _{L}  )\nonumber \\
&&+\sum _{l=1}^{\infty } \frac{V_{L}V_{0}}{2}   [ \cos  (  (2 l \pi f-\omega_{L} ) t+\varphi -\theta _{L}   )\nonumber \\
&&-\cos  ( (2 l \pi f+\omega_{L}) t+\varphi +\theta _{L}   )   ],
\end{eqnarray}
\begin{eqnarray}
V_{out}&=&\frac{{ I}_{0}}{2} V_{L} \cos  ( \omega _{L} t +\theta _{L}  )\nonumber \\
&&+\sum _{l=1}^{\infty } \frac{V_{L}V_{0}}{2}   [ \sin  (  (2 l \pi f+\omega_{L} ) t+\varphi +\theta _{L}   )\nonumber \\
&&+\sin  ( (2 l \pi f-\omega_{L}) t+\varphi -\theta _{L}   )   ].
\end{eqnarray}
where $V_{L}$ is the amplitude of the reference signals. After these signals pass through a low pass filter, only the non-AC signals will remain. And the frequency of the internal reference signal is fixed such that it has the same phase as the external reference signal. Thanks to this condition, $\omega _{L} \approx 2 \pi f$, the in-phase and out-of-phase signals become
\begin{equation}\label{eq:in}
V_{in}=\frac{V_{01}}{2}  \cos  ( \varphi_{1}  )=\frac{1}{2} I_{s1},
\end{equation}
\begin{equation}\label{eq:out}
V_{out}=\frac{V_{01}}{2}  \sin  ( \varphi_{1}  )=\frac{1}{2} I_{c1}
\end{equation}
where $V_{01}=V_{0}$, $I_{s1}=I_{s}$, $I_{c1}=I_{c}$, and $\varphi_{1}=\varphi$ at $l=1$, $\theta_{L}$ is usually set to zero, and $V_{L}$ is set to 1.

\section{Parameters used for calculation}
The parameters used for all data presented in the figures are listed in Table~\ref{tab:all_parameters}.
\begingroup
\squeezetable
\begin{table*}[htbp]
  \centering
  \caption{Parameters used for the calculation of some of the data in this report. All values have a unit of $s^{-1}$ except for $F$ which is unit-less.}
    \begin{tabular*}{0.75\textwidth}{@{\extracolsep{\fill}}ccccccccccccc}
    \toprule\toprule
    Figure number &  & $r_{\textrm{s}}$ & $r_{\textrm{s,nr}}$ & $r_{\textrm{t}}$ & $d_{\textrm{s}}$ & $d_{\textrm{t}}$ & $k_{\textrm{ISC}}$ & $\alpha$ & $F$ & $G_{\textrm{s}}$ & $G_{\textrm{t}}$ & f \\
    \midrule
    \ref{fig:decomposition} &       & $10^{4}$ & 0     & 1     & $10^{2}$ & $10^{6}$ & $10^{-2}$ & $10^{5}$ & 0.25  & $10^{23}$ & $10^{20}$ & $10^{4}$ \\
    \midrule
    \multirow{3}[6]{*}{\ref{fig:ambiguity}} & (a)   & $10^{2}$ & 0     & 1     & $10^{4}$ & $10^{6}$ & $10^{-2}$ & $10^{3}$ & 0.25  & $10^{23}$ & $10^{20}$ & - \\
          & (b)   & $10^{4}$ & 0     & $10^{-1}$ & 10    & $10^{2}$ & $10^{-2}$ & $10^{-1}$ & 0.25  & $10^{25}$ & $10^{20}$ & - \\
          & (c)   & $10^{4}$ & 0     & 1     & $10^{2}$ & $10^{6}$ & $10^{-2}$ & $10^{3}$ & 0.25  & $10^{20}$/3 & $10^{20}$ & - \\
    \midrule
    \multirow{7}[14]{*}{\ref{fig:patterns}} & (a)   & $10^{4}$ & 0     & 1     & $10^{2}$ & $10^{3}$ & $10^{-2}$ & $10^{-3}$ & 0.25  & $10^{24}$ & $10^{20}$ & - \\
          & (b)   & $10^{6}$ & 0     & 1     & $10^{2}$ & $10^{4}$ & $10^{-2}$ & $10^{7}$ & 0.25  & $10^{22}$ & $10^{20}$ & - \\
          & (c)   & $10^{4}$ & 0     & 1     & $10^{2}$ & $10^{3}$ & $10^{-2}$ & $10^{-3}$ & 0.25  & $10^{20}$/3 & $10^{20}$ & - \\
          & (d)   & $10^{6}$ & 0     & 1     & $10^{2}$ & $10^{4}$ & $10^{-2}$ & $10^{7}$ & 0.25  & $10^{20}$/3 & $10^{20}$ & - \\
          & (e)   & $10^{6}$ & 0     & $10^{4}$ & 1     & $10^{2}$ & $10^{4}$ & $10^{-3}$ & 0.25  & $10^{24}$ & $10^{20}$ & - \\          & (f)   & $10^{6}$ & 0     & $10^{4}$ & 1     & $10^{2}$ & $10^{4}$ & $10^{-3}$ & 0.25  & $10^{20}$/3 & $10^{20}$ & - \\
          & (g)   & 1     & 0     & $10^{-1}$ & $10^{2}$ & $10^{4}$ & $10^{6}$ & $10^{-3}$ & 0.25  & $10^{20}$/3 & $10^{20}$ & - \\
    \midrule
    \multirow{4}[8]{*}{\ref{fig:rs}} & (a)   & $10^{2}$ & 0     & 1     & $10^{4}$ & $10^{6}$ & $10^{-2}$ & $10^{-3}$ & 0.25  & $10^{22}$ & $10^{20}$ & - \\
          & (b)   & $10^{7}$ & 0     & 1     & $10^{4}$ & $10^{6}$ & $10^{-2}$ & $10^{-3}$ & 0.25  & $10^{22}$ & $10^{20}$ & - \\
          & (c)   & $10^{2}$ & 0     & 1     & $10^{4}$ & $10^{6}$ & $10^{-2}$ & $10^{8}$ & 0.25  & $10^{22}$ & $10^{20}$ & - \\
          & (d)   & $10^{7}$ & 0     & 1     & $10^{4}$ & $10^{6}$ & $10^{-2}$ & $10^{8}$ & 0.25  & $10^{22}$ & $10^{20}$ & - \\
    \midrule
    \multirow{4}[8]{*}{\ref{fig:kSL}} & (a)   & $10^{2}$ & 0     & 1     & $10^{4}$ & $10^{6}$ & $10^{2}$ & $10^{1}$ & 0.25  & $10^{22}$ & $10^{20}$ & - \\
          & (b)   & $10^{7}$ & 0     & 1     & $10^{4}$ & $10^{6}$ & $10^{-2}$ & $10^{1}$ & 0.25  & $10^{22}$ & $10^{20}$ & - \\
          & (c)   & $10^{2}$ & 0     & 1     & $10^{4}$ & $10^{6}$ & $10^{8}$ & $10^{1}$ & 0.25  & $10^{22}$ & $10^{20}$ & - \\
          & (d)   & $10^{7}$ & 0     & 1     & $10^{4}$ & $10^{6}$ & $10^{8}$ & $10^{1}$ & 0.25  & $10^{22}$ & $10^{20}$ & - \\
    \midrule
    \multirow{3}[6]{*}{\ref{fig:power}} & (a)   & $10^{6}$ & 0     & 1     & $10^{2}$ & $10^{4}$ & $10^{-2}$ & -     & 0.25  & $10^{22}$ & $10^{20}$ & $10^{3}$ \\
          & (b)   & $10^{6}$ & 0     & 1     & $10^{2}$ & $10^{4}$ & $10^{-2}$ & -     & 0.25  & $10^{22}$ & $10^{20}$ & $10^{4}$ \\
          & (c)   & $10^{6}$ & 0     & 1     & $10^{2}$ & $10^{4}$ & $10^{-2}$ & -     & 0.25  & $10^{22}$ & $10^{20}$ & $10^{7}$ \\
    \midrule
    \multirow{2}[4]{*}{\ref{fig:zero}} & (a)   & $10^{6}$ & 0     & 1     & $10^{2}$ & $10^{4}$ & $10^{-2}$ & $10^{1}$ & 0.25  & $10^{22}$ & $10^{20}$ & - \\
          & (b)   & $10^{6}$ & 0     & $10^{6}$ & $10^{2}$ & $10^{4}$ & $10^{-2}$ & $10^{1}$ & 0.25  & $10^{22}$ & $10^{20}$ & - \\
    \midrule
    \ref{fig:signs} &       & $10^{4}$ & 0     & 1     & -     & -     & 1     & 1     & 0.25  & \multicolumn{2}{c}{$G_{s}+G_{t}=10^{16}$} & - \\
    \midrule
    \multirow{4}[7]{*}{\ref{fig:radiative}} & (a)   & $10^{4}$ & 1     & $10^{-1}$ & 10    & $10^{2}$ & $10^{-2}$ & $10^{-1}$ & 0.25  & $10^{25}$ & $10^{20}$ & - \\
          & (b)   & $10^{4}$ & 1     & $10^{-1}$ & 10    & $10^{2}$ & $10^{-2}$ & $10^{-1}$ & 0.25  & $10^{25}$ & $10^{20}$ & - \\
          & (c)   & 1     & $10^{4}$ & $10^{-1}$ & 10    & $10^{2}$ & $10^{-2}$ & $10^{-1}$ & 0.25  & $10^{25}$ & $10^{20}$ & - \\
          & (d)   & 1     & $10^{4}$ & $10^{-1}$ & 10    & $10^{2}$ & $10^{-2}$ & $10^{-1}$ & 0.25  & $10^{25}$ & $10^{20}$ & - \\
    \bottomrule\bottomrule
    \end{tabular*}%
  \label{tab:all_parameters}%
\end{table*}%
\endgroup

\newpage
%\bibliography{ref_2012-06-24}

\begin{thebibliography}{81}%
\makeatletter
\providecommand \@ifxundefined [1]{%
 \@ifx{#1\undefined}
}%
\providecommand \@ifnum [1]{%
 \ifnum #1\expandafter \@firstoftwo
 \else \expandafter \@secondoftwo
 \fi
}%
\providecommand \@ifx [1]{%
 \ifx #1\expandafter \@firstoftwo
 \else \expandafter \@secondoftwo
 \fi
}%
\providecommand \natexlab [1]{#1}%
\providecommand \enquote  [1]{``#1''}%
\providecommand \bibnamefont  [1]{#1}%
\providecommand \bibfnamefont [1]{#1}%
\providecommand \citenamefont [1]{#1}%
\providecommand \href@noop [0]{\@secondoftwo}%
\providecommand \href [0]{\begingroup \@sanitize@url \@href}%
\providecommand \@href[1]{\@@startlink{#1}\@@href}%
\providecommand \@@href[1]{\endgroup#1\@@endlink}%
\providecommand \@sanitize@url [0]{\catcode `\\12\catcode `\$12\catcode
  `\&12\catcode `\#12\catcode `\^12\catcode `\_12\catcode `\%12\relax}%
\providecommand \@@startlink[1]{}%
\providecommand \@@endlink[0]{}%
\providecommand \url  [0]{\begingroup\@sanitize@url \@url }%
\providecommand \@url [1]{\endgroup\@href {#1}{\urlprefix }}%
\providecommand \urlprefix  [0]{URL }%
\providecommand \Eprint [0]{\href }%
\providecommand \doibase [0]{http://dx.doi.org/}%
\providecommand \selectlanguage [0]{\@gobble}%
\providecommand \bibinfo  [0]{\@secondoftwo}%
\providecommand \bibfield  [0]{\@secondoftwo}%
\providecommand \translation [1]{[#1]}%
\providecommand \BibitemOpen [0]{}%
\providecommand \bibitemStop [0]{}%
\providecommand \bibitemNoStop [0]{.\EOS\space}%
\providecommand \EOS [0]{\spacefactor3000\relax}%
\providecommand \BibitemShut  [1]{\csname bibitem#1\endcsname}%
\let\auto@bib@innerbib\@empty
%</preamble>
\bibitem [{\citenamefont {Lepine}(1972)}]{Lepine:prb1972}%
  \BibitemOpen
  \bibfield  {author} {\bibinfo {author} {\bibfnamefont {D.~J.}\ \bibnamefont
  {Lepine}},\ }\href@noop {} {\bibfield  {journal} {\bibinfo  {journal}
  {Physical Review B}\ }\textbf {\bibinfo {volume} {6}},\ \bibinfo {pages}
  {436} (\bibinfo {year} {1972})}\BibitemShut {NoStop}%
\bibitem [{\citenamefont {Stutzmann}\ \emph {et~al.}(2000)\citenamefont
  {Stutzmann}, \citenamefont {Brandt},\ and\ \citenamefont
  {Bayerl}}]{Stutzmann:jncs2000}%
  \BibitemOpen
  \bibfield  {author} {\bibinfo {author} {\bibfnamefont {M.}~\bibnamefont
  {Stutzmann}}, \bibinfo {author} {\bibfnamefont {M.~S.}\ \bibnamefont
  {Brandt}}, \ and\ \bibinfo {author} {\bibfnamefont {M.~W.}\ \bibnamefont
  {Bayerl}},\ }\href@noop {} {\bibfield  {journal} {\bibinfo  {journal}
  {Journal of Non-Crystalline Solids}\ }\textbf {\bibinfo {volume} {266-269}},\
  \bibinfo {pages} {22} (\bibinfo {year} {2000})}\BibitemShut {NoStop}%
\bibitem [{\citenamefont {Geschwind}\ \emph {et~al.}(1959)\citenamefont
  {Geschwind}, \citenamefont {Collins},\ and\ \citenamefont
  {Schawlow}}]{Geschwind:prl1959}%
  \BibitemOpen
  \bibfield  {author} {\bibinfo {author} {\bibfnamefont {S.}~\bibnamefont
  {Geschwind}}, \bibinfo {author} {\bibfnamefont {R.~J.}\ \bibnamefont
  {Collins}}, \ and\ \bibinfo {author} {\bibfnamefont {A.~L.}\ \bibnamefont
  {Schawlow}},\ }\href@noop {} {\bibfield  {journal} {\bibinfo  {journal}
  {Physical Review Letters}\ }\textbf {\bibinfo {volume} {3}},\ \bibinfo
  {pages} {545} (\bibinfo {year} {1959})}\BibitemShut {NoStop}%
\bibitem [{\citenamefont {Brossel}\ \emph {et~al.}(1959)\citenamefont
  {Brossel}, \citenamefont {Geschwind},\ and\ \citenamefont
  {Schawlow}}]{Brossel:prl1959}%
  \BibitemOpen
  \bibfield  {author} {\bibinfo {author} {\bibfnamefont {J.}~\bibnamefont
  {Brossel}}, \bibinfo {author} {\bibfnamefont {G.}~\bibnamefont {Geschwind}},
  \ and\ \bibinfo {author} {\bibfnamefont {A.~L.}\ \bibnamefont {Schawlow}},\
  }\href@noop {} {\bibfield  {journal} {\bibinfo  {journal} {Physical Review
  Letters}\ }\textbf {\bibinfo {volume} {3}},\ \bibinfo {pages} {548} (\bibinfo
  {year} {1959})}\BibitemShut {NoStop}%
\bibitem [{\citenamefont {Cavenett}(1981)}]{Cavenett:ap1981}%
  \BibitemOpen
  \bibfield  {author} {\bibinfo {author} {\bibfnamefont {B.~C.}\ \bibnamefont
  {Cavenett}},\ }\href@noop {} {\bibfield  {journal} {\bibinfo  {journal}
  {Advances in Physics}\ }\textbf {\bibinfo {volume} {30}},\ \bibinfo {pages}
  {475} (\bibinfo {year} {1981})}\BibitemShut {NoStop}%
\bibitem [{\citenamefont {Street}(1982)}]{Street:prb1982}%
  \BibitemOpen
  \bibfield  {author} {\bibinfo {author} {\bibfnamefont {R.~A.}\ \bibnamefont
  {Street}},\ }\href@noop {} {\bibfield  {journal} {\bibinfo  {journal}
  {Physical Review B}\ }\textbf {\bibinfo {volume} {26}},\ \bibinfo {pages}
  {3588} (\bibinfo {year} {1982})}\BibitemShut {NoStop}%
\bibitem [{\citenamefont {Depinna}\ \emph
  {et~al.}(1982{\natexlab{a}})\citenamefont {Depinna}, \citenamefont
  {Cavenett}, \citenamefont {Austin}, \citenamefont {Searle}, \citenamefont
  {Thompson}, \citenamefont {Allison},\ and\ \citenamefont
  {Comberd}}]{Depinna:pmb1982_1st}%
  \BibitemOpen
  \bibfield  {author} {\bibinfo {author} {\bibfnamefont {S.}~\bibnamefont
  {Depinna}}, \bibinfo {author} {\bibfnamefont {B.~C.}\ \bibnamefont
  {Cavenett}}, \bibinfo {author} {\bibfnamefont {I.~G.}\ \bibnamefont
  {Austin}}, \bibinfo {author} {\bibfnamefont {T.~M.}\ \bibnamefont {Searle}},
  \bibinfo {author} {\bibfnamefont {M.~J.}\ \bibnamefont {Thompson}}, \bibinfo
  {author} {\bibfnamefont {J.}~\bibnamefont {Allison}}, \ and\ \bibinfo
  {author} {\bibfnamefont {P.~G.~L.}\ \bibnamefont {Comberd}},\ }\href@noop {}
  {\bibfield  {journal} {\bibinfo  {journal} {Philosophical Magazine B}\
  }\textbf {\bibinfo {volume} {46}},\ \bibinfo {pages} {473 } (\bibinfo {year}
  {1982}{\natexlab{a}})}\BibitemShut {NoStop}%
\bibitem [{\citenamefont {Lifshitz}\ \emph {et~al.}(2004)\citenamefont
  {Lifshitz}, \citenamefont {Fradkin}, \citenamefont {Glozman},\ and\
  \citenamefont {Langof}}]{Lifshitz:arpc2004}%
  \BibitemOpen
  \bibfield  {author} {\bibinfo {author} {\bibfnamefont {E.}~\bibnamefont
  {Lifshitz}}, \bibinfo {author} {\bibfnamefont {L.}~\bibnamefont {Fradkin}},
  \bibinfo {author} {\bibfnamefont {A.}~\bibnamefont {Glozman}}, \ and\
  \bibinfo {author} {\bibfnamefont {L.}~\bibnamefont {Langof}},\ }\href@noop {}
  {\bibfield  {journal} {\bibinfo  {journal} {Annual Review of Physical
  Chemistry}\ }\textbf {\bibinfo {volume} {55}},\ \bibinfo {pages} {509}
  (\bibinfo {year} {2004})}\BibitemShut {NoStop}%
\bibitem [{\citenamefont {Chen}(2000)}]{Chen:tsf2000}%
  \BibitemOpen
  \bibfield  {author} {\bibinfo {author} {\bibfnamefont {W.~M.}\ \bibnamefont
  {Chen}},\ }\href@noop {} {\bibfield  {journal} {\bibinfo  {journal} {Thin
  Solid Films}\ }\textbf {\bibinfo {volume} {364}},\ \bibinfo {pages} {45}
  (\bibinfo {year} {2000})}\BibitemShut {NoStop}%
\bibitem [{\citenamefont {Boulitrop}(1983)}]{Boulitrop:prb1983}%
  \BibitemOpen
  \bibfield  {author} {\bibinfo {author} {\bibfnamefont {F.}~\bibnamefont
  {Boulitrop}},\ }\href@noop {} {\bibfield  {journal} {\bibinfo  {journal}
  {Physical Review B}\ }\textbf {\bibinfo {volume} {28}},\ \bibinfo {pages}
  {6192} (\bibinfo {year} {1983})}\BibitemShut {NoStop}%
\bibitem [{\citenamefont {McCamey}\ \emph {et~al.}(2006)\citenamefont
  {McCamey}, \citenamefont {Huebl}, \citenamefont {Brandt}, \citenamefont
  {Hutchison}, \citenamefont {McCallum}, \citenamefont {Clark},\ and\
  \citenamefont {Hamilton}}]{McCamey:APL2006}%
  \BibitemOpen
  \bibfield  {author} {\bibinfo {author} {\bibfnamefont {D.~R.}\ \bibnamefont
  {McCamey}}, \bibinfo {author} {\bibfnamefont {H.}~\bibnamefont {Huebl}},
  \bibinfo {author} {\bibfnamefont {M.~S.}\ \bibnamefont {Brandt}}, \bibinfo
  {author} {\bibfnamefont {W.~D.}\ \bibnamefont {Hutchison}}, \bibinfo {author}
  {\bibfnamefont {J.~C.}\ \bibnamefont {McCallum}}, \bibinfo {author}
  {\bibfnamefont {R.~G.}\ \bibnamefont {Clark}}, \ and\ \bibinfo {author}
  {\bibfnamefont {A.~R.}\ \bibnamefont {Hamilton}},\ }\href@noop {} {\bibfield
  {journal} {\bibinfo  {journal} {Applied Physics Letters}\ }\textbf {\bibinfo
  {volume} {89}},\ \bibinfo {pages} {182115} (\bibinfo {year}
  {2006})}\BibitemShut {NoStop}%
\bibitem [{\citenamefont {Dunstan}\ and\ \citenamefont
  {Davies}(1979)}]{Dunstan:jpc1979}%
  \BibitemOpen
  \bibfield  {author} {\bibinfo {author} {\bibfnamefont {D.~J.}\ \bibnamefont
  {Dunstan}}\ and\ \bibinfo {author} {\bibfnamefont {J.~J.}\ \bibnamefont
  {Davies}},\ }\href@noop {} {\bibfield  {journal} {\bibinfo  {journal}
  {Journal of Physics C: Solid State Physics}\ }\textbf {\bibinfo {volume}
  {12}},\ \bibinfo {pages} {2927} (\bibinfo {year} {1979})}\BibitemShut
  {NoStop}%
\bibitem [{\citenamefont {Depinna}\ \emph
  {et~al.}(1982{\natexlab{b}})\citenamefont {Depinna}, \citenamefont
  {Cavenett}, \citenamefont {Austin}, \citenamefont {Searle}, \citenamefont
  {Thompson}, \citenamefont {Allison},\ and\ \citenamefont
  {Comberd}}]{Depinna:pmb1982_2nd}%
  \BibitemOpen
  \bibfield  {author} {\bibinfo {author} {\bibfnamefont {S.}~\bibnamefont
  {Depinna}}, \bibinfo {author} {\bibfnamefont {B.~C.}\ \bibnamefont
  {Cavenett}}, \bibinfo {author} {\bibfnamefont {I.~G.}\ \bibnamefont
  {Austin}}, \bibinfo {author} {\bibfnamefont {T.~M.}\ \bibnamefont {Searle}},
  \bibinfo {author} {\bibfnamefont {M.~J.}\ \bibnamefont {Thompson}}, \bibinfo
  {author} {\bibfnamefont {J.}~\bibnamefont {Allison}}, \ and\ \bibinfo
  {author} {\bibfnamefont {P.~G.~L.}\ \bibnamefont {Comberd}},\ }\href@noop {}
  {\bibfield  {journal} {\bibinfo  {journal} {Philosophical Magazine B}\
  }\textbf {\bibinfo {volume} {46}},\ \bibinfo {pages} {501 } (\bibinfo {year}
  {1982}{\natexlab{b}})}\BibitemShut {NoStop}%
\bibitem [{\citenamefont {Morigaki}\ \emph {et~al.}(1978)\citenamefont
  {Morigaki}, \citenamefont {Dunstan}, \citenamefont {Cavenett}, \citenamefont
  {Dawson}, \citenamefont {Nicholls}, \citenamefont {Nitta},\ and\
  \citenamefont {Shimakawa}}]{Morigaki:ssc1978}%
  \BibitemOpen
  \bibfield  {author} {\bibinfo {author} {\bibfnamefont {K.}~\bibnamefont
  {Morigaki}}, \bibinfo {author} {\bibfnamefont {D.~J.}\ \bibnamefont
  {Dunstan}}, \bibinfo {author} {\bibfnamefont {B.~C.}\ \bibnamefont
  {Cavenett}}, \bibinfo {author} {\bibfnamefont {P.}~\bibnamefont {Dawson}},
  \bibinfo {author} {\bibfnamefont {J.~E.}\ \bibnamefont {Nicholls}}, \bibinfo
  {author} {\bibfnamefont {S.}~\bibnamefont {Nitta}}, \ and\ \bibinfo {author}
  {\bibfnamefont {K.}~\bibnamefont {Shimakawa}},\ }\href@noop {} {\bibfield
  {journal} {\bibinfo  {journal} {Solid State Communications}\ }\textbf
  {\bibinfo {volume} {26}},\ \bibinfo {pages} {981} (\bibinfo {year}
  {1978})}\BibitemShut {NoStop}%
\bibitem [{\citenamefont {Lenahan}\ and\ \citenamefont
  {Schubert}(1984)}]{Lenahan:prb1984}%
  \BibitemOpen
  \bibfield  {author} {\bibinfo {author} {\bibfnamefont {P.~M.}\ \bibnamefont
  {Lenahan}}\ and\ \bibinfo {author} {\bibfnamefont {W.~K.}\ \bibnamefont
  {Schubert}},\ }\href@noop {} {\bibfield  {journal} {\bibinfo  {journal}
  {Physical Review B}\ }\textbf {\bibinfo {volume} {30}},\ \bibinfo {pages} {3}
  (\bibinfo {year} {1984})}\BibitemShut {NoStop}%
\bibitem [{\citenamefont {Dersch}\ \emph {et~al.}(1983)\citenamefont {Dersch},
  \citenamefont {Schweitzer},\ and\ \citenamefont {Stuke}}]{Dersch:prb1983}%
  \BibitemOpen
  \bibfield  {author} {\bibinfo {author} {\bibfnamefont {H.}~\bibnamefont
  {Dersch}}, \bibinfo {author} {\bibfnamefont {L.}~\bibnamefont {Schweitzer}},
  \ and\ \bibinfo {author} {\bibfnamefont {J.}~\bibnamefont {Stuke}},\
  }\href@noop {} {\bibfield  {journal} {\bibinfo  {journal} {Physical Review
  B}\ }\textbf {\bibinfo {volume} {28}},\ \bibinfo {pages} {4678} (\bibinfo
  {year} {1983})}\BibitemShut {NoStop}%
\bibitem [{\citenamefont {Morishita}\ \emph {et~al.}(2009)\citenamefont
  {Morishita}, \citenamefont {Vlasenko}, \citenamefont {Tanaka}, \citenamefont
  {Semba}, \citenamefont {Sawano}, \citenamefont {Shiraki}, \citenamefont
  {Eto},\ and\ \citenamefont {Itoh}}]{Morishita:prb2009}%
  \BibitemOpen
  \bibfield  {author} {\bibinfo {author} {\bibfnamefont {H.}~\bibnamefont
  {Morishita}}, \bibinfo {author} {\bibfnamefont {L.~S.}\ \bibnamefont
  {Vlasenko}}, \bibinfo {author} {\bibfnamefont {H.}~\bibnamefont {Tanaka}},
  \bibinfo {author} {\bibfnamefont {K.}~\bibnamefont {Semba}}, \bibinfo
  {author} {\bibfnamefont {K.}~\bibnamefont {Sawano}}, \bibinfo {author}
  {\bibfnamefont {Y.}~\bibnamefont {Shiraki}}, \bibinfo {author} {\bibfnamefont
  {M.}~\bibnamefont {Eto}}, \ and\ \bibinfo {author} {\bibfnamefont {K.~M.}\
  \bibnamefont {Itoh}},\ }\href@noop {} {\bibfield  {journal} {\bibinfo
  {journal} {Physical Review B}\ }\textbf {\bibinfo {volume} {80}},\ \bibinfo
  {pages} {205206} (\bibinfo {year} {2009})}\BibitemShut {NoStop}%
\bibitem [{\citenamefont {Lifshitz}\ \emph {et~al.}(1997)\citenamefont
  {Lifshitz}, \citenamefont {Bykov}, \citenamefont {Yassen},\ and\
  \citenamefont {Chen-Esterlit}}]{Lifshitz:cpl1997}%
  \BibitemOpen
  \bibfield  {author} {\bibinfo {author} {\bibfnamefont {E.}~\bibnamefont
  {Lifshitz}}, \bibinfo {author} {\bibfnamefont {L.}~\bibnamefont {Bykov}},
  \bibinfo {author} {\bibfnamefont {M.}~\bibnamefont {Yassen}}, \ and\ \bibinfo
  {author} {\bibfnamefont {Z.}~\bibnamefont {Chen-Esterlit}},\ }\href@noop {}
  {\bibfield  {journal} {\bibinfo  {journal} {Chemical Physics Letters}\
  }\textbf {\bibinfo {volume} {273}},\ \bibinfo {pages} {381} (\bibinfo {year}
  {1997})}\BibitemShut {NoStop}%
\bibitem [{\citenamefont {Lifshitz}\ \emph {et~al.}(2000)\citenamefont
  {Lifshitz}, \citenamefont {Glozman}, \citenamefont {Litvin},\ and\
  \citenamefont {Porteanu}}]{Lifshitz:jpc2000}%
  \BibitemOpen
  \bibfield  {author} {\bibinfo {author} {\bibfnamefont {E.}~\bibnamefont
  {Lifshitz}}, \bibinfo {author} {\bibfnamefont {A.}~\bibnamefont {Glozman}},
  \bibinfo {author} {\bibfnamefont {I.~D.}\ \bibnamefont {Litvin}}, \ and\
  \bibinfo {author} {\bibfnamefont {H.}~\bibnamefont {Porteanu}},\ }\href@noop
  {} {\bibfield  {journal} {\bibinfo  {journal} {The Journal of Physical
  Chemistry B}\ }\textbf {\bibinfo {volume} {104}},\ \bibinfo {pages} {10449}
  (\bibinfo {year} {2000})}\BibitemShut {NoStop}%
\bibitem [{\citenamefont {Langof}\ \emph {et~al.}(2002)\citenamefont {Langof},
  \citenamefont {Ehrenfreund}, \citenamefont {Lifshitz}, \citenamefont
  {Micic},\ and\ \citenamefont {Nozik}}]{Langof:jpcb2002}%
  \BibitemOpen
  \bibfield  {author} {\bibinfo {author} {\bibfnamefont {L.}~\bibnamefont
  {Langof}}, \bibinfo {author} {\bibfnamefont {E.}~\bibnamefont {Ehrenfreund}},
  \bibinfo {author} {\bibfnamefont {E.}~\bibnamefont {Lifshitz}}, \bibinfo
  {author} {\bibfnamefont {O.~I.}\ \bibnamefont {Micic}}, \ and\ \bibinfo
  {author} {\bibfnamefont {A.~J.}\ \bibnamefont {Nozik}},\ }\href@noop {}
  {\bibfield  {journal} {\bibinfo  {journal} {Journal of Physical Chemistry B}\
  }\textbf {\bibinfo {volume} {106}},\ \bibinfo {pages} {1606} (\bibinfo {year}
  {2002})}\BibitemShut {NoStop}%
\bibitem [{\citenamefont {Kanschat}\ \emph {et~al.}(2000)\citenamefont
  {Kanschat}, \citenamefont {Lips},\ and\ \citenamefont
  {Fuhs}}]{Kanschat:jncs2000}%
  \BibitemOpen
  \bibfield  {author} {\bibinfo {author} {\bibfnamefont {P.}~\bibnamefont
  {Kanschat}}, \bibinfo {author} {\bibfnamefont {K.}~\bibnamefont {Lips}}, \
  and\ \bibinfo {author} {\bibfnamefont {W.}~\bibnamefont {Fuhs}},\ }\href@noop
  {} {\bibfield  {journal} {\bibinfo  {journal} {Journal of Non-Crystalline
  Solids}\ }\textbf {\bibinfo {volume} {266-269}},\ \bibinfo {pages} {5}
  (\bibinfo {year} {2000})}\BibitemShut {NoStop}%
\bibitem [{\citenamefont {Glaser}\ \emph {et~al.}(1995)\citenamefont {Glaser},
  \citenamefont {Kennedy}, \citenamefont {Doverspike}, \citenamefont {Rowland},
  \citenamefont {Gaskill}, \citenamefont {Freitas}, \citenamefont {Asif~Khan},
  \citenamefont {Olson}, \citenamefont {Kuznia},\ and\ \citenamefont
  {Wickenden}}]{Glaser:prb1995}%
  \BibitemOpen
  \bibfield  {author} {\bibinfo {author} {\bibfnamefont {E.~R.}\ \bibnamefont
  {Glaser}}, \bibinfo {author} {\bibfnamefont {T.~A.}\ \bibnamefont {Kennedy}},
  \bibinfo {author} {\bibfnamefont {K.}~\bibnamefont {Doverspike}}, \bibinfo
  {author} {\bibfnamefont {L.~B.}\ \bibnamefont {Rowland}}, \bibinfo {author}
  {\bibfnamefont {D.~K.}\ \bibnamefont {Gaskill}}, \bibinfo {author}
  {\bibfnamefont {J.~A.}\ \bibnamefont {Freitas}}, \bibinfo {author}
  {\bibfnamefont {M.}~\bibnamefont {Asif~Khan}}, \bibinfo {author}
  {\bibfnamefont {D.~T.}\ \bibnamefont {Olson}}, \bibinfo {author}
  {\bibfnamefont {J.~N.}\ \bibnamefont {Kuznia}}, \ and\ \bibinfo {author}
  {\bibfnamefont {D.~K.}\ \bibnamefont {Wickenden}},\ }\href@noop {} {\bibfield
   {journal} {\bibinfo  {journal} {Physical Review B}\ }\textbf {\bibinfo
  {volume} {51}},\ \bibinfo {pages} {13326} (\bibinfo {year}
  {1995})}\BibitemShut {NoStop}%
\bibitem [{\citenamefont {Glaser}\ \emph {et~al.}(2003)\citenamefont {Glaser},
  \citenamefont {Freitas}, \citenamefont {Shanabrook}, \citenamefont {Koleske},
  \citenamefont {Lee}, \citenamefont {Park},\ and\ \citenamefont
  {Han}}]{Glaser:prb2003}%
  \BibitemOpen
  \bibfield  {author} {\bibinfo {author} {\bibfnamefont {E.~R.}\ \bibnamefont
  {Glaser}}, \bibinfo {author} {\bibfnamefont {J.~A.}\ \bibnamefont {Freitas}},
  \bibinfo {author} {\bibfnamefont {B.~V.}\ \bibnamefont {Shanabrook}},
  \bibinfo {author} {\bibfnamefont {D.~D.}\ \bibnamefont {Koleske}}, \bibinfo
  {author} {\bibfnamefont {S.~K.}\ \bibnamefont {Lee}}, \bibinfo {author}
  {\bibfnamefont {S.~S.}\ \bibnamefont {Park}}, \ and\ \bibinfo {author}
  {\bibfnamefont {J.~Y.}\ \bibnamefont {Han}},\ }\href@noop {} {\bibfield
  {journal} {\bibinfo  {journal} {Physical Review B}\ }\textbf {\bibinfo
  {volume} {68}},\ \bibinfo {pages} {195201} (\bibinfo {year}
  {2003})}\BibitemShut {NoStop}%
\bibitem [{\citenamefont {Son}\ \emph {et~al.}(1997)\citenamefont {Son},
  \citenamefont {S$\textrm{\"{o}}$rman}, \citenamefont {Chen}, \citenamefont
  {Hallin}, \citenamefont {Kordina}, \citenamefont {Monemar},\ and\
  \citenamefont {Janz$\textrm{\'{e}}$n}}]{Son:prb1997}%
  \BibitemOpen
  \bibfield  {author} {\bibinfo {author} {\bibfnamefont {N.~T.}\ \bibnamefont
  {Son}}, \bibinfo {author} {\bibfnamefont {E.}~\bibnamefont
  {S$\textrm{\"{o}}$rman}}, \bibinfo {author} {\bibfnamefont {W.~M.}\
  \bibnamefont {Chen}}, \bibinfo {author} {\bibfnamefont {C.}~\bibnamefont
  {Hallin}}, \bibinfo {author} {\bibfnamefont {O.}~\bibnamefont {Kordina}},
  \bibinfo {author} {\bibfnamefont {B.}~\bibnamefont {Monemar}}, \ and\
  \bibinfo {author} {\bibfnamefont {E.}~\bibnamefont {Janz$\textrm{\'{e}}$n}},\
  }\href@noop {} {\bibfield  {journal} {\bibinfo  {journal} {Physical Review
  B}\ }\textbf {\bibinfo {volume} {55}},\ \bibinfo {pages} {2863} (\bibinfo
  {year} {1997})}\BibitemShut {NoStop}%
\bibitem [{\citenamefont {Oort}\ \emph {et~al.}(1988)\citenamefont {Oort},
  \citenamefont {Manson},\ and\ \citenamefont {Glasbeek}}]{Oort:jpc1988}%
  \BibitemOpen
  \bibfield  {author} {\bibinfo {author} {\bibfnamefont {E.~v.}\ \bibnamefont
  {Oort}}, \bibinfo {author} {\bibfnamefont {N.~B.}\ \bibnamefont {Manson}}, \
  and\ \bibinfo {author} {\bibfnamefont {M.}~\bibnamefont {Glasbeek}},\
  }\href@noop {} {\bibfield  {journal} {\bibinfo  {journal} {Journal of Physics
  C: Solid State Physics}\ }\textbf {\bibinfo {volume} {21}},\ \bibinfo {pages}
  {4385} (\bibinfo {year} {1988})}\BibitemShut {NoStop}%
\bibitem [{\citenamefont {Nazare}\ \emph {et~al.}(1995)\citenamefont {Nazare},
  \citenamefont {Mason}, \citenamefont {Watkins},\ and\ \citenamefont
  {Kanda}}]{Nazare:prb1995}%
  \BibitemOpen
  \bibfield  {author} {\bibinfo {author} {\bibfnamefont {M.~H.}\ \bibnamefont
  {Nazare}}, \bibinfo {author} {\bibfnamefont {P.~W.}\ \bibnamefont {Mason}},
  \bibinfo {author} {\bibfnamefont {G.~D.}\ \bibnamefont {Watkins}}, \ and\
  \bibinfo {author} {\bibfnamefont {H.}~\bibnamefont {Kanda}},\ }\href@noop {}
  {\bibfield  {journal} {\bibinfo  {journal} {Physical Review B}\ }\textbf
  {\bibinfo {volume} {51}},\ \bibinfo {pages} {16741} (\bibinfo {year}
  {1995})}\BibitemShut {NoStop}%
\bibitem [{\citenamefont {Nizovtsev}\ \emph {et~al.}(2001)\citenamefont
  {Nizovtsev}, \citenamefont {Kilin}, \citenamefont {Tietz}, \citenamefont
  {Jelezko},\ and\ \citenamefont {Wrachtrup}}]{Nizovtsev:pb2001}%
  \BibitemOpen
  \bibfield  {author} {\bibinfo {author} {\bibfnamefont {A.~P.}\ \bibnamefont
  {Nizovtsev}}, \bibinfo {author} {\bibfnamefont {S.~Y.}\ \bibnamefont
  {Kilin}}, \bibinfo {author} {\bibfnamefont {C.}~\bibnamefont {Tietz}},
  \bibinfo {author} {\bibfnamefont {F.}~\bibnamefont {Jelezko}}, \ and\
  \bibinfo {author} {\bibfnamefont {J.}~\bibnamefont {Wrachtrup}},\ }\href@noop
  {} {\bibfield  {journal} {\bibinfo  {journal} {Physica B: Condensed Matter}\
  }\textbf {\bibinfo {volume} {308-310}},\ \bibinfo {pages} {608} (\bibinfo
  {year} {2001})}\BibitemShut {NoStop}%
\bibitem [{\citenamefont {Swanson}\ \emph {et~al.}(1990)\citenamefont
  {Swanson}, \citenamefont {Shinar},\ and\ \citenamefont
  {Yoshino}}]{Swanson:prl1990}%
  \BibitemOpen
  \bibfield  {author} {\bibinfo {author} {\bibfnamefont {L.~S.}\ \bibnamefont
  {Swanson}}, \bibinfo {author} {\bibfnamefont {J.}~\bibnamefont {Shinar}}, \
  and\ \bibinfo {author} {\bibfnamefont {K.}~\bibnamefont {Yoshino}},\
  }\href@noop {} {\bibfield  {journal} {\bibinfo  {journal} {Physical Review
  Letters}\ }\textbf {\bibinfo {volume} {65}},\ \bibinfo {pages} {1140}
  (\bibinfo {year} {1990})}\BibitemShut {NoStop}%
\bibitem [{\citenamefont {Swanson}\ \emph {et~al.}(1991)\citenamefont
  {Swanson}, \citenamefont {Lane}, \citenamefont {Shinar},\ and\ \citenamefont
  {Wudl}}]{Swanson:prb1991}%
  \BibitemOpen
  \bibfield  {author} {\bibinfo {author} {\bibfnamefont {L.~S.}\ \bibnamefont
  {Swanson}}, \bibinfo {author} {\bibfnamefont {P.~A.}\ \bibnamefont {Lane}},
  \bibinfo {author} {\bibfnamefont {J.}~\bibnamefont {Shinar}}, \ and\ \bibinfo
  {author} {\bibfnamefont {F.}~\bibnamefont {Wudl}},\ }\href@noop {} {\bibfield
   {journal} {\bibinfo  {journal} {Physical Review B}\ }\textbf {\bibinfo
  {volume} {44}},\ \bibinfo {pages} {10617} (\bibinfo {year}
  {1991})}\BibitemShut {NoStop}%
\bibitem [{\citenamefont {Swanson}\ \emph {et~al.}(1992)\citenamefont
  {Swanson}, \citenamefont {Shinar}, \citenamefont {Brown}, \citenamefont
  {Bradley}, \citenamefont {Friend}, \citenamefont {Burn}, \citenamefont
  {Kraft},\ and\ \citenamefont {Holmes}}]{Swanson:prb1992}%
  \BibitemOpen
  \bibfield  {author} {\bibinfo {author} {\bibfnamefont {L.~S.}\ \bibnamefont
  {Swanson}}, \bibinfo {author} {\bibfnamefont {J.}~\bibnamefont {Shinar}},
  \bibinfo {author} {\bibfnamefont {A.~R.}\ \bibnamefont {Brown}}, \bibinfo
  {author} {\bibfnamefont {D.~D.~C.}\ \bibnamefont {Bradley}}, \bibinfo
  {author} {\bibfnamefont {R.~H.}\ \bibnamefont {Friend}}, \bibinfo {author}
  {\bibfnamefont {P.~L.}\ \bibnamefont {Burn}}, \bibinfo {author}
  {\bibfnamefont {A.}~\bibnamefont {Kraft}}, \ and\ \bibinfo {author}
  {\bibfnamefont {A.~B.}\ \bibnamefont {Holmes}},\ }\href@noop {} {\bibfield
  {journal} {\bibinfo  {journal} {Physical Review B}\ }\textbf {\bibinfo
  {volume} {46}},\ \bibinfo {pages} {15072} (\bibinfo {year}
  {1992})}\BibitemShut {NoStop}%
\bibitem [{\citenamefont {Graupner}\ \emph {et~al.}(1996)\citenamefont
  {Graupner}, \citenamefont {Partee}, \citenamefont {Shinar}, \citenamefont
  {Leising},\ and\ \citenamefont {Scherf}}]{Graupner:prl1996}%
  \BibitemOpen
  \bibfield  {author} {\bibinfo {author} {\bibfnamefont {W.}~\bibnamefont
  {Graupner}}, \bibinfo {author} {\bibfnamefont {J.}~\bibnamefont {Partee}},
  \bibinfo {author} {\bibfnamefont {J.}~\bibnamefont {Shinar}}, \bibinfo
  {author} {\bibfnamefont {G.}~\bibnamefont {Leising}}, \ and\ \bibinfo
  {author} {\bibfnamefont {U.}~\bibnamefont {Scherf}},\ }\href@noop {}
  {\bibfield  {journal} {\bibinfo  {journal} {Physical Review Letters}\
  }\textbf {\bibinfo {volume} {77}},\ \bibinfo {pages} {2033} (\bibinfo {year}
  {1996})}\BibitemShut {NoStop}%
\bibitem [{\citenamefont {Greenham}\ \emph {et~al.}(1996)\citenamefont
  {Greenham}, \citenamefont {Shinar}, \citenamefont {Partee}, \citenamefont
  {Lane}, \citenamefont {Amir}, \citenamefont {Lu},\ and\ \citenamefont
  {Friend}}]{Greenham:prb1996}%
  \BibitemOpen
  \bibfield  {author} {\bibinfo {author} {\bibfnamefont {N.~C.}\ \bibnamefont
  {Greenham}}, \bibinfo {author} {\bibfnamefont {J.}~\bibnamefont {Shinar}},
  \bibinfo {author} {\bibfnamefont {J.}~\bibnamefont {Partee}}, \bibinfo
  {author} {\bibfnamefont {P.~A.}\ \bibnamefont {Lane}}, \bibinfo {author}
  {\bibfnamefont {O.}~\bibnamefont {Amir}}, \bibinfo {author} {\bibfnamefont
  {F.}~\bibnamefont {Lu}}, \ and\ \bibinfo {author} {\bibfnamefont {R.~H.}\
  \bibnamefont {Friend}},\ }\href@noop {} {\bibfield  {journal} {\bibinfo
  {journal} {Physical Review B}\ }\textbf {\bibinfo {volume} {53}},\ \bibinfo
  {pages} {13528} (\bibinfo {year} {1996})}\BibitemShut {NoStop}%
\bibitem [{\citenamefont {Lane}\ \emph {et~al.}(1997)\citenamefont {Lane},
  \citenamefont {Wei},\ and\ \citenamefont {Vardeny}}]{Lane:prb1997}%
  \BibitemOpen
  \bibfield  {author} {\bibinfo {author} {\bibfnamefont {P.~A.}\ \bibnamefont
  {Lane}}, \bibinfo {author} {\bibfnamefont {X.}~\bibnamefont {Wei}}, \ and\
  \bibinfo {author} {\bibfnamefont {Z.~V.}\ \bibnamefont {Vardeny}},\
  }\href@noop {} {\bibfield  {journal} {\bibinfo  {journal} {Physical Review
  B}\ }\textbf {\bibinfo {volume} {56}},\ \bibinfo {pages} {4626} (\bibinfo
  {year} {1997})}\BibitemShut {NoStop}%
\bibitem [{\citenamefont {Silva}\ \emph {et~al.}(2001)\citenamefont {Silva},
  \citenamefont {Santos}, \citenamefont {Faria},\ and\ \citenamefont
  {Graeff}}]{Silva:pb2001}%
  \BibitemOpen
  \bibfield  {author} {\bibinfo {author} {\bibfnamefont {G.~B.}\ \bibnamefont
  {Silva}}, \bibinfo {author} {\bibfnamefont {L.~F.}\ \bibnamefont {Santos}},
  \bibinfo {author} {\bibfnamefont {R.~M.}\ \bibnamefont {Faria}}, \ and\
  \bibinfo {author} {\bibfnamefont {C.~F.~O.}\ \bibnamefont {Graeff}},\
  }\href@noop {} {\bibfield  {journal} {\bibinfo  {journal} {Physica B:
  Condensed Matter}\ }\textbf {\bibinfo {volume} {308-310}},\ \bibinfo {pages}
  {1078} (\bibinfo {year} {2001})}\BibitemShut {NoStop}%
\bibitem [{\citenamefont {List}\ \emph {et~al.}(2002)\citenamefont {List},
  \citenamefont {Scherf}, \citenamefont {M$\textrm{\"{u}}$llen}, \citenamefont
  {Graupner}, \citenamefont {Kim},\ and\ \citenamefont
  {Shinar}}]{List:prb2002}%
  \BibitemOpen
  \bibfield  {author} {\bibinfo {author} {\bibfnamefont {E.~J.~W.}\
  \bibnamefont {List}}, \bibinfo {author} {\bibfnamefont {U.}~\bibnamefont
  {Scherf}}, \bibinfo {author} {\bibfnamefont {K.}~\bibnamefont
  {M$\textrm{\"{u}}$llen}}, \bibinfo {author} {\bibfnamefont {W.}~\bibnamefont
  {Graupner}}, \bibinfo {author} {\bibfnamefont {C.~H.}\ \bibnamefont {Kim}}, \
  and\ \bibinfo {author} {\bibfnamefont {J.}~\bibnamefont {Shinar}},\
  }\href@noop {} {\bibfield  {journal} {\bibinfo  {journal} {Physical Review
  B}\ }\textbf {\bibinfo {volume} {66}},\ \bibinfo {pages} {235203} (\bibinfo
  {year} {2002})}\BibitemShut {NoStop}%
\bibitem [{\citenamefont {$\textrm{\"{O}}$sterbacka}\ \emph
  {et~al.}(2003)\citenamefont {$\textrm{\"{O}}$sterbacka}, \citenamefont
  {Wohlgenannt}, \citenamefont {Shkunov}, \citenamefont {Chinn},\ and\
  \citenamefont {Vardeny}}]{Osterbacka:jcp2003}%
  \BibitemOpen
  \bibfield  {author} {\bibinfo {author} {\bibfnamefont {R.}~\bibnamefont
  {$\textrm{\"{O}}$sterbacka}}, \bibinfo {author} {\bibfnamefont
  {M.}~\bibnamefont {Wohlgenannt}}, \bibinfo {author} {\bibfnamefont
  {M.}~\bibnamefont {Shkunov}}, \bibinfo {author} {\bibfnamefont
  {D.}~\bibnamefont {Chinn}}, \ and\ \bibinfo {author} {\bibfnamefont {Z.~V.}\
  \bibnamefont {Vardeny}},\ }\href@noop {} {\bibfield  {journal} {\bibinfo
  {journal} {The Journal of Chemical Physics}\ }\textbf {\bibinfo {volume}
  {118}},\ \bibinfo {pages} {8905} (\bibinfo {year} {2003})}\BibitemShut
  {NoStop}%
\bibitem [{\citenamefont {Lee}\ \emph {et~al.}(2005)\citenamefont {Lee},
  \citenamefont {Segal}, \citenamefont {Soos}, \citenamefont {Shinar},\ and\
  \citenamefont {Baldo}}]{lee:prl2005}%
  \BibitemOpen
  \bibfield  {author} {\bibinfo {author} {\bibfnamefont {M.~K.}\ \bibnamefont
  {Lee}}, \bibinfo {author} {\bibfnamefont {M.}~\bibnamefont {Segal}}, \bibinfo
  {author} {\bibfnamefont {Z.~G.}\ \bibnamefont {Soos}}, \bibinfo {author}
  {\bibfnamefont {J.}~\bibnamefont {Shinar}}, \ and\ \bibinfo {author}
  {\bibfnamefont {M.~A.}\ \bibnamefont {Baldo}},\ }\href@noop {} {\bibfield
  {journal} {\bibinfo  {journal} {Physical Review Letters}\ }\textbf {\bibinfo
  {volume} {94}},\ \bibinfo {pages} {137403} (\bibinfo {year}
  {2005})}\BibitemShut {NoStop}%
\bibitem [{\citenamefont {Segal}\ \emph {et~al.}(2005)\citenamefont {Segal},
  \citenamefont {Baldo}, \citenamefont {Lee}, \citenamefont {Shinar},\ and\
  \citenamefont {Soos}}]{Segal:prb2005}%
  \BibitemOpen
  \bibfield  {author} {\bibinfo {author} {\bibfnamefont {M.}~\bibnamefont
  {Segal}}, \bibinfo {author} {\bibfnamefont {M.~A.}\ \bibnamefont {Baldo}},
  \bibinfo {author} {\bibfnamefont {M.~K.}\ \bibnamefont {Lee}}, \bibinfo
  {author} {\bibfnamefont {J.}~\bibnamefont {Shinar}}, \ and\ \bibinfo {author}
  {\bibfnamefont {Z.~G.}\ \bibnamefont {Soos}},\ }\href@noop {} {\bibfield
  {journal} {\bibinfo  {journal} {Physical Review B}\ }\textbf {\bibinfo
  {volume} {71}},\ \bibinfo {pages} {245201} (\bibinfo {year}
  {2005})}\BibitemShut {NoStop}%
\bibitem [{\citenamefont {Yang}\ \emph {et~al.}(2008)\citenamefont {Yang},
  \citenamefont {Ehrenfreund}, \citenamefont {Wang}, \citenamefont {Drori},\
  and\ \citenamefont {Vardeny}}]{Yang:prb2008}%
  \BibitemOpen
  \bibfield  {author} {\bibinfo {author} {\bibfnamefont {C.~G.}\ \bibnamefont
  {Yang}}, \bibinfo {author} {\bibfnamefont {E.}~\bibnamefont {Ehrenfreund}},
  \bibinfo {author} {\bibfnamefont {F.}~\bibnamefont {Wang}}, \bibinfo {author}
  {\bibfnamefont {T.}~\bibnamefont {Drori}}, \ and\ \bibinfo {author}
  {\bibfnamefont {Z.~V.}\ \bibnamefont {Vardeny}},\ }\href@noop {} {\bibfield
  {journal} {\bibinfo  {journal} {Physical Review B}\ }\textbf {\bibinfo
  {volume} {78}},\ \bibinfo {pages} {6} (\bibinfo {year} {2008})}\BibitemShut
  {NoStop}%
\bibitem [{\citenamefont {Castro}\ \emph {et~al.}(2007)\citenamefont {Castro},
  \citenamefont {Silva}, \citenamefont {N$\textrm{\"{u}}$esch}, \citenamefont
  {Zuppiroli},\ and\ \citenamefont {Graeff}}]{Castro:oe2007}%
  \BibitemOpen
  \bibfield  {author} {\bibinfo {author} {\bibfnamefont {F.~A.}\ \bibnamefont
  {Castro}}, \bibinfo {author} {\bibfnamefont {G.~B.}\ \bibnamefont {Silva}},
  \bibinfo {author} {\bibfnamefont {F.}~\bibnamefont {N$\textrm{\"{u}}$esch}},
  \bibinfo {author} {\bibfnamefont {L.}~\bibnamefont {Zuppiroli}}, \ and\
  \bibinfo {author} {\bibfnamefont {C.~F.~O.}\ \bibnamefont {Graeff}},\
  }\href@noop {} {\bibfield  {journal} {\bibinfo  {journal} {Organic
  Electronics}\ }\textbf {\bibinfo {volume} {8}},\ \bibinfo {pages} {249}
  (\bibinfo {year} {2007})}\BibitemShut {NoStop}%
\bibitem [{\citenamefont {Li}\ \emph {et~al.}(2004)\citenamefont {Li},
  \citenamefont {Kim}, \citenamefont {Lane},\ and\ \citenamefont
  {Shinar}}]{Li:prb2004}%
  \BibitemOpen
  \bibfield  {author} {\bibinfo {author} {\bibfnamefont {G.}~\bibnamefont
  {Li}}, \bibinfo {author} {\bibfnamefont {C.~H.}\ \bibnamefont {Kim}},
  \bibinfo {author} {\bibfnamefont {P.~A.}\ \bibnamefont {Lane}}, \ and\
  \bibinfo {author} {\bibfnamefont {J.}~\bibnamefont {Shinar}},\ }\href@noop {}
  {\bibfield  {journal} {\bibinfo  {journal} {Physical Review B}\ }\textbf
  {\bibinfo {volume} {69}},\ \bibinfo {pages} {165311} (\bibinfo {year}
  {2004})}\BibitemShut {NoStop}%
\bibitem [{\citenamefont {Castro}\ \emph {et~al.}(2004)\citenamefont {Castro},
  \citenamefont {Silva}, \citenamefont {Santos}, \citenamefont {Faria},
  \citenamefont {N$\textrm{\"{u}}$esch}, \citenamefont {Zuppiroli},\ and\
  \citenamefont {Graeff}}]{Castro:jncs2004}%
  \BibitemOpen
  \bibfield  {author} {\bibinfo {author} {\bibfnamefont {F.~A.}\ \bibnamefont
  {Castro}}, \bibinfo {author} {\bibfnamefont {G.~B.}\ \bibnamefont {Silva}},
  \bibinfo {author} {\bibfnamefont {L.~F.}\ \bibnamefont {Santos}}, \bibinfo
  {author} {\bibfnamefont {R.~M.}\ \bibnamefont {Faria}}, \bibinfo {author}
  {\bibfnamefont {F.}~\bibnamefont {N$\textrm{\"{u}}$esch}}, \bibinfo {author}
  {\bibfnamefont {L.}~\bibnamefont {Zuppiroli}}, \ and\ \bibinfo {author}
  {\bibfnamefont {C.~F.~O.}\ \bibnamefont {Graeff}},\ }\href@noop {} {\bibfield
   {journal} {\bibinfo  {journal} {Journal of Non-Crystalline Solids}\ }\textbf
  {\bibinfo {volume} {338-340}},\ \bibinfo {pages} {622} (\bibinfo {year}
  {2004})}\BibitemShut {NoStop}%
\bibitem [{\citenamefont {Hiromitsu}\ \emph {et~al.}(1999)\citenamefont
  {Hiromitsu}, \citenamefont {Kaimori}, \citenamefont {Kitano},\ and\
  \citenamefont {Ito}}]{Hiromitsu:prb1999}%
  \BibitemOpen
  \bibfield  {author} {\bibinfo {author} {\bibfnamefont {I.}~\bibnamefont
  {Hiromitsu}}, \bibinfo {author} {\bibfnamefont {Y.}~\bibnamefont {Kaimori}},
  \bibinfo {author} {\bibfnamefont {M.}~\bibnamefont {Kitano}}, \ and\ \bibinfo
  {author} {\bibfnamefont {T.}~\bibnamefont {Ito}},\ }\href@noop {} {\bibfield
  {journal} {\bibinfo  {journal} {Physical Review B}\ }\textbf {\bibinfo
  {volume} {59}},\ \bibinfo {pages} {2151} (\bibinfo {year}
  {1999})}\BibitemShut {NoStop}%
\bibitem [{\citenamefont {Scharber}\ \emph {et~al.}(2003)\citenamefont
  {Scharber}, \citenamefont {Schultz}, \citenamefont {Sariciftci},\ and\
  \citenamefont {Brabec}}]{Scharber:prb2003}%
  \BibitemOpen
  \bibfield  {author} {\bibinfo {author} {\bibfnamefont {M.~C.}\ \bibnamefont
  {Scharber}}, \bibinfo {author} {\bibfnamefont {N.~A.}\ \bibnamefont
  {Schultz}}, \bibinfo {author} {\bibfnamefont {N.~S.}\ \bibnamefont
  {Sariciftci}}, \ and\ \bibinfo {author} {\bibfnamefont {C.~J.}\ \bibnamefont
  {Brabec}},\ }\href@noop {} {\bibfield  {journal} {\bibinfo  {journal}
  {Physical Review B}\ }\textbf {\bibinfo {volume} {67}},\ \bibinfo {pages}
  {085202} (\bibinfo {year} {2003})}\BibitemShut {NoStop}%
\bibitem [{\citenamefont {Nguyen}\ \emph {et~al.}(2010)\citenamefont {Nguyen},
  \citenamefont {Hukic-Markosian}, \citenamefont {Wang}, \citenamefont
  {Wojcik}, \citenamefont {Li}, \citenamefont {Ehrenfreund},\ and\
  \citenamefont {Vardeny}}]{Nguyen:nm2010}%
  \BibitemOpen
  \bibfield  {author} {\bibinfo {author} {\bibfnamefont {T.~D.}\ \bibnamefont
  {Nguyen}}, \bibinfo {author} {\bibfnamefont {G.}~\bibnamefont
  {Hukic-Markosian}}, \bibinfo {author} {\bibfnamefont {F.}~\bibnamefont
  {Wang}}, \bibinfo {author} {\bibfnamefont {L.}~\bibnamefont {Wojcik}},
  \bibinfo {author} {\bibfnamefont {X.-G.}\ \bibnamefont {Li}}, \bibinfo
  {author} {\bibfnamefont {E.}~\bibnamefont {Ehrenfreund}}, \ and\ \bibinfo
  {author} {\bibfnamefont {Z.~V.}\ \bibnamefont {Vardeny}},\ }\href@noop {}
  {\bibfield  {journal} {\bibinfo  {journal} {Nature Materials}\ }\textbf
  {\bibinfo {volume} {9}},\ \bibinfo {pages} {345} (\bibinfo {year}
  {2010})}\BibitemShut {NoStop}%
\bibitem [{\citenamefont {Yang}\ \emph
  {et~al.}(2007{\natexlab{a}})\citenamefont {Yang}, \citenamefont
  {Ehrenfreund},\ and\ \citenamefont {Vardeny}}]{Yang:prl2007}%
  \BibitemOpen
  \bibfield  {author} {\bibinfo {author} {\bibfnamefont {C.~G.}\ \bibnamefont
  {Yang}}, \bibinfo {author} {\bibfnamefont {E.}~\bibnamefont {Ehrenfreund}}, \
  and\ \bibinfo {author} {\bibfnamefont {Z.~V.}\ \bibnamefont {Vardeny}},\
  }\href@noop {} {\bibfield  {journal} {\bibinfo  {journal} {Physical Review
  Letters}\ }\textbf {\bibinfo {volume} {99}},\ \bibinfo {pages} {4} (\bibinfo
  {year} {2007}{\natexlab{a}})}\BibitemShut {NoStop}%
\bibitem [{\citenamefont {Dyakonov}\ and\ \citenamefont
  {Frankevich}(1998)}]{Dyakonov:cp1998}%
  \BibitemOpen
  \bibfield  {author} {\bibinfo {author} {\bibfnamefont {V.}~\bibnamefont
  {Dyakonov}}\ and\ \bibinfo {author} {\bibfnamefont {E.}~\bibnamefont
  {Frankevich}},\ }\href@noop {} {\bibfield  {journal} {\bibinfo  {journal}
  {Chemical Physics}\ }\textbf {\bibinfo {volume} {227}},\ \bibinfo {pages}
  {203} (\bibinfo {year} {1998})}\BibitemShut {NoStop}%
\bibitem [{\citenamefont {Lifshitz}\ \emph {et~al.}(1995)\citenamefont
  {Lifshitz}, \citenamefont {Bykov},\ and\ \citenamefont
  {Yassen}}]{Lifshitz:jpc1995}%
  \BibitemOpen
  \bibfield  {author} {\bibinfo {author} {\bibfnamefont {E.}~\bibnamefont
  {Lifshitz}}, \bibinfo {author} {\bibfnamefont {L.}~\bibnamefont {Bykov}}, \
  and\ \bibinfo {author} {\bibfnamefont {M.}~\bibnamefont {Yassen}},\
  }\href@noop {} {\bibfield  {journal} {\bibinfo  {journal} {The Journal of
  Physical Chemistry}\ }\textbf {\bibinfo {volume} {99}},\ \bibinfo {pages}
  {15262} (\bibinfo {year} {1995})}\BibitemShut {NoStop}%
\bibitem [{\citenamefont {List}\ \emph {et~al.}(2001)\citenamefont {List},
  \citenamefont {Kim}, \citenamefont {Naik}, \citenamefont {Scherf},
  \citenamefont {Leising}, \citenamefont {Graupner},\ and\ \citenamefont
  {Shinar}}]{List:prb2001}%
  \BibitemOpen
  \bibfield  {author} {\bibinfo {author} {\bibfnamefont {E.~J.~W.}\
  \bibnamefont {List}}, \bibinfo {author} {\bibfnamefont {C.~H.}\ \bibnamefont
  {Kim}}, \bibinfo {author} {\bibfnamefont {A.~K.}\ \bibnamefont {Naik}},
  \bibinfo {author} {\bibfnamefont {U.}~\bibnamefont {Scherf}}, \bibinfo
  {author} {\bibfnamefont {G.}~\bibnamefont {Leising}}, \bibinfo {author}
  {\bibfnamefont {W.}~\bibnamefont {Graupner}}, \ and\ \bibinfo {author}
  {\bibfnamefont {J.}~\bibnamefont {Shinar}},\ }\href@noop {} {\bibfield
  {journal} {\bibinfo  {journal} {Physical Review B}\ }\textbf {\bibinfo
  {volume} {64}},\ \bibinfo {pages} {155204} (\bibinfo {year}
  {2001})}\BibitemShut {NoStop}%
\bibitem [{\citenamefont {Dyakonov}\ \emph {et~al.}(1997)\citenamefont
  {Dyakonov}, \citenamefont {R$\textrm{\"{o}}$sler}, \citenamefont
  {Schwoerer},\ and\ \citenamefont {Frankevich}}]{Dyakonov:prb1997}%
  \BibitemOpen
  \bibfield  {author} {\bibinfo {author} {\bibfnamefont {V.}~\bibnamefont
  {Dyakonov}}, \bibinfo {author} {\bibfnamefont {G.}~\bibnamefont
  {R$\textrm{\"{o}}$sler}}, \bibinfo {author} {\bibfnamefont {M.}~\bibnamefont
  {Schwoerer}}, \ and\ \bibinfo {author} {\bibfnamefont {E.~L.}\ \bibnamefont
  {Frankevich}},\ }\href@noop {} {\bibfield  {journal} {\bibinfo  {journal}
  {Physical Review B}\ }\textbf {\bibinfo {volume} {56}},\ \bibinfo {pages}
  {3852} (\bibinfo {year} {1997})}\BibitemShut {NoStop}%
\bibitem [{\citenamefont {Lifshitz}\ and\ \citenamefont
  {Bykov}(1993)}]{Lifshitz:jpc1993}%
  \BibitemOpen
  \bibfield  {author} {\bibinfo {author} {\bibfnamefont {E.}~\bibnamefont
  {Lifshitz}}\ and\ \bibinfo {author} {\bibfnamefont {L.}~\bibnamefont
  {Bykov}},\ }\href@noop {} {\bibfield  {journal} {\bibinfo  {journal} {The
  Journal of Physical Chemistry}\ }\textbf {\bibinfo {volume} {97}},\ \bibinfo
  {pages} {9288} (\bibinfo {year} {1993})}\BibitemShut {NoStop}%
\bibitem [{\citenamefont {Stich}\ \emph {et~al.}(1995)\citenamefont {Stich},
  \citenamefont {Greulich-Weber},\ and\ \citenamefont
  {Spaeth}}]{Stich:jap1995}%
  \BibitemOpen
  \bibfield  {author} {\bibinfo {author} {\bibfnamefont {B.}~\bibnamefont
  {Stich}}, \bibinfo {author} {\bibfnamefont {S.}~\bibnamefont
  {Greulich-Weber}}, \ and\ \bibinfo {author} {\bibfnamefont {J.~M.}\
  \bibnamefont {Spaeth}},\ }\href@noop {} {\bibfield  {journal} {\bibinfo
  {journal} {Journal of Applied Physics}\ }\textbf {\bibinfo {volume} {77}},\
  \bibinfo {pages} {1546} (\bibinfo {year} {1995})}\BibitemShut {NoStop}%
\bibitem [{\citenamefont {Biegelsen}\ \emph {et~al.}(1978)\citenamefont
  {Biegelsen}, \citenamefont {Knights}, \citenamefont {Street}, \citenamefont
  {Tsang},\ and\ \citenamefont {White}}]{Biegelsen:pmb1978}%
  \BibitemOpen
  \bibfield  {author} {\bibinfo {author} {\bibfnamefont {D.~K.}\ \bibnamefont
  {Biegelsen}}, \bibinfo {author} {\bibfnamefont {J.~C.}\ \bibnamefont
  {Knights}}, \bibinfo {author} {\bibfnamefont {R.~A.}\ \bibnamefont {Street}},
  \bibinfo {author} {\bibfnamefont {C.}~\bibnamefont {Tsang}}, \ and\ \bibinfo
  {author} {\bibfnamefont {R.~M.}\ \bibnamefont {White}},\ }\href@noop {}
  {\bibfield  {journal} {\bibinfo  {journal} {Philosophical Magazine Part B}\
  }\textbf {\bibinfo {volume} {37}},\ \bibinfo {pages} {477} (\bibinfo {year}
  {1978})}\BibitemShut {NoStop}%
\bibitem [{\citenamefont {Xiong}\ and\ \citenamefont
  {Miller}(1993)}]{Xiong:apl1993}%
  \BibitemOpen
  \bibfield  {author} {\bibinfo {author} {\bibfnamefont {Z.}~\bibnamefont
  {Xiong}}\ and\ \bibinfo {author} {\bibfnamefont {D.~J.}\ \bibnamefont
  {Miller}},\ }\href@noop {} {\bibfield  {journal} {\bibinfo  {journal}
  {Applied Physics Letters}\ }\textbf {\bibinfo {volume} {63}},\ \bibinfo
  {pages} {352} (\bibinfo {year} {1993})}\BibitemShut {NoStop}%
\bibitem [{\citenamefont {Kawachi}\ \emph {et~al.}(1997)\citenamefont
  {Kawachi}, \citenamefont {Graeff}, \citenamefont {Brandt},\ and\
  \citenamefont {Stutzmann}}]{Kawachi:jjap1997}%
  \BibitemOpen
  \bibfield  {author} {\bibinfo {author} {\bibfnamefont {G.}~\bibnamefont
  {Kawachi}}, \bibinfo {author} {\bibfnamefont {C.~F.~O.}\ \bibnamefont
  {Graeff}}, \bibinfo {author} {\bibfnamefont {M.~S.}\ \bibnamefont {Brandt}},
  \ and\ \bibinfo {author} {\bibfnamefont {M.}~\bibnamefont {Stutzmann}},\
  }\href@noop {} {\bibfield  {journal} {\bibinfo  {journal} {Japanese Journal
  of Applied Physics}\ }\textbf {\bibinfo {volume} {36}},\ \bibinfo {pages} {5}
  (\bibinfo {year} {1997})}\BibitemShut {NoStop}%
\bibitem [{\citenamefont {Baldo}\ \emph {et~al.}(2007)\citenamefont {Baldo},
  \citenamefont {Segal}, \citenamefont {Shinar},\ and\ \citenamefont
  {Soos}}]{Baldo:prb2007}%
  \BibitemOpen
  \bibfield  {author} {\bibinfo {author} {\bibfnamefont {M.~A.}\ \bibnamefont
  {Baldo}}, \bibinfo {author} {\bibfnamefont {M.}~\bibnamefont {Segal}},
  \bibinfo {author} {\bibfnamefont {J.}~\bibnamefont {Shinar}}, \ and\ \bibinfo
  {author} {\bibfnamefont {Z.~G.}\ \bibnamefont {Soos}},\ }\href@noop {}
  {\bibfield  {journal} {\bibinfo  {journal} {Physical Review B}\ }\textbf
  {\bibinfo {volume} {75}},\ \bibinfo {pages} {3} (\bibinfo {year}
  {2007})}\BibitemShut {NoStop}%
\bibitem [{\citenamefont {Mur$\textrm{\'{a}}$nyi}\ \emph
  {et~al.}(2004)\citenamefont {Mur$\textrm{\'{a}}$nyi}, \citenamefont {Simon},
  \citenamefont {F$\textrm{\"{u}}$l$\textrm{\"{o}}$p},\ and\ \citenamefont
  {J$\textrm{\'{a}}$nossy}}]{Murany:jmr2004}%
  \BibitemOpen
  \bibfield  {author} {\bibinfo {author} {\bibfnamefont {F.}~\bibnamefont
  {Mur$\textrm{\'{a}}$nyi}}, \bibinfo {author} {\bibfnamefont {F.}~\bibnamefont
  {Simon}}, \bibinfo {author} {\bibfnamefont {F.}~\bibnamefont
  {F$\textrm{\"{u}}$l$\textrm{\"{o}}$p}}, \ and\ \bibinfo {author}
  {\bibfnamefont {A.}~\bibnamefont {J$\textrm{\'{a}}$nossy}},\ }\href@noop {}
  {\bibfield  {journal} {\bibinfo  {journal} {Journal of Magnetic Resonance}\
  }\textbf {\bibinfo {volume} {167}},\ \bibinfo {pages} {221} (\bibinfo {year}
  {2004})}\BibitemShut {NoStop}%
\bibitem [{\citenamefont {Simon}\ \emph {et~al.}(2007)\citenamefont {Simon},
  \citenamefont {Mur$\textrm{\'{a}}$nyi}, \citenamefont {Feh$\textrm{\'{e}}$r},
  \citenamefont {J$\textrm{\'{a}}$nossy}, \citenamefont {Forr$\textrm{\'{o}}$},
  \citenamefont {Petrovic}, \citenamefont {Budko},\ and\ \citenamefont
  {Canfield}}]{Simon:prb2007}%
  \BibitemOpen
  \bibfield  {author} {\bibinfo {author} {\bibfnamefont {F.}~\bibnamefont
  {Simon}}, \bibinfo {author} {\bibfnamefont {F.}~\bibnamefont
  {Mur$\textrm{\'{a}}$nyi}}, \bibinfo {author} {\bibfnamefont {T.}~\bibnamefont
  {Feh$\textrm{\'{e}}$r}}, \bibinfo {author} {\bibfnamefont {A.}~\bibnamefont
  {J$\textrm{\'{a}}$nossy}}, \bibinfo {author} {\bibfnamefont {L.}~\bibnamefont
  {Forr$\textrm{\'{o}}$}}, \bibinfo {author} {\bibfnamefont {C.}~\bibnamefont
  {Petrovic}}, \bibinfo {author} {\bibfnamefont {S.~L.}\ \bibnamefont {Budko}},
  \ and\ \bibinfo {author} {\bibfnamefont {P.~C.}\ \bibnamefont {Canfield}},\
  }\href@noop {} {\bibfield  {journal} {\bibinfo  {journal} {Physical Review
  B}\ }\textbf {\bibinfo {volume} {76}},\ \bibinfo {pages} {024519} (\bibinfo
  {year} {2007})}\BibitemShut {NoStop}%
\bibitem [{\citenamefont {Yang}\ \emph
  {et~al.}(2007{\natexlab{b}})\citenamefont {Yang}, \citenamefont
  {Ehrenfreund}, \citenamefont {Wohlgenannt},\ and\ \citenamefont
  {Vardeny}}]{Yang:prb2007}%
  \BibitemOpen
  \bibfield  {author} {\bibinfo {author} {\bibfnamefont {C.~G.}\ \bibnamefont
  {Yang}}, \bibinfo {author} {\bibfnamefont {E.}~\bibnamefont {Ehrenfreund}},
  \bibinfo {author} {\bibfnamefont {M.}~\bibnamefont {Wohlgenannt}}, \ and\
  \bibinfo {author} {\bibfnamefont {Z.~V.}\ \bibnamefont {Vardeny}},\
  }\href@noop {} {\bibfield  {journal} {\bibinfo  {journal} {Physical Review
  B}\ }\textbf {\bibinfo {volume} {75}},\ \bibinfo {pages} {5} (\bibinfo {year}
  {2007}{\natexlab{b}})}\BibitemShut {NoStop}%
\bibitem [{\citenamefont {Depinna}(1983)}]{Depinna:prb1983}%
  \BibitemOpen
  \bibfield  {author} {\bibinfo {author} {\bibfnamefont {S.~P.}\ \bibnamefont
  {Depinna}},\ }\href@noop {} {\bibfield  {journal} {\bibinfo  {journal}
  {Physical Review B}\ }\textbf {\bibinfo {volume} {28}},\ \bibinfo {pages}
  {5327} (\bibinfo {year} {1983})}\BibitemShut {NoStop}%
\bibitem [{\citenamefont {Lifshitz}\ \emph {et~al.}(1998)\citenamefont
  {Lifshitz}, \citenamefont {Litvin}, \citenamefont {Porteanu},\ and\
  \citenamefont {Lipovskii}}]{Lifshitz:cpl1998}%
  \BibitemOpen
  \bibfield  {author} {\bibinfo {author} {\bibfnamefont {E.}~\bibnamefont
  {Lifshitz}}, \bibinfo {author} {\bibfnamefont {I.~D.}\ \bibnamefont
  {Litvin}}, \bibinfo {author} {\bibfnamefont {H.}~\bibnamefont {Porteanu}}, \
  and\ \bibinfo {author} {\bibfnamefont {A.~A.}\ \bibnamefont {Lipovskii}},\
  }\href@noop {} {\bibfield  {journal} {\bibinfo  {journal} {Chemical Physics
  Letters}\ }\textbf {\bibinfo {volume} {295}},\ \bibinfo {pages} {249}
  (\bibinfo {year} {1998})}\BibitemShut {NoStop}%
\bibitem [{\citenamefont {Carlos}\ and\ \citenamefont
  {Nakamura}(1998)}]{Carlos:jcg1998}%
  \BibitemOpen
  \bibfield  {author} {\bibinfo {author} {\bibfnamefont {W.~E.}\ \bibnamefont
  {Carlos}}\ and\ \bibinfo {author} {\bibfnamefont {S.}~\bibnamefont
  {Nakamura}},\ }\href@noop {} {\bibfield  {journal} {\bibinfo  {journal}
  {Journal of Crystal Growth}\ }\textbf {\bibinfo {volume} {189-190}},\
  \bibinfo {pages} {794} (\bibinfo {year} {1998})}\BibitemShut {NoStop}%
\bibitem [{\citenamefont {Kaplan}\ \emph {et~al.}(1978)\citenamefont {Kaplan},
  \citenamefont {Solomon},\ and\ \citenamefont {Mott}}]{Kaplan:jpl1978}%
  \BibitemOpen
  \bibfield  {author} {\bibinfo {author} {\bibfnamefont {D.}~\bibnamefont
  {Kaplan}}, \bibinfo {author} {\bibfnamefont {I.}~\bibnamefont {Solomon}}, \
  and\ \bibinfo {author} {\bibfnamefont {N.~F.}\ \bibnamefont {Mott}},\
  }\href@noop {} {\bibfield  {journal} {\bibinfo  {journal} {J. Physique
  Lett.}\ }\textbf {\bibinfo {volume} {39}},\ \bibinfo {pages} {51} (\bibinfo
  {year} {1978})}\BibitemShut {NoStop}%
\bibitem [{\citenamefont {Brandt}\ \emph {et~al.}(1998)\citenamefont {Brandt},
  \citenamefont {Bayerl}, \citenamefont {Stutzmann},\ and\ \citenamefont
  {Graeff}}]{Brandt:jncs1998}%
  \BibitemOpen
  \bibfield  {author} {\bibinfo {author} {\bibfnamefont {M.~S.}\ \bibnamefont
  {Brandt}}, \bibinfo {author} {\bibfnamefont {M.~W.}\ \bibnamefont {Bayerl}},
  \bibinfo {author} {\bibfnamefont {M.}~\bibnamefont {Stutzmann}}, \ and\
  \bibinfo {author} {\bibfnamefont {C.~F.~O.}\ \bibnamefont {Graeff}},\
  }\href@noop {} {\bibfield  {journal} {\bibinfo  {journal} {Journal of
  Non-Crystalline Solids}\ }\textbf {\bibinfo {volume} {227-230}},\ \bibinfo
  {pages} {343} (\bibinfo {year} {1998})}\BibitemShut {NoStop}%
\bibitem [{\citenamefont {Stutzmann}\ and\ \citenamefont
  {Brandt}(1992)}]{Stutzmann:jncs1992}%
  \BibitemOpen
  \bibfield  {author} {\bibinfo {author} {\bibfnamefont {M.}~\bibnamefont
  {Stutzmann}}\ and\ \bibinfo {author} {\bibfnamefont {M.~S.}\ \bibnamefont
  {Brandt}},\ }\href@noop {} {\bibfield  {journal} {\bibinfo  {journal}
  {Journal of Non-Crystalline Solids}\ }\textbf {\bibinfo {volume} {141}},\
  \bibinfo {pages} {97} (\bibinfo {year} {1992})}\BibitemShut {NoStop}%
\bibitem [{\citenamefont {Boehme}\ and\ \citenamefont
  {Lips}(2003)}]{Boehme:prb2003}%
  \BibitemOpen
  \bibfield  {author} {\bibinfo {author} {\bibfnamefont {C.}~\bibnamefont
  {Boehme}}\ and\ \bibinfo {author} {\bibfnamefont {K.}~\bibnamefont {Lips}},\
  }\href@noop {} {\bibfield  {journal} {\bibinfo  {journal} {Physical Review
  B}\ }\textbf {\bibinfo {volume} {68}},\ \bibinfo {pages} {245105} (\bibinfo
  {year} {2003})}\BibitemShut {NoStop}%
\bibitem [{\citenamefont {McNaught}\ \emph {et~al.}(1997)\citenamefont
  {McNaught}, \citenamefont {Wilkinson}, \citenamefont {of~Pure},\ and\
  \citenamefont {Chemistry.}}]{McNaught:book1997}%
  \BibitemOpen
  \bibfield  {author} {\bibinfo {author} {\bibfnamefont {A.~D.}\ \bibnamefont
  {McNaught}}, \bibinfo {author} {\bibfnamefont {A.}~\bibnamefont {Wilkinson}},
  \bibinfo {author} {\bibfnamefont {I.~U.}\ \bibnamefont {of~Pure}}, \ and\
  \bibinfo {author} {\bibfnamefont {A.}~\bibnamefont {Chemistry.}},\
  }\href@noop {} {\emph {\bibinfo {title} {Compendium of chemical terminology :
  IUPAC recommendations}}},\ \bibinfo {edition} {2nd}\ ed.\ (\bibinfo
  {publisher} {Blackwell Science},\ \bibinfo {address} {Oxford England ;
  Malden, MA, USA},\ \bibinfo {year} {1997})\BibitemShut {NoStop}%
\bibitem [{\citenamefont {Eickelkamp}\ \emph {et~al.}(1998)\citenamefont
  {Eickelkamp}, \citenamefont {Roth},\ and\ \citenamefont
  {Mehring}}]{Eickelkamp:mp1998}%
  \BibitemOpen
  \bibfield  {author} {\bibinfo {author} {\bibfnamefont {T.}~\bibnamefont
  {Eickelkamp}}, \bibinfo {author} {\bibfnamefont {S.}~\bibnamefont {Roth}}, \
  and\ \bibinfo {author} {\bibfnamefont {M.}~\bibnamefont {Mehring}},\
  }\href@noop {} {\bibfield  {journal} {\bibinfo  {journal} {Molecular Physics:
  An International Journal at the Interface Between Chemistry and Physics}\
  }\textbf {\bibinfo {volume} {95}},\ \bibinfo {pages} {967} (\bibinfo {year}
  {1998})}\BibitemShut {NoStop}%
\bibitem [{\citenamefont {Aoki}(2006)}]{Aoki:jncs2006}%
  \BibitemOpen
  \bibfield  {author} {\bibinfo {author} {\bibfnamefont {T.}~\bibnamefont
  {Aoki}},\ }\href@noop {} {\bibfield  {journal} {\bibinfo  {journal} {Journal
  of Non-Crystalline Solids}\ }\textbf {\bibinfo {volume} {352}},\ \bibinfo
  {pages} {1138} (\bibinfo {year} {2006})}\BibitemShut {NoStop}%
\bibitem [{\citenamefont {Mihailetchi}\ \emph {et~al.}(2006)\citenamefont
  {Mihailetchi}, \citenamefont {Xie}, \citenamefont {Boer}, \citenamefont
  {Koster},\ and\ \citenamefont {Blom}}]{Mihailetch:afm2006}%
  \BibitemOpen
  \bibfield  {author} {\bibinfo {author} {\bibfnamefont {V.~D.}\ \bibnamefont
  {Mihailetchi}}, \bibinfo {author} {\bibfnamefont {H.}~\bibnamefont {Xie}},
  \bibinfo {author} {\bibfnamefont {B.~d.}\ \bibnamefont {Boer}}, \bibinfo
  {author} {\bibfnamefont {L.~J.~A.}\ \bibnamefont {Koster}}, \ and\ \bibinfo
  {author} {\bibfnamefont {P.~W.~M.}\ \bibnamefont {Blom}},\ }\href@noop {}
  {\bibfield  {journal} {\bibinfo  {journal} {Advanced Functional Materials}\
  }\textbf {\bibinfo {volume} {16}},\ \bibinfo {pages} {699} (\bibinfo {year}
  {2006})}\BibitemShut {NoStop}%
\bibitem [{\citenamefont {Romanovskii}\ \emph {et~al.}(1999)\citenamefont
  {Romanovskii}, \citenamefont {Gerhard}, \citenamefont {Schweitzer},
  \citenamefont {Personov},\ and\ \citenamefont
  {B$\textrm{\"{a}}$ssler}}]{Romanovski:cp1999}%
  \BibitemOpen
  \bibfield  {author} {\bibinfo {author} {\bibfnamefont {Y.~V.}\ \bibnamefont
  {Romanovskii}}, \bibinfo {author} {\bibfnamefont {A.}~\bibnamefont
  {Gerhard}}, \bibinfo {author} {\bibfnamefont {B.}~\bibnamefont {Schweitzer}},
  \bibinfo {author} {\bibfnamefont {R.~I.}\ \bibnamefont {Personov}}, \ and\
  \bibinfo {author} {\bibfnamefont {H.}~\bibnamefont
  {B$\textrm{\"{a}}$ssler}},\ }\href@noop {} {\bibfield  {journal} {\bibinfo
  {journal} {Chemical Physics}\ }\textbf {\bibinfo {volume} {249}},\ \bibinfo
  {pages} {29} (\bibinfo {year} {1999})}\BibitemShut {NoStop}%
\bibitem [{\citenamefont {Tachibana}\ \emph {et~al.}(1996)\citenamefont
  {Tachibana}, \citenamefont {Moser}, \citenamefont {Gratzel}, \citenamefont
  {Klug},\ and\ \citenamefont {Durrant}}]{Tachibana:jpc1996}%
  \BibitemOpen
  \bibfield  {author} {\bibinfo {author} {\bibfnamefont {Y.}~\bibnamefont
  {Tachibana}}, \bibinfo {author} {\bibfnamefont {J.~E.}\ \bibnamefont
  {Moser}}, \bibinfo {author} {\bibfnamefont {M.}~\bibnamefont {Gratzel}},
  \bibinfo {author} {\bibfnamefont {D.~R.}\ \bibnamefont {Klug}}, \ and\
  \bibinfo {author} {\bibfnamefont {J.~R.}\ \bibnamefont {Durrant}},\
  }\href@noop {} {\bibfield  {journal} {\bibinfo  {journal} {The Journal of
  Physical Chemistry}\ }\textbf {\bibinfo {volume} {100}},\ \bibinfo {pages}
  {20056} (\bibinfo {year} {1996})}\BibitemShut {NoStop}%
\bibitem [{\citenamefont {Nollau}\ \emph {et~al.}(2000)\citenamefont {Nollau},
  \citenamefont {Hoffmann}, \citenamefont {Fritz},\ and\ \citenamefont
  {Leo}}]{Nollau:tsf2000}%
  \BibitemOpen
  \bibfield  {author} {\bibinfo {author} {\bibfnamefont {A.}~\bibnamefont
  {Nollau}}, \bibinfo {author} {\bibfnamefont {M.}~\bibnamefont {Hoffmann}},
  \bibinfo {author} {\bibfnamefont {T.}~\bibnamefont {Fritz}}, \ and\ \bibinfo
  {author} {\bibfnamefont {K.}~\bibnamefont {Leo}},\ }\href@noop {} {\bibfield
  {journal} {\bibinfo  {journal} {Thin Solid Films}\ }\textbf {\bibinfo
  {volume} {368}},\ \bibinfo {pages} {130} (\bibinfo {year}
  {2000})}\BibitemShut {NoStop}%
\bibitem [{\citenamefont {McCamey}\ \emph {et~al.}(2010)\citenamefont
  {McCamey}, \citenamefont {Lee}, \citenamefont {Paik}, \citenamefont
  {Lupton},\ and\ \citenamefont {Boehme}}]{McCamey:prb2010}%
  \BibitemOpen
  \bibfield  {author} {\bibinfo {author} {\bibfnamefont {D.~R.}\ \bibnamefont
  {McCamey}}, \bibinfo {author} {\bibfnamefont {S.~Y.}\ \bibnamefont {Lee}},
  \bibinfo {author} {\bibfnamefont {S.~Y.}\ \bibnamefont {Paik}}, \bibinfo
  {author} {\bibfnamefont {J.~M.}\ \bibnamefont {Lupton}}, \ and\ \bibinfo
  {author} {\bibfnamefont {C.}~\bibnamefont {Boehme}},\ }\href@noop {}
  {\bibfield  {journal} {\bibinfo  {journal} {Physical Review B}\ }\textbf
  {\bibinfo {volume} {82}},\ \bibinfo {pages} {125206} (\bibinfo {year}
  {2010})}\BibitemShut {NoStop}%
\bibitem [{\citenamefont {Street}(1991)}]{Street1991}%
  \BibitemOpen
  \bibfield  {author} {\bibinfo {author} {\bibfnamefont {R.~A.}\ \bibnamefont
  {Street}},\ }\href@noop {} {\emph {\bibinfo {title} {Hydrogenated Amorphous
  Silicon}}}\ (\bibinfo  {publisher} {Cambridge University Press},\ \bibinfo
  {address} {New York},\ \bibinfo {year} {1991})\BibitemShut {NoStop}%
\bibitem [{\citenamefont {Pankove}(1971)}]{Pankove:1971}%
  \BibitemOpen
  \bibfield  {author} {\bibinfo {author} {\bibfnamefont {J.~I.}\ \bibnamefont
  {Pankove}},\ }\href@noop {} {\emph {\bibinfo {title} {Optical processes in
  semiconductors}}}\ (\bibinfo  {publisher} {Dover Publications},\ \bibinfo
  {address} {New York},\ \bibinfo {year} {1971})\BibitemShut {NoStop}%
\bibitem [{\citenamefont {Bradley}\ and\ \citenamefont
  {Friend}(1989)}]{Bradley:jpcm1981}%
  \BibitemOpen
  \bibfield  {author} {\bibinfo {author} {\bibfnamefont {D.~D.~C.}\
  \bibnamefont {Bradley}}\ and\ \bibinfo {author} {\bibfnamefont {R.~H.}\
  \bibnamefont {Friend}},\ }\href@noop {} {\bibfield  {journal} {\bibinfo
  {journal} {Journal of Physics: Condensed Matter}\ }\textbf {\bibinfo {volume}
  {1}},\ \bibinfo {pages} {3671} (\bibinfo {year} {1989})}\BibitemShut
  {NoStop}%
\bibitem [{\citenamefont {Wohlgenannt}\ \emph {et~al.}(2002)\citenamefont
  {Wohlgenannt}, \citenamefont {Yang},\ and\ \citenamefont
  {Vardeny}}]{Wohlgenannt:prb2002}%
  \BibitemOpen
  \bibfield  {author} {\bibinfo {author} {\bibfnamefont {M.}~\bibnamefont
  {Wohlgenannt}}, \bibinfo {author} {\bibfnamefont {C.}~\bibnamefont {Yang}}, \
  and\ \bibinfo {author} {\bibfnamefont {Z.~V.}\ \bibnamefont {Vardeny}},\
  }\href@noop {} {\bibfield  {journal} {\bibinfo  {journal} {Physical Review
  B}\ }\textbf {\bibinfo {volume} {66}},\ \bibinfo {pages} {4} (\bibinfo {year}
  {2002})}\BibitemShut {NoStop}%
\bibitem [{\citenamefont {McCamey}\ \emph {et~al.}(2008)\citenamefont
  {McCamey}, \citenamefont {Seipel}, \citenamefont {Paik}, \citenamefont
  {Walter}, \citenamefont {Borys}, \citenamefont {Lupton},\ and\ \citenamefont
  {Boehme}}]{McCamey:nm2008}%
  \BibitemOpen
  \bibfield  {author} {\bibinfo {author} {\bibfnamefont {D.~R.}\ \bibnamefont
  {McCamey}}, \bibinfo {author} {\bibfnamefont {H.~A.}\ \bibnamefont {Seipel}},
  \bibinfo {author} {\bibfnamefont {S.-Y.}\ \bibnamefont {Paik}}, \bibinfo
  {author} {\bibfnamefont {M.~J.}\ \bibnamefont {Walter}}, \bibinfo {author}
  {\bibfnamefont {N.~J.}\ \bibnamefont {Borys}}, \bibinfo {author}
  {\bibfnamefont {J.~M.}\ \bibnamefont {Lupton}}, \ and\ \bibinfo {author}
  {\bibfnamefont {C.}~\bibnamefont {Boehme}},\ }\href@noop {} {\bibfield
  {journal} {\bibinfo  {journal} {Nature Materials}\ }\textbf {\bibinfo
  {volume} {7}},\ \bibinfo {pages} {723} (\bibinfo {year} {2008})}\BibitemShut
  {NoStop}%
\bibitem [{\citenamefont {Boehme}\ and\ \citenamefont
  {Lips}(2001)}]{Boehme:apl2001}%
  \BibitemOpen
  \bibfield  {author} {\bibinfo {author} {\bibfnamefont {C.}~\bibnamefont
  {Boehme}}\ and\ \bibinfo {author} {\bibfnamefont {K.}~\bibnamefont {Lips}},\
  }\href@noop {} {\bibfield  {journal} {\bibinfo  {journal} {Applied Physics
  Letters}\ }\textbf {\bibinfo {volume} {79}},\ \bibinfo {pages} {4363}
  (\bibinfo {year} {2001})}\BibitemShut {NoStop}%
\bibitem [{\citenamefont {Atkins}\ and\ \citenamefont
  {Friedman}(1999)}]{MQM:Atkins}%
  \BibitemOpen
  \bibfield  {author} {\bibinfo {author} {\bibfnamefont {P.~W.}\ \bibnamefont
  {Atkins}}\ and\ \bibinfo {author} {\bibfnamefont {R.~S.}\ \bibnamefont
  {Friedman}},\ }\href@noop {} {\emph {\bibinfo {title} {Molecular quantum
  mechanics}}},\ \bibinfo {edition} {3rd}\ ed.\ (\bibinfo  {publisher} {Oxford
  University Press},\ \bibinfo {address} {Oxford},\ \bibinfo {year}
  {1999})\BibitemShut {NoStop}%
\end{thebibliography}
%

\end{document}